\documentclass[10pt]{article}

\usepackage[super,compress]{cite}
\usepackage{hyperref}
\usepackage{breakurl}
\usepackage{amsfonts}
\usepackage{amscd}
\usepackage{amssymb}
\usepackage{amsmath,bm}

\usepackage{graphicx} 

\usepackage{mathrsfs}

\usepackage{url}
\usepackage{youngtab}

\newcommand{\slashed}{{\bf\not}}

\numberwithin{equation}{section}

\newcommand{\nn}{\nonumber}
\newcommand{\mat}[1]{\begin{pmatrix} #1 \end{pmatrix}}

\newcommand{\be}{\begin{equation}} 
\newcommand{\ee}{\end{equation}}
\newcommand{\bea}{\begin{equation} \begin{aligned}} \newcommand{\eea}{\end{aligned} \end{equation}}

\newcommand{\bit}{\begin{itemize}} 
\newcommand{\eit}{\end{itemize}}

\newcommand{\cN}{\mathcal{N}}

\newcommand{\Q}{\mathbb{Q}}
\newcommand{\Z}{\mathbb{Z}}
\newcommand{\C}{\mathbb{C}}
\newcommand{\R}{\mathbb{R}}

\renewcommand{\t}{\widetilde }
\renewcommand{\d}{\partial }

\renewcommand{\b}{\bar }

\newcommand{\half}{{1\over 2}}

\newcommand{\bz}{{\b z}}

\newcommand{\CA}{\mathcal{A}}
\newcommand{\CB}{\mathcal{B}}
\newcommand{\CC}{\mathcal{C}}

\newcommand{\CF}{\mathcal{F}}
\newcommand{\CG}{\mathcal{G}}
\newcommand{\CH}{\mathcal{H}}
\newcommand{\CI}{\mathcal{I}}

\newcommand{\CK}{\mathcal{K}}
\newcommand{\CL}{\mathcal{L}}
\newcommand{\CM}{\mathcal{M}}
\newcommand{\CN}{\mathcal{N}}
\newcommand{\CO}{\mathcal{O}}

\newcommand{\CQ}{\mathcal{Q}}

\newcommand{\CS}{\mathcal{S}}
\newcommand{\CT}{\mathcal{T}}
\newcommand{\CU}{\mathcal{U}}
\newcommand{\CV}{\mathcal{V}}
\newcommand{\CW}{\mathcal{W}}

\newcommand{\FR}{\mathfrak{R}}
\newcommand{\Fg}{\mathfrak{g}}

\newcommand{\GG}{\mathbf{G}}

\newcommand{\rk}{{{\rm rk}(\GG)}}

\newcommand{\m}{\mathfrak{m}}
\newcommand{\n}{\mathfrak{n}}

\newcommand{\ba}{{\b a}}
\newcommand{\bb}{{\bb b}}

\newcommand{\h}{\hat}

\newcommand{\oneloop}{\text{1-loop}}

\newcommand{\fr}{r } 

\DeclareMathOperator{\Tr}{Tr}
\DeclareMathOperator{\tr}{tr}

\DeclareMathOperator{\sign}{sign}

\newcommand{\eps}{\epsilon}

\newcommand{\ep}{\varepsilon}

\newcommand{\SL}{{\mathscr L}}

\newcommand{\ov}{\over}

\newcommand{\fM}{\mathfrak{M}}

\newcommand{\dilog}{{\text{Li}_2}}

\newcommand{\Pic}{{\rm Pic}}
\newcommand{\bd}{{\bf d}}

\begin{document}
\markboth{Cyril Closset and Heeyeon Kim}{Three-dimensional $\CN=2$ gauge theories on Seifert manifolds}

\title{Three-dimensional $\CN=2$ supersymmetric gauge theories\\ and partition functions on Seifert manifolds: A review}

\author{Cyril Closset and Heeyeon Kim}

\date{\it\small{Mathematical Institute, University of Oxford\\ Woodstock Road, Oxford, OX2 6GG, United Kingdom}}


\maketitle

\begin{abstract}
We give a pedagogical introduction to the study of supersymmetric partition functions of 3D $\mathcal{N}{=}2$ supersymmetric Chern-Simons-matter theories (with an $R$-symmetry) on half-BPS closed three-manifolds---including $S^3$, $S^2 \times S^1$, and any Seifert three-manifold. Three-dimensional gauge theories can flow to non-trivial fixed points in the infrared. In the presence of 3D $\mathcal{N}{=}2$ supersymmetry, many exact results are known about the strongly-coupled infrared, due in good part to powerful localization techniques. We review some of these  techniques and emphasize some more recent developments, which provide a simple and comprehensive formalism for the exact computation of half-BPS observables on closed three-manifolds (partition functions and correlation functions of line operators). Along the way, we also review simple examples of 3D infrared dualities. The computation of supersymmetric partition functions provides exceedingly precise tests of these dualities.\\
\\ {\it Review for the International Journal of Modern Physics A.}
\end{abstract}

\tableofcontents

\section{Introduction}	
Supersymmetry allows us to perform exact computations of many observables in quantum field theory (and in string theory). This, of course, has been recurring theme in high-energy theoretical physics for many years. When considering the relevant supersymmetric observables from the path integral point of view, the common phenomenon is that of {\it supersymmetric localization,} which is the very non-trivial statement that the saddle point approximation to a given observable is actually the exact answer. Morally speaking, supersymmetric localization in quantum field theory (QFT) is an infinite-dimensional path-integral generalization of rigorous mathematical results such as the Atiyah-Bott formula for certain finite-dimensional equivariant integrals.~\cite{ATIYAH19841, Witten:1988xj, Witten:1991zz}

In these notes, we review a number of interesting ``localization-related results'' in three-dimensional gauge theories with $\CN=2$ supersymmetry and a $U(1)_R$ $R$-symmetry. These supersymmetric QFTs are defined by a free Lagrangian in the ultraviolet (UV), and they often flow to strongly-coupled 3D $\CN=2$ superconformal field theories (SCFTs) in the infrared (IR)---one important example, for instance, is the SCFT that lives at low energy on a stack of M2-branes in M-theory.~\cite{Aharony:2008ug}

Our main interest will be in {\it supersymmetric partition functions on closed three-manifolds}, $\CM_3$, which are simply defined as path integrals of the 3D $\CN=2$ gauge theory on $\CM_3$:
\be\label{ZM3 intro}
Z_{\CM_3}(y)= \int [D\varphi]_{\CM_3} \, e^{-S[\varphi; y]}~.
\ee
These objects depend on the choice of three-manifold $\CM_3$, as well on other discrete choices consistent with supersymmetry. They also depend analytically on some continuous complex parameters, $y$, which couple to the conserved flavor symmetry currents of the 3D $\CN=2$ QFT.

The study of the 3D partition functions \eqref{ZM3 intro}  was initiated 10 years ago by Kapustin, Willett and Yaakov,~\cite{Kapustin:2009kz} who computed the three-sphere partition function, $Z_{S^3}$, for theories with at least $\CN=3$ supersymmetry. Various generalizations soon followed.~\cite{Jafferis:2010un,Hama:2010av,Hama:2011ea,Imamura:2011wg}
Around the same time, S.~Kim initiated the study of the 3D superconformal index,~\cite{Kim:2009wb} which corresponds to a supersymmetric partition function on $S^2 \times S^1$.~\cite{Imamura:2011su,Kapustin:2011jm}
See also Refs.~\citen{Beem:2012mb,Alday:2013lba,Yoshida:2014ssa,Gukov:2015sna,Benini:2015noa,Assel:2016pgi,Benini:2016hjo,Closset:2016arn} for related developments.
 The explicit computation of \eqref{ZM3 intro} for three-manifolds more general than $S^3$ or $S^2 \times S^1$ has turned out to be more challenging, however. One of the main technical complications, when attempting to apply localization techniques directly, is that one needs to understand an increasingly complicated sum over topological sectors, corresponding roughly to all possible gauge bundles over $\CM_3$.  
 
 On the other hand, more recently, using methods inspired by the study of topological quantum field theories (TQFT)~\cite{Witten:1988hf, Blau:1993tv, Beasley:2005vf, Blau:2006gh, Nekrasov:2014xaa, Blau:2013oha} and by the Nekrasov-Shatashvili Bethe/gauge correspondence~\cite{Nekrasov:2009ui, Nekrasov:2009uh}, a systematic computation of the supersymmetric partition function, $Z_{\CM_3}$, was carried out for essentially any $\CM_3$ compatible with a certain half-BPS condition~\cite{Closset:2017zgf, Closset:2018ghr} by Willett and the present authors. This TQFT-like approach unifies and subsumes many previous results; it can also be generalized to other types of theories and space-time dimensions.~\cite{Closset:2017bse, Hosseini:2018uzp, Crichigno:2018adf, Benini:2018mlo}

Before computing the supersymmetric partition function on a three-manifold $\CM_3$, we must first understand what it means to have rigid supersymmetry on curved space. We choose a Riemannian metric on $\CM_3$  and follow the so-called Festuccia-Seiberg approach~\cite{Festuccia:2011ws}, that views curved-space supersymmetry as a ``rigid'' limit of supergravity---recall that, in supergravity, supersymmetry is a gauge invariance; sending the Planck mass to infinity under certain conditions, one can recover some global supersymmetries. The condition for curved-space supersymmetry is that there exists some generalized Killing spinor(s), $\zeta$, such that a supersymmetric supergravity background exists. In that framework, we have a curved-space ``supercharge'' for each independent Killing spinor---for 3D $\CN=2$ supersymmetry, we can have up to four supercharges, the amount also preserved by flat space.

The three-manifold $\CM_3$ admits (at least) one supercharge if and only if it admits a transversely holomorphic foliation (THF).~\cite{Closset:2012ru, Klare:2012gn} A THF on $\CM_3$ provides a covering of the three-manifold by local patches $\R \times \C$ with adapted local coordinates $(\tau, z)$, such that the transition functions between patches are of the form: 
$$(\tau', z')= (\tau+ t(z, \bz), f(z))~.$$
 In particular, the transition function $z'= f(z)$ is holomorphic. The existence of a THF on $\CM_3$ is very restrictive.~\cite{Brunella1996, Ghys1996, Brunella1995}

In order to compute the supersymmetric partition function with current techniques, we need two supercharges of opposite $R$-charges, corresponding to two generalized Killing spinors, $\zeta$ and $\t\zeta$. We call the corresponding supersymmetric backgrounds on $\CM_3$ the ``half-BPS geometry.'' All these half-BPS backgrounds correspond to either Seifert manifolds---a THF in which the foliation is also a fibration by circles---, continuous deformations thereof, or else a special type of THF on $S^2\times S^1$, which we discuss separately. We will then give an overview of the current techniques to compute the partition function on any half-BPS geometry. 

The partition functions, $Z_{\CM_3}$, encode a lot of information about supersymmetric theories. For instance, the $S^3$ partition function gives the quantity $F_{S^3}$ of the IR SCFT, an effective measure of the number of degrees of freedoms.~\cite{Jafferis:2011zi, Casini:2012ei} The (untwisted) $S^2 \times S^1$ partition functions compute a 3D $\CN=2$ superconformal index~\cite{Kim:2009wb, Imamura:2011su, Kapustin:2011jm}---it counts local operators in the IR SCFT, with signs. More general Seifert manifolds partition functions also encode a rich algebra of half-BPS line operators, including interesting defect operators. They most likely also encode further non-perturbative information, which is still to be discovered.

A common application of localization techniques is to check infrared dualities between distinct  UV gauge theories.  Conversely, checking field-theory dualities at the level of the supersymmetric partition functions provides exceedingly precise independent checks of any claimed exact result for $Z_{\CM_3}$. We will see several examples of this in these notes.

\subsection*{Related developments}
There are many more related developments that benefitted from advances in localization techniques in 3D, in recent years, and which we will {\it not} cover. Let us mention some of them briefly, without any claim to exhaustivity.

\paragraph{The 3D/3D correspondence.}
The 3D/3D correspondance assigns a 3D $\CN=2$ supersymmetric theory, denoted by $T[{\bf M}_3,\GG]$, to {\it any} three manifold ${\bf M}_3$, for a given choice of the gauge group $\GG$ whose Lie algebra is $\Fg$ of ADE type.~\cite{Dimofte:2011ju, Cecotti:2011iy, Dimofte:2011py, Dimofte:2014ija} The theory $T[{\bf M}_3,\GG]$ is expected to arise by compatifying the 6D $\CN=(2,0)$ SCFT of type $\Fg$ on $M_3$---or, in other words, by wrapping M5-branes on $M_3$:
\be
\text{M5-branes on}\; \; \R^3 \times {\bf M}_3\qquad   \rightsquigarrow \qquad  T[{\bf M}_3] \; \; \text{on} \;\; \R^3~.
\ee  
The full 3D $\CN=2$ theory $T[{\bf M}_3]$ should be considered as a ``quantum invariant'' of the three-manifold ${\bf M}_3$. 

This is a fascinating and active subject, with many fundamental questions still open. For some more recent developments, see {\it e.g.} Refs.~\citen{Gang:2015wya, Gukov:2017kmk, Alday:2017yxk, Gang:2017lsr, Gang:2018hjd, Gang:2019uay, Cheng:2018vpl}. One can, in principle, compute the supersymmetric partition functions of $T[{\bf M}_3]$ on supersymmetric three-manifolds $\CM_3$, which should give us interesting numerical invariants of the three manifold ${\bf M}_3$; so far, only the case when $\CM_3$ is a Lens space or $\Sigma_g\times S^1$ has been considered in any detail.

\paragraph{Precision tests of the $AdS_4$/CFT$_3$ correspondence and supersymmetric black hole counting.}
Since supersymmetric partition functions $Z_{\CM_3}$ can be computed exactly, they can be used to test the $AdS$/CFT correspondence at strong coupling---in 3D, this generally means at large $K$ in some $1/K$ expansion, with $K$ a Chern-Simons level of the 3D $\CN=2$ theory.

The prime example of that, and by-far the most studied example, is the 3D $\CN=6$ supersymmetric ABJM Chern-Simons matter theory,~\cite{Aharony:2008ug} which is dual to M-theory on $AdS_4 \times S^7$. Many more Chern-Simons matter theories with 3D $\CN=2$ supersymmetry are known to be dual to specific $AdS_4\times X_7$ vacua of M-theory, with $X_7$ a Sasaki-Einstein seven-manifold.~\cite{Benna:2008zy, Jafferis:2008qz, Imamura:2008nn, Gaiotto:2009tk,Aganagic:2009zk, Benini:2009qs} One can compute the exact partition functions of these $\CN=2$ theories (generally CS quivers with $U(N)$ gauge groups) on $S^3$ and match to the prediction from 11D supergravity,~\cite{Drukker:2010nc, Herzog:2010hf, Drukker:2011zy, Jafferis:2011zi, Gulotta:2011si, Marino:2011eh}
\be
F_{S^3} \equiv -{ \rm Re}\left[ \log Z_{S^3}^{\rm SCFT}\right] = N^{3\ov 2} \sqrt{2 \pi^6\ov 27 {\rm Vol}(X_7)}~,
\ee
at leading order in $1/N$ in the appropriate large $N$ and large $K$ limit. Historically, this kind of approach has provided some of the most detailed and precise checks of the AdS/CFT correspondence at strong coupling on the CFT side.  One can also consider different boundaries in Euclidean $AdS_4$ (instead of the round $S^3$), and consider various other deformations (such as mass term), and then match  the corresponding supergravity solutions to the exact localization results on the boundary.~\cite{Martelli:2011fw, Martelli:2012sz, Alday:2012au, Freedman:2013ryh, Farquet:2014kma, Genolini:2016ecx, Toldo:2017qsh, Bobev:2018wbt}

More recently, Benini, Hristov and Zaffaroni~\cite{Benini:2015eyy} used a  different supersymmetric partition function on $S^2 \times S^1$, the so-called topologically twisted index,~\cite{Benini:2015noa} to account for the microstates of certain BPS black holes in $AdS_4$, at least at first order in $1/N$. See Refs.~\citen{Benini:2016rke, Hosseini:2016tor, Hosseini:2016ume, Hosseini:2017mds, Azzurli:2017kxo, Benini:2017oxt, Hosseini:2017fjo, Hosseini:2019ddy} for further developments, and also Refs.~\citen{Liu:2017vbl, Jeon:2017aif, Liu:2017vll, Aguilera-Damia:2018bam, Binder:2019mpb} for recent computations beyond the leading order in $1/N$.

\paragraph{Partition functions on manifolds with boundaries.} In these notes, we focus on the computation of partition functions on {\it closed} three-manifolds $\CM_3$. It is of obvious interest to also consider manifolds with boundaries. The prototypical case is $D_2\times S^1$, a solid torus. Now, to compute the partition function, one must specify some boundary conditions. One important early attempt to that effect was provided by Beem, Dimofte and Pasquetti, leading to the so-called holomorphic blocks $Z_{D^2 \times S^1}^\alpha$,~\cite{Beem:2012mb, Pasquetti:2011fj} which are indexed by the so-called Bethe vacua, $|\alpha \rangle$, the 2D vacua of the theory on $\R^2 \times S^1$---those vacua will also play an important role for us. A different prescription for the $D_2\times S^1$ partition functions was discussed in Refs.~\citen{Gukov:2017kmk, Cheng:2018vpl} in the context of the 3D/3D correspondence. See also Refs.~\citen{Dimofte:2010tz, Yoshida:2014ssa, Nieri:2015yia, Assel:2016pgi, Dimofte:2017tpi, Dedushenko:2018tgx, Bade:2018qrl} for related developments and computations.

 \subsection*{Organization of the review}
We begin in section~\ref{sec:2} by reviewing 3D $\CN=2$ supersymmetric gauge theories in flat space, to set the stage and our notations. We also introduce a few explicit examples of supersymmetric gauge theories, which we will use to illustrate our discussion in the rest of the review. In section~\ref{sec:3}, we discuss in detail the formalism of rigid curved-space supersymmetry, and the classification of ``supersymmetric three-manifolds.'' In section~\ref{sec: susy loc gen},  we give a brief account of ``traditional'' supersymmetric localization techniques, emphasizing the computation of one-loop determinants. In section~\ref{sec: lens space}, we discuss the half-BPS partition function on the squashed three-sphere, $S^3_b$. In section~\ref{sec: S2 top index}, we discuss the topologically twisted index. In section~\ref{sec: bethe}, we introduce the 3D $A$-model, which is simply the effective 2D low-energy theory of the 3D theory on a circle, topologically twisted. In section~\ref{subsec: seifert operators}, we use the $A$-model picture to compute supersymmetric partition functions on any Seifert manifold. Finally, in section~\ref{sec: S2S1 index}, we discuss the 3D superconformal index as a non-Seifert partition function on $S^2 \times S^1$.

  
\section{Three-dimensional $\CN=2$ gauge theories in flat space}\label{sec:2}

Let us first consider 3D $\CN=2$ supersymmetric field theories on $\R^3$. (We will work in Euclidean signature throughout.)
The three-dimensional $\CN=2$ supersymmetry algebra has four real supercharges, $Q_\alpha$ and $\t Q_\alpha$,~\footnote{In Euclidean signature, $\psi$ and $\t \psi$ should be viewed as independent Dirac spinors, not complex conjugate. We denote by $\psi^\dagger= (\psi^\alpha)^\ast$ the Hermitian conjugate of $\psi=(\psi_\alpha)$.} which satisfy the anti-commutation relations:
\bea
&\{Q_\alpha, \t Q_\beta\} = 2 \gamma^\mu_{\alpha\beta} P_\mu + 2i \epsilon_{\alpha\beta} Z~,\cr
& \{Q_\alpha, Q_\beta\} =0~,\cr
&\{\t Q_\alpha, \t Q_\beta\} =0~.
\eea
Here, $P_\mu$ is the 3D momentum and $Z$ is a real central charge.

\vskip0.3cm
\paragraph{Spinor conventions.} Three-dimensional Dirac spinors are denoted by $\psi_\alpha$, $\eta_\alpha, \cdots$, with $\alpha= \mp$ the Dirac index. We choose our 3D $\gamma$-matrices to be:
\be
{(\gamma^\mu)_\alpha}^\beta = \{\sigma^3,  -\sigma^1, -\sigma^2\}= \left\{ \mat{1 &0 \\ 0 & -1}~, \, \mat{0 & -1 \\ -1 & 0}~, \, \mat{0 & i \\ -i & 0} \right\}~,
\ee
 in terms of the 3D Pauli matrices. We then have $\gamma^\mu \gamma^\nu = \delta^{\mu\nu} + i \epsilon^{\mu\nu\rho} \gamma_\rho$. The Dirac spinor representation of $SO(3)$ is given by:
 \be
 \Sigma^{\mu\nu} = {i \ov 4}[\gamma^\mu, \gamma^\nu]= -\half \epsilon^{\mu\nu\rho} \gamma_\rho~.
 \ee
 One can raise and lower Dirac indices with the antisymmetric tensors $\eps^{\alpha\beta}$ and $\eps_{\alpha\beta}$, defined as $\eps^{12}= \eps_{21}=1$.
 Dirac indices are generally left implicit, with the northwest-southeast convention understood, $\psi \eta\equiv \psi^\alpha \eta_\alpha$.

\subsection{Supermultiplets and supersymmetric Lagrangians}
Given a compact gauge group $\GG$ with Lie algebra $\Fg= {\rm Lie}(\GG)$, an $\CN=2$ supersymmetric gauge theory consists of a vector multiplet $\CV$ in the adjoint of representation of $\Fg$, and of a chiral multiplet $\Phi$ in a given (generally reducible) representation $\FR$ of $\Fg$.

Given the supermultiplets, one can then easily build supersymmetric Lagrangians in flat space. All the formulas reviewed in this section can be obtained by a straightforward dimensional reduction of the standard 4D $\CN=1$ supersymmetry formulas, in the standard Wess-and-Bagger~\cite{Wess:1992cp} notation. We spell them out in components, instead of using the (perhaps more familiar) superfield notation, because this is the form that is most easily generalized to rigid supersymmetry on curved space.

\subsubsection{Vector multiplet}
In the so-called Wess-Zumino (WZ) gauge, the vector multiplet contains the fields:
\be
\CV= (\sigma~, \, A_\mu~, \, \lambda~, \, \b\lambda~, \, D)~,
\ee
with the real scalar $\sigma$, the 3D gauge field $A_\mu$, the gaugini $\lambda, \t\lambda$, and the auxiliary field $D$. Their supersymmetry transformations read:
\bea
& \delta \sigma = - \zeta \t \lambda + \t \zeta \lambda~,\cr
& \delta A_\mu = - i \big(\zeta \gamma_\mu \t \lambda + \t \zeta \gamma_\mu \lambda\big)~,\cr
& \delta \lambda = i \zeta D-i \gamma^\mu \zeta\big( D_\mu \sigma+ \half \ep_{\mu\nu\rho} F^{\nu\rho}\big)~,\cr
&\delta \t \lambda = - i \t \zeta D +i \gamma^\mu \t\zeta\big( D_\mu \sigma- \half \ep_{\mu\nu\rho} F^{\nu\rho}\big)~,\cr
&\delta D =\zeta \gamma^\mu D_\mu \t \lambda- \zeta [\sigma, \t\lambda] - \t \zeta \gamma^\mu D_\mu\lambda- \t\zeta [\sigma, \lambda]~. 
\eea
with $D_\mu= \d_\mu - i A_\mu$ the gauge-covariant derivative, and $F_{\mu\nu}$ the field strength:
\be
F_{\mu\nu}=\d_\mu A_\nu- \d_\nu A_\mu - i [A_\mu, A_\nu]~.
\ee
Note that $A_\mu$ is chosen to be Hermitian, using the standard physics conventions. The supersymmetry parameters $\zeta_\alpha$, $\t\zeta_\alpha$ are arbitrary constant spinors in $\R^3$.

\subsubsection{Chiral multiplet} Matter fields charged under the gauge group sit in chiral multiplets:
\be\label{chiral mult flat}
\Phi = (\phi~, \, \psi~,\, F)~, 
\ee
with the supersymmetry transformations:
\bea\label{chiral mult flat var}
& \delta \phi = \sqrt 2 \zeta \psi~,\cr
& \delta \psi = \sqrt 2 \zeta F + \sqrt 2 i  \sigma \t \zeta \phi - \sqrt 2 i \gamma^\mu \t \zeta D_\mu \phi~,\cr
& \delta F =- \sqrt 2 i  \sigma \t \zeta \psi + 2 i \phi  \t \zeta \t \lambda - \sqrt2 i \t \zeta \gamma^\mu D_\mu \psi~.
\eea
The chiral multiplet is valued in some representation $\FR$ of $\Fg$, and the vector-multiplet fields act on 
$\phi$ and $\psi$ accordingly (including through the gauge-covariant derivative $D_\mu$). The CPT conjugate of $\Phi$ is the anti-chiral multiplet $\t\Phi = (\t\phi,  \t\psi, \t F)$, valued in the conjugate representation $\b\FR$, with the supersymmetry transformations:
\bea
& \delta \t \phi = - \sqrt 2 \t \zeta \t \psi~,\cr
& \delta \t \psi = \sqrt 2 \t \zeta \t F - \sqrt 2 i \sigma \zeta \t \phi + \sqrt 2 i \gamma^\mu \zeta \, D_\mu \t \phi~,\cr
& \delta \t F =- \sqrt 2 i \sigma \zeta \t \psi + 2 i\t \phi  \zeta  \lambda - \sqrt2 i \zeta \gamma^\mu D_\mu  \t \psi~.
\eea
One can check that the matter field transformations realize the (gauge-covariant) supersymmetry algebra:
\be
\delta_\zeta^2= 0~, \qquad \delta_{\t\zeta}^2= 0~, \qquad
\{\delta_\zeta, \delta_{\t\zeta}\}=- 2 i \zeta\gamma^\mu \t\zeta D_\mu - 2 i \zeta \t \zeta \sigma~,
\ee
Note that, in WZ gauge, the real scalar $\sigma$ enters as a field-dependent central charge, with $Z= -\sigma$, acting in the appropriate representation $\FR$ or $\t\FR$.

\subsubsection{Global symmetries and background gauge fields}
The field theories we consider often have some non-trivial continuous global symmetry group $\GG_F$. To keep track of these symmetries, it is useful to turn on background vector multiplets $\CV_F$ coupling to the conserved currents, consistently with supersymmetry. This corresponds to a bosonic background:
\be
\CV_F= (\sigma^{(F)}~,\,  A_\mu^{(F)}~,\, D^{(F)})~,
\ee
such that the gaugino variations vanish, $\delta \lambda^{(F)}=\delta \t\lambda^{(F)}=0$. In flat space and to preserve the four supercharges, we take:
\be\label{mF flat space}
\sigma^{(F)}= m_F \in i \Fg_F~, \qquad A_\mu^{(F)}=0~, \qquad D^{(F)}=0~,
\ee
where $\sigma^{(F)}$ takes a constant VEV.
It is often useful to choose a maximal torus of the global symmetry group:
\be
\prod_\alpha U(1)_\alpha \subset \GG_F~,
\ee
with $\sigma_\alpha= m_\alpha \in \R$ for each $U(1)_\alpha$ vector multiplet.
The real parameters $m_F$ are called ``real masses'' in 3D, since they induce the mass terms:
\be\label{SLm}
\SL_m =  \t\phi m^2_F \phi - i \t \psi m_F \psi \; \subset \SL_{\t\Phi\Phi}
\ee
in the chiral-multiplet Lagrangian (to be discussed below).

In 3D, an important and somewhat peculiar symmetry occurs in the presence of abelian gauge fields. For every $U(1)$ gauge field $A_\mu$, one can define a current:
\be
j^{\mu}_T = {1\ov 4\pi} \epsilon^{\mu\nu\rho} F_{\mu\nu}~,
\ee
which is conserved due to the Bianchi identity. 
The corresponding global symmetry, $U(1)_T$, is often called a ``topological symmetry.'' The states charged under $U(1)_T$ carry non-trivial $U(1)$ magnetic charge. The corresponding local operators are known as ``monopole operators.''

\subsubsection{Supersymmetric Lagrangians}
The supersymmetric Yang-Mills-Chern-Simons-matter Lagrangian is given by:
\be
 \SL= \SL_{\rm YM} + \SL_{\rm CS} + \SL_{\t\Phi\Phi}+ \SL_{W}+ \SL_{\t W}~,
\ee
schematically. The various contributions are as follows:


\paragraph{SYM terms.} 
The super-Yang-Mills term reads:
\be
\SL_{\rm YM}={1\ov e_0^2}\tr\left( {1\ov 4} F_{\mu\nu}F^{\mu\nu} + \half D_\mu \sigma D^\mu \sigma - i \t \lambda \gamma^\mu D_\mu \lambda - i \t\lambda [\sigma, \lambda] -\half D^2  \right)~.
\ee
Recall that the  YM gauge coupling squared has  dimension of mass, $[e_0^2]=1$.

\paragraph{CS terms.}
The Chern-Simons (CS) term takes the form:
\be\label{susy CS}
\SL_{\rm CS}= {k \over 4 \pi} \tr\left(i \ep^{\mu\nu\rho} \left(A_\mu \d_\nu A_\rho- {2i\over 3} A_\mu A_\nu A_\rho \right) - 2 D \sigma + 2 i \t \lambda \lambda\right)~.
\ee
This is schematic. For a gauge group:
\be
\GG= \prod_\gamma \GG_\gamma \times  \prod_I U(1)_I~,
\ee
with $\GG_\gamma$ a simple gauge group, we have an independent Chern-Simons level $k_\gamma$ for each simple factor, while we can have a matrix of CS levels $k_{IJ}$ for the abelian sector, with the mixed CS terms:
\be\nn
\SL_{\rm CS}^{ I \neq J}= {k_{IJ} \over 2 \pi} \left(i \ep^{\mu\nu\rho} A^{(I)}_\mu \d_\nu A^{(J)}_\rho -  D^{(I)} \sigma^{(J)} -  D^{(J)} \sigma^{(I)} +  i \t \lambda^{(I)} \lambda^{(J)}+i \t \lambda^{(J)} \lambda^{(I)}\right)~.
\ee
Assuming that each $\GG_\gamma$ is compact and simply-connected, the CS levels are integer quantized:
\be
k_{\gamma} \in \Z~, \qquad k_{IJ} \in \Z~.
\ee
We will also consider $U(N)$ gauge groups; in that case, one can in principle choose different CS levels for the $U(1)$ and $SU(N)$ factors, but we will take them to be equal in what follows.

We can also have Chern-Simons terms mixing some gauge $U(1)_I$ with some  global symmetry $U(1)_\alpha$, with the background vector multiplet $\CV_\alpha$ set to its supersymmetric value~\eqref{mF flat space}. An important example is a CS term mixing $U(1)_I$ with the associated topological symmetry, $U(1)_{T_I}$, with level $k_{I, T_I}=1$. This gives:
\be
\SL_{{\rm FI}, I} = - \xi_I D_I~, \qquad \xi_I \equiv  {1\ov 2 \pi} m_{T_I}~.
\ee
In other words, the 3D Fayet-Iliopoulos (FI) term is a real mass for the topological symmetry.

\paragraph{Matter Lagrangians.} Consider a chiral multiplet $\Phi$ coupled to a vector multiplet $\CV$. The standard kinetic term reads:
\be
\SL_{\t\Phi \Phi} = D_\mu \t\phi D^\mu \phi - i \t \psi \gamma^\mu D_\mu \psi - \t F F + \t\phi D \phi +\t\phi \sigma^2 \phi -i \t \psi \sigma \psi + \sqrt2 i (\t \phi \lambda \psi + \t \psi \t\lambda \phi)~.
\ee
This implicitly includes the mass term \eqref{SLm}, in the case of a background vector multiplet with $\sigma^{(F)}=m_F$. We also have interaction terms encoded in the 3D $\CN=2$ superpotential $W(\phi)$, an holomorphic function of the chiral multiplet scalars. This gives:
\be\label{Lag W}
\SL_W= F^i { \d W\ov \d \phi^i} -\half \psi^i \psi^j {\d W\ov \d \phi^i\d \phi^j}~,
\ee
and similarly for the conjugate superpotential $\t W(\t\phi)$.

\subsection{Parity anomaly and Chern-Simons contact terms}\label{subsec: partity}
Three-dimensional ``parity'' is the $\Z_2$ symmetry that reverse the sign of a single spacial coordinate. In this work, we consider theories in Euclidean signature only, and therefore parity inverses the sign of any of the three directions of $\R^3$, say:
\be
P \; : \; x^3 \rightarrow - x^3~.
\ee
All the Chern-Simons interactions, as well as the fermion real-mass term in \eqref{SLm}, are odd under parity. (Equivalently, using the background-field language, $m_F$ transform as $m_F\rightarrow -m_F$ under $P$.)

It is well-known that massless Dirac fermions in 3D can suffer from a parity anomaly---that is, although the classical Lagrangian is parity-even, the quantum theory might not be \cite{Redlich:1983dv, Niemi:1983rq, AlvarezGaume:1984nf}. The term ``parity anomaly'' is a bit of a misnomer, since the parity anomaly is more precisely a mixed parity-gauge anomaly.~\footnote{In the presence of a background metric, it can also be a mixed anomaly between parity and diffeomorphism. In that case, the 3D Majorana fermion ``carries'' the minimal parity anomaly, and a Dirac fermion carries twice that amount; this corresponds to $\kappa_g=-\half$ and $\kappa_g=-1$, respectively. (This so-called gravitational contact term, $\kappa_g$, will be discussed momentarily.)} Consider a single free Dirac fermion $\psi_\alpha$ coupled to a background $U(1)$ gauge field $A_\mu$ with unit charge. The free theory:
\be
\SL_{\psi}= - i \t \psi \gamma^\mu D_\mu \psi= - i \t \psi \gamma^\mu (\d_\mu - i A_\mu) \psi~.
\ee
can be easily quantized. In the path integral language, we have an effective action for $A_\mu$ given by the determinant of the 3D Dirac operator:
\be\label{SeffA}
S_{\rm eff}[A] = - \log \det( - i \slashed D_A)~.
\ee
The parity anomaly, in this case, is the statement that $S_{\rm eff}[A]$ cannot preserve both parity and $U(1)$ gauge invariance. (This is a somewhat formal in flat space, but this discussion can be made completely rigorous by replacing $\R^3$ by a closed three-manifold $\CM_3$.) Since we always want to preserve gauge invariance (for instance, because we will later consider making $A_\mu$ dynamical), this means that $S_{\rm eff}[A]$ must break parity.

Let $j^\mu= - \t\psi \gamma^\mu \psi$ denote the conserved current which couples to $A_\mu$. 
The parity-breaking term in \eqref{SeffA} is conveniently captured by the two-point function of $j^\mu$. For any $U(1)$ current, the two-point function takes the form:
\be\label{jjcor}
\langle j_\mu(p) j_\nu(-p)\rangle =\tau { p_\mu p_\nu - p^2 \delta_{\mu\nu} \over 16|p|}\   +\kappa {\ep_{\mu\nu\rho} p^\rho\over 2\pi}~.
\ee
The coefficients $\tau$ and $\kappa$ are themselves functions of $p^2/\mu^2$ (with $\mu$ the RG scale), but they become constant at fixed points. The quantity $\kappa$ is parity-odd. For a single free Dirac fermion, one finds:
\be\label{kappa psi}
\kappa^{\rm UV}= -\half + k~, \qquad k \in \Z~,
\ee
where $k$ corresponds to a scheme ambiguity. (The meaning of the superscript ``UV'' will become clear momentarily.) This ambiguity has a simple origin. Note that $\kappa$ multiplies a local term in \eqref{jjcor}---that is, it is a contact term. In general, contact terms are completely scheme-dependent, and therefore unphysical, since they correspond to local terms in the sources.  The corresponding local term in this case, however, is the $U(1)$ CS action itself:
\be\label{SCS A}
S_{\rm CS}[A]={k \ov4\pi} \int d^3 x \, i \ep^{\mu\nu\rho}A_\mu \d_\nu A_\rho~.
\ee
Its coefficient, the CS level $k$, is integer-quantized due to gauge invariance.~\footnote{More precisely, the CS action itself is not really well-defined when $A_\mu$ is a topologically non-trivial gauge field. Then, on any closed (spin) three-manifold $\CM_3$, one defines $S_{\rm CS}$ in terms of a four-manifold $\CM_4$ with boundary $\CM_3$. The quantization condition $k\in \Z$ stems from the requirement that the CS action be independent of the choice of $\CM_4$.}
This gives us the ambiguity in \eqref{kappa psi}. Thus for a single free fermion coupled to $A_\mu$, the ``parity anomaly'' can be ascribed to the ``CS contact term'' $\kappa = -\half$ (mod $1$).~\footnote{Strictly speaking, the expression ``CS contact term'' would denote the {\it classical} term \protect\eqref{SCS A} in the effective action, with its quantized coefficient. By abuse of notation, we also use the expression to denote the full $\kappa$ (at fixed points) as well, including its non-integer physical part.}  More generally, for a collection of Dirac fermions $\psi^i$ with $U(1)$ charges $Q_i\in \Z$, we would have:
\be\label{kappa eff UV}
\kappa^{\rm UV} = -\half \sum_i Q_i^2 + k~.
\ee
Whenever $\sum_i Q_i^2$ is even, we can choose $k\in \Z$ such that $\kappa^{\rm UV}=0$, thus ``canceling the parity anomaly'' to obtain a parity-even effective action. In general, the parity anomaly---or, equivalently, the presence of non-zero CS contact terms---is an interesting fact of life in 3D.

For each Dirac fermion in our microscopic (UV) gauge theory, we should choose a ``quantization scheme'' consistent with gauge invariance---that is, we must specify all the CS contact terms in the UV---in general, we must do so for both the dynamical and background gauge fields. Unless otherwise stated, we will always choose the ``$U(1)_{-\half}$ quantization.'' For a single Dirac fermion $\psi$ of unit charge, this corresponds to $k=0$ in \eqref{kappa psi}; for $\psi$ of $U(1)$ charge $Q\in \Z$, we choose $\kappa=-\half Q^2$.

More generally, the notation ``$\GG_{\kappa}$,'' with $\GG$ the gauge group and $\kappa\in \half \Z$, is common in the literature. Here $\kappa$ always denotes the ``effective CS level'' in the UV, as in \eqref{kappa eff UV}.

\subsubsection{Quantizing 3D supersymmetric multiplets}\label{subsec: quantize susy mult}
Given the above discussion, we should specify our conventions for quantising supersymmetric multiplets, consistently with gauge invariance. We also should keep track of the CS contact terms for the $U(1)_R$ R-symmetry, and of an analogous gravitational CS contact term, denoted by $\kappa_g$. These CS contact terms, and the corresponding 3D $\CN=2$ supersymmetric Chern-Simons actions, were studied in Ref.~\citen{Closset:2012vp}, to which we refer for further discussion. Our conventions for the UV quantization of the 3D $\cN=2$ gauge theory follow Ref.~\citen{Closset:2018ghr}.

Consider any chiral multiplet $\Phi$ of charge $Q_a\in \Z$ under some $U(1)_a$ gauge fields, of $R$-charge $r$. In our choice of quantization, we have:
\bea\label{kappaUV Phi def}
&\kappa_{ab}^{\rm UV}= -\half Q_a Q_b~, \qquad && \kappa_{aR}^{\rm UV} = - \half Q_a (r-1)~, \cr
&\kappa_g^{\rm UV}= -1~, \qquad
&& \kappa_{RR}^{\rm UV}= - \half (r-1)^2~.
\eea
Here $\kappa_{ab}$, $\kappa_{aR}$ and $\kappa_{RR}$ denote the mixed $U(1)_a{-}U(1)_b$, mixed $U(1)_a{-}U(1)_R$ and pure $U(1)_R$ CS contact terms, respectively.

When quantizing the vector multiplet, we generally also need to introduce a non-zero contact term for the R-symmetry. We choose:~\cite{Closset:2018ghr}
\be\label{kappaRR App vec}
\kappa_{RR}^{\rm UV}= \half {\rm dim}(\GG)~, \qquad \qquad
\kappa_{g}^{\rm UV}={\rm dim}(\GG)~.
\ee
In any gauge theory, the UV contact terms are obtained by adding the contributions \eqref{kappaUV Phi def} from all the chiral multiplets to the vector-multiplet contribution \eqref{kappaRR App vec}.

\subsubsection{Decoupling in the limit of large real mass.}
A 3D Dirac fermion (with $U(1)_a$ charges $Q_a$) can be given a real mass,  
\be
\Delta\SL_m = -i m \b\psi \psi~,\qquad\qquad m\in \R~.
\ee
 Integrating out the fermion $\psi$ with a large real mass, $m \rightarrow \pm \infty$, shifts the CS contact terms according to:
\be\label{shift kappa intro}
\delta \kappa_{ab}= \half \sign(m)\, Q^a Q^b~, \qquad\qquad \delta \kappa_g = \sign(m)~,
\ee
and similarly for the $R$-symmetry. 
In particular, a 3D $\CN=2$ chiral multiplet $\Phi$ can be given a real mass by coupling it to a background vector multiplet with non-zero VEV for $\sigma_F$, as in \eqref{SLm}. The choice of UV contact terms \eqref{kappaUV Phi def} can be equivalently stated as the requirement that, as we integrate out $\Phi$ with a large positive mass, $m \rightarrow \infty$, the contribution of $\Phi$ to every $\kappa$ become trivial. (Conversely, integrating out $\Phi$ with $m \rightarrow -\infty$ leaves behind some integer-quantized CS levels.)

\subsection{Supersymmetric gauge theories and infrared dualities: Examples}\label{subsec: 3 examples}
In this review and for concreteness, we will focus on three examples of 3D $\CN=2$ supersymmetric gauge theories, which together contain most of the ingredients of the general case:
\begin{itemize}
\item[(1)] The $U(N)_k$ gauge theory---a supersymmetric CS theory with gauge group $U(N)$ at level $k$.
\item[(2)] The $U(1)_{\half}{+}\Phi$ gauge theory, a $U(1)$ theory coupled to a single chiral multiplet $\Phi$ of unit charge, with effective CS level $\half$.
\item[(3)] 3D SQCD with unitary gauge group: a $U(N_c)$ theory with $N_f$ ``flavors,'' namely $N_f$ pairs of chiral multiplets in the fundamental and anti-fundamental representations, with vanishing effective CS level.
\end{itemize}
The case (3) subsumes the first two, which can be obtained from it by appropriate decoupling limits. Nonetheless, it is interesting to consider each example in turn, to focus on their distinctive aspects.

These theories also enjoy infrared dualities---namely, there exists two different ``dual'' gauge theory descriptions in the UV which flow to the same IR fixed point. A subtle and interesting aspect of these dualities is that, in order to make the duality statement precise, we often have to specify {\it relative Chern-Simons contact terms.}~\cite{Closset:2012vp} That is, given a theory $\CT$ and its dual $\CT^D$, we could add to either theory some Chern-Simons terms for the flavor symmetries, and for the gravitational CS term, with properly quantized CS levels $k_F(\CT), k_F(\CT^D)$ and $k_g(\CT), k_g(\CT^D)$, respectively. As we explained above, the presence of a Chern-Simons contact term in a given theory does not affect the physics, since it just shift the coefficient $\kappa_F$ in \eqref{jjcor} (and similarly $\kappa_g$) by an integer. Nonetheless, in giving the precise statement of IR dualities, we also need to specify the {\it relative} CS levels:~\cite{Benini:2011mf, Closset:2012vp}
\be\label{def Delta kF}
\Delta k_F = k_F(\CT^D) - k_F(\CT)~, \qquad \quad
\Delta k_g = k_g(\CT^D) - k_g(\CT)~,
\ee
which ensure that all two-point functions of conserved currents agree between the two descriptions, in the infrared. 
The precise meaning of the UV levels $k_F, k_g$ itself depends on our conventions in quantizing fermions, which we stated above. Once these conventions are chosen, the relative CS levels \eqref{def Delta kF} are unambiguous.

Let us now review the above gauge theories and their infrared dualities, in our conventions.

\subsubsection{$U(N)_k$ CS theory and supersymmetric level/rank duality}\label{subsec: review UNk levelrk}
Consider the $U(N)$ vector multiplet with a SYM Lagrangian and a supersymmetric CS term at level $k\in \Z$:
\be
S= \int d^3 x (\SL_{\rm SYM} + \SL_{\rm CS})~.
\ee
 For $|k| >N$, the theories flows to supersymmetric Chern-Simons theory; that is, the YM term decouples. The supersymmetric CS action \eqref{susy CS} effectively gives a mass to the gauginos. In the far infrared, the theory flows to a topological quantum field theory,  pure Chern-Simons theory with gauge group $U(N)$
 \be
 U(N) \cong {SU(N) \times U(1)\ov \Z_N}
 \ee
 with the levels:
\be
{\bf k}= k - {\rm sign}(k) N~, \qquad  k^{U(1)}= N k~,
\ee
for the $SU(N)$ and $U(1)$ factors, respectively. The IR CS theory is generally denoted by $U(N)_{{\bf k}, k}$.
The shift of the $SU(N)$ CS level arises from integrating out the gauginos. The theory has a global symmetry $U(1)_T\times U(1)_R$, with $U(1)_T$ the topological symmetry.

Consider $k>0$ for simplicity. The $U(N)_k$ supersymmetric Chern-Simons theory is IR-dual to a $U(k-N)_{-k}$ CS theory:
\be\label{levelrk susy}
U(N)_k \qquad \longleftrightarrow \qquad U(k-N)_{-k}~,
\ee
with the following relative CS levels for global symmetries:
\be\label{rel level susy levelrk}
\Delta k_{TT} = -1~, \qquad \Delta k_{RR}= - (k-N)^2~, \qquad \Delta k_g = - 2 k (k-N) - 2~.
\ee
In terms of the infrared pure CS theory, the duality \eqref{levelrk susy} corresponds to:
\be
U(N)_{{\bf k}, {\bf k}+N} \qquad \longleftrightarrow \qquad U({\bf k})_{-N, -{\bf k}-N}~,
\ee 
with ${\bf k}= k-N$, and the relative CS levels:
\be\label{rel cct level rk}
\Delta k_{TT}=-1~, \qquad  \Delta k_g = - 2 N {\bf k} - 2~,
\ee
which follow from \eqref{rel level susy levelrk}.
 This {\it level-rank duality} was recently revisited in Ref.~\citen{Hsin:2016blu} where the relative contact terms \eqref{rel cct level rk} were derived. Note that, upon integrating out the gauginos, the $R$-symmetry decouples entirely; with our choice of quantization~\eqref{kappaRR App vec}, with vanishing bare $U(1)_R$ CS level in the $U(N)_k$ UV description, and with $k_{RR}=\Delta k_{RR}$ as in \eqref{rel level susy levelrk} in the $U(k-N)_{-k}$ description, we have that $\kappa_{RR}^{\rm IR}= 0$.

\subsubsection{Elementary mirror symmetry}\label{subsec: EMS}
Possibly the simplest 3D $\CN=2$ duality with chiral multiplets is between a $U(1)_\half$ gauge theory with a single chiral multiplet $\Phi$ of electric charge $Q=1$. In our convention for the quantization of $\Phi$, we have a bare CS level $k=1$, so that $\kappa^{\rm UV}$ in \eqref{kappa eff UV} is equal to $\half$. The theory has a global symmetry $U(1)_T \times U(1)_R$.
The dual description is in terms of a single chiral multiplet $T^+$, which is identified with a gauge-invariant monopole operator (of $U(1)_T$ charge $1$) in the supersymmetric field theory:
\be\label{EMS duality}
\CT\; : \; U(1)_\half + \Phi\;\; \text{(gauge theory)}  \qquad \longleftrightarrow \qquad \CT^D\; : \; \;\;  T^+ \;\; \text{(free chiral)}~.
\ee
Let $r\in \R$ denote the $R$-charge of $\Phi$ (this choice is not physical, since one can always set $r$ to zero by mixing the R-charge with the gauge symmetry). Then the gauge-invariant monopole operator $T^+$ has global charges:
\be
Q_T[T^+]=1~, \qquad R[T^+] = 1-r~.
\ee
The duality \eqref{EMS duality} holds with the following non-trivial relative CS levels:
\be\label{kTR kRR rel mirrsym 1}
\Delta k_{TR}= -r~, \qquad\qquad \Delta k_{RR}=r^2~.
\ee
As a consistency check, one can integrate out the chiral multiplet $T^+$ in $\CT^D$ with a real mass $m_T$ for $U(1)_T$. For $m_T$ large and negative, $m_T \rightarrow -\infty$,  the IR theory is empty with only a $U(1)_T$ and a gravitational CS contact term, with levels $k_{TT}=-1$ and $k_g=-2$, which arise from integrating out the chiral multiplet $T^+$. In the gauge-theory description $\CT$, $m_T=\xi$ is the FI parameter, and one should also redefine the dynamical field $\sigma$ so that the effective FI term $\xi_{\rm eff} = \xi + \sigma$ remain finite. This gives a large positive real mass to $\Phi$, so that the low-energy theory is a $U(1)_1$ CS theory without any other CS contact terms. This precisely reproduces the level/rank duality \eqref{levelrk susy} in the special case $N=k=1$. 

The duality \eqref{EMS duality} is the simplest example of a 3D $\CN=2$ mirror symmetry,~\cite{Dorey:1999rb} but also of a Seiberg-like duality.~\cite{Benini:2011mf} In recent literature, the $U(1)_\half{+} \Phi$ and its duality are also known as the  ``tetrahedron'' theory and duality, due to its role in the 3D/3D correspondence.~\cite{Dimofte:2011ju}

\subsubsection{$U(N_c)$ SQCD and the Aharony duality}\label{subsec: Aharony duality}
The Aharony duality~\cite{Aharony:1997gp} is a three-dimensional generalization of four-dimensional Seiberg duality for 4D supersymmetric QCD.~\cite{Seiberg:1994pq}  It relates a $U(N_c)$ gauge theory with $N_f$ ``flavors'' with a $U(N_f-N_c)$ dual gauge theory:
\be\label{Aharony duality}
\CT\; : \; U(N_c) \; +\, Q, \t Q \qquad \longleftrightarrow \qquad \CT^D\; : \; \;\;  U(N_f-N_c) \; +\,  q, \t q \;+ \;  M, T_+, T_-~.
\ee
More precisely, the two theories are defined as follows:


\paragraph{Theory $\CT$:} A $U(N_c)$ vector multiplet with vanishing effective CS level, $\kappa^{\rm UV} =0$, coupled to $N_f$ chiral multiplets $Q_i$ in the fundamental and $N_f$ chiral multiplets $\t Q^j$ in the anti-fundamental  (with the flavor indices $i, j=1, \cdots, N_f$). In our conventions (with the ``$U(1)_{-\half}$ quantization'' of every chiral multiplet), we must have a bare CS level $k=N_f$ for the $U(N_c)$ gauge group.

\paragraph{Theory $\CT^D$:} A $U(N_f-N_c)$ vector multiplet with vanishing effective CS level and $N_f$ ``flavor'' pairs of chiral multiplets $q_j$ and $\t q^i$. The theory also contains $N_f^2$ gauge-singlet chiral multiplets, called ${M^j}_i$, and two additional gauge-singlet multiplets, $T^\pm$, all coupled to the gauge sector through the superpotential:
\be
W= M^j_i \t q^i q_j + T_+ t_- + T_- t_+~,
\ee
with $t_\pm$ the elementary monopole operators of $U(N_f-N_c)$.

\medskip
\noindent The gauge-singlet fields in theory $\CT^D$ are identified with the gauge-invariant mesons ${M^j}_i= \t Q^j Q_i$ and  with the monopole operators $T^\pm$ in the theory $\CT$, respectively. The global symmetry of either theory is:
\be
SU(N_f)\times SU(N_f) \times U(1)_A \times U(1)_T \times U(1)_R~.
\ee
The gauge and global charges of the fields in the two theories are as follows:
\be\nn
\begin{array}{c|c|c|ccccc}
    &  U(N_c)&U(N_f-N_c)& SU(N_f) & SU(N_f)  & U(1)_A &  U(1)_T & U(1)_R  \\
\hline
Q_i        & \bm{N_c}&\bm{1}& \bm{\overline{N_f}} & \bm{1}& 1   & 0   &r \\
\t Q^j   & \bm{\overline{N_c}}&\bm{1}  &  \bm{1}& \bm{N_f}  & 1   & 0   &r \\
\hline
q_j   &\bm{1}&\bm{N_f-N_c}&    \bm{1} & \bm{\overline{N_f}} & -1   & 0   &1-r \\
\t q^i     &\bm{1} & \bm{\overline{N_f-N_c}}& \bm{N_f}    & \bm{1}    & -1   & 0   &1-r \\
{M ^j}_i    &\bm{1}  & \bm{1} &\bm{\overline{N_f}}& \bm{N_f} &  2   & 0   &2 r \\
T^+   &\bm{1}    & \bm{1} & \bm{1} & \bm{1} & -N_f   & 1   & -N_f(r-1) -N_c +1\\
T^-   &\bm{1}  & \bm{1} & \bm{1} & \bm{1} & -N_f   & -1   & -N_f(r-1) -N_c +1
\end{array}
\ee
Here, $r\in \R$ denotes the $R$-charge of the fundamental fields $Q$ and $\t Q$. It could be set to zero by mixing $U(1)_R$ with $U(1)_A$. 

Finally, to fully specify the duality, we should give the relative CS level for the global symmetries. The non-zero levels are:
\bea\label{AD CS ct1}
&\Delta k_{SU(N_f)}=\Delta \t k_{SU(N_f)}= N_f-N_c~, \cr
&\Delta k_{TT}=1~,\cr
&\Delta k_{AA}=  4 N_f^2 -2 N_c N_f~,\cr
&\Delta k_{AR}=2N_f^2 +(4N_f^2 - 2 N_c N_f)(r-1)~,\cr
&\Delta k_{RR}= N_c ^2 + N_f^2 + 4 N_f^2 (r-1) + (4N_f^2 - 2 N_c N_f)(r-1)^2~,\cr
&\Delta k_g = 2 N_f (N_f-N_c)+2~,
\eea
with the first line giving the levels for the $SU(N_f)\times SU(N_f)$ symmetry.~\footnote{As explained in detail in Ref~\protect\cite{Closset:2018ghr}, for Aharony duality it is usually stated that all flavor CS levels vanish, but this is true only in a quantization scheme for the fermions that preserves parity and violates background gauge invariance.}

The special case $N_c=N_f=1$ is worth commenting on. In that case, theory $\CT$ is known as SQED. The dual gauge sector is empty, and the theory $\CT^D$, sometimes known as the $XYZ$ model, consists of only three chiral multiplets $X=M$, $Y=T^+$ and $Z=T^-$, with the non-trivial superpotential:
\be
W= XYZ~.
\ee
One can easily check that this superpotential is allowed by the symmetries. (An effective superpotential $W={\rm det}(M)T^+ T^-$ appears for $N_c= N_f\geq 1$, more generally.~\cite{Aharony:1997bx})

\section{Curved-space supersymmetry and half-BPS geometry}\label{sec:3}
In the previous section, we discussed 3D $\CN=2$ supersymmetric gauge theory on $\R^3$. While QFT is naturally formulated in flat space, its is also possible to consider a non-trivial Riemannian metric on (Euclidean) space-time. Recall that, for any internal $U(1)_F$ symmetry of our theory, we can introduce a {\it background gauge field} $A_\mu^{(F)}$ that couples to the conserved current. We then have the coupling:
\be
\SL_{jA} = j^\mu A_\mu^{(F)}~,
\ee
at first order in the ``source'' $A_\mu^{(F)}$; the conservation of the current $j^\mu$ is equivalent to the $U(1)_F$ gauge-invariance of this Lagrangian, up to a total derivative.
Similarly, any Poincar\'e-invariant QFT possesses a conserved energy-momentum tensor $T_{\mu\nu}=T_{\nu\mu}$, whose corresponding source is a {\it background metric} $g_{\mu\nu}$. At first order around flat space, $g_{\mu\nu}= \delta_{\mu\nu}+ \Delta g_{\mu\nu}$, with the graviton $ \Delta g_{\mu\nu}$, we have:
\be\label{Tg coupling}
\SL_{Tg}= -T^{\mu\nu} \Delta g_{\mu\nu}~. 
\ee
The higher-order terms are constrained by diff invariance. From \eqref{Tg coupling}, it should come as no surprise that coupling the field theory to a non-trivial metric can give us direct access to some of its most important and universal features.

Thinking more geometrically, we would like to consider our quantum field theory on a non-trivial Riemannian three-manifold $\CM_3$, with metric $g$. For simplicity, we choose $\CM_3$ to be closed and oriented. In the presence of continuous internal symmetries $\GG_F$, the background gauge field $A^{(F)}$ should be interpreted as a connection on a principal $\GG_F$-bundle $P$ over $\CM_3$.

In that context, a very natural observable is the partition function of the field theory on a given $\CM_3$ (with its $\GG_F$-bundle $P$). This is given by:
\be\label{ZM3 general}
Z_{\CM_3}[g, A^{(F)}] = \int [D\varphi]_{\CM_3} \, e^{-S[\varphi, g, A^{(F)}]}~,
\ee
schematically. Here, the right-hand-side is the full Feynman path integral over the various dynamical fields $\varphi$ on $\CM_3$, at fixed sources $g$ and $A^{(F)}$. In general, of course, computing this explicitly would too daunting a task. With the help of 3D $\CN=2$ supersymmetry, however, we can often compute \eqref{ZM3 general} {\it exactly,} irrespective of any perturbation theory, by giving some specific supersymmetry-preserving values to the sources $g=g_0$, $A^{(F)}=A^{(F)}_0$. In particular, this gives us very precise and powerful tools to explore the strongly-coupled infrared of 3D supersymmetric gauge theories. 

\subsection{A first example: the flavored 3D Witten index.}
Before explaining this point more in detail, it is useful to consider a first example of the setup we just introduced. Let us choose the three-manifold to be a real 3-torus:
\be
\CM_3 = T^3~,
\ee
and let $P$ be a flat bundle, which is determined by its flat connections along $T^3$. In order to preserve supersymmetry, we choose all the fermions to have periodic boundary conditions~\footnote{This is not strictly necessary, as we will discuss later on. One can choose any spin structure as long as one turns on appropriate $U(1)_R$ holonomies so that the Killing spinors remain periodic.} along the one-cycles of $T^3$. It is useful to single out the third direction of $T^3$; then, by a standard argument, the partition function is a {\it Witten index,} which counts states of the theory quantized on $T^2 \times \R$, with signs:
\be\label{ZT3 0}
Z_{T^3} = \Tr_{T^2} (-1)^{F} e^{-\beta H}~.
\ee
Here, $F$ is the fermion number, $H$ denotes the Hamiltonian, and we choose $P$ to be trivial. This quantity is only well-defined when $H$ has a discrete spectrum. For instance, this is expected to be the case for 3D $\CN=2$ supersymmetric CS theory. For the $\CN=2$ Chern-Simons theory  $U(N)_k$, in particular, we have:~\cite{Witten:1999ds, Ohta:1999iv}
\be\label{ZT3 expls1}
Z_{T^3}^{U(N)_k}= \mat{k \\  N}~.
\ee

Most of the theories of interest to us will have a moduli space of vacua and a continuous spectrum, in which case \eqref{ZT3 0} is ill-defined. Nonetheless, one can define a useful Witten index for a deformation of the theory, obtained after turning on the non-trivial flat connections mentioned above. We also turn on the real mass $\sigma^{(F)}$. The torus $T^3$ is a flat three-manifold, which preserves four supercharges, and one can show that $Z_{T^3}$ cannot depend on small deformations of $A_\mu^{(F)}$ and $\sigma^{(F)}$, therefore it is a constant. Nonetheless, the case with non-zero background fields is qualitatively different from the case when they vanish, since they make the Witten index well-defined~\cite{Intriligator:2013lca} (for the gauge theories of interest in this review, at least). This is often called the ``flavored Witten index.'' One finds:
\be\label{ZT3 expls2}
Z_{T^3}^{U(1)_\half{+}\Phi} = 1~, \qquad  \qquad
Z_{T^3}^{U(N)\, {\rm SQCD}} = \mat{N_f \\ N}~,
\ee
for the theories (2) and (3) of section~\ref{subsec: 3 examples}.  Note that the results \eqref{ZT3 expls1} and \eqref{ZT3 expls2} are consistent with the respective infrared dualities. We will later give a derivation of \eqref{ZT3 expls1}-\eqref{ZT3 expls2} as a special case of a more general formalism.

\subsection{Rigid supersymmetric in curved space}
Consider a supersymmetric field theory on a Riemannian three-manifold $\CM_3$. In general, the background metric breaks supersymmetry completely. Indeed, supersymmetry is an extension of the Poincar\'e symmetry group, the isometry group of flat space, which is also completely broken on an arbitrary manifold with a generic metric. On the other hand, a particular $\CM_3$ might admit  Killing vector fields $K^\mu$, which generate non-trivial isometries. Similarly, it might also admit (generalized) Killing spinors,  denoted by $\zeta_\alpha$ and $\t \zeta_\alpha$, which generate curved-space supersymmetries. The Killing vectors and spinors then generate a ``rigid supersymmetric algebra'' in curved space.

Curved-space rigid supersymmetry can be analysed as a limit of supergravity.~\cite{Festuccia:2011ws} See Ref.~\citen{Dumitrescu:2016ltq} for a thorough introduction to that approach. In our study of 3D $\CN=2$ gauge theories, we are interested in theories with an exact $U(1)_R$ symmetry. Such theories can be coupled to the so-called 3D $\CN=2$ ``new-minimal'' supergravity,~\cite{Festuccia:2011ws,Closset:2012ru,Kuzenko:2013uya}  whose bosonic field content is:
\be\label{sugra mult}
g_{\mu\nu}~, \qquad A_\mu^{(R)}~, \qquad H~, \qquad V_\mu~,
\ee
with $g_{\mu\nu}$ the metric, $A_\mu^{(R)}$ a background gauge field which couples to the $R$-symmetry current, and $H$ and $V_\mu$ some additional background auxiliary fields, with the constraint $\nabla_\mu V^\mu=0$.

For 3D $\CN=2$ theories with an $R$-symmetry, the generalized Killing spinor equations are:~\cite{Klare:2012gn, Closset:2012ru}
\bea\label{KSE}
&(\nabla_\mu -i A_\mu^{(R)})\zeta= - \half  H \gamma_\mu \zeta+ {i\ov 2}V_\mu\zeta - \half \epsilon_{\mu\nu\rho}V^\nu \gamma^\rho\zeta~, \cr
&(\nabla_\mu +i A_\mu^{(R)})\t\zeta= - \half  H \gamma_\mu \t\zeta - {i\ov 2}V_\mu\t\zeta + \half \epsilon_{\mu\nu\rho}V^\nu \gamma^\rho\t\zeta~.
\eea
This is the condition for the gravitino variation to vanish. 
A supersymmetric background:
\be\label{susy backg geom}
(\CM_3~,\, {\bf L}_R~; \; g~,\, A^{(R)}~,\, H~,\, V)~,
\ee
is a choice of supergravity background fields such that \eqref{KSE} admits some non-trivial solutions $\zeta$ and/or $\t\zeta$. In addition to the choice of Riemannian three-manifold $\CM_3$ with a metric $g_{\mu\nu}$, the background  \eqref{susy backg geom} includes a choice of a $U(1)_R$ line bundle ${\bf L}_R$ with connection $A_\mu^{(R)}$.~\footnote{We should note that $A_\mu^{(R)}$ corresponds to $A_\mu- {3\ov 2} V_\mu$ in the notation of Ref.\cite{Closset:2012ru}.} 
Since the first equation in \eqref{KSE} is linear and homogeneous, any non-trivial Killing spinor $\zeta$ must be nowhere-vanishing (and similarly for $\t \zeta$).
Note that $\zeta$ and $\t\zeta$ have $R$-charges:
\be
R[\zeta]=1~, \qquad \quad R[\t\zeta]=-1~.
\ee
Given two Killing spinors $\zeta$ and $\t\zeta$ of opposite $R$-charge, one can define a nowhere-vanishing Killing vector:
\be\label{K def}
K^\mu \equiv \t\zeta \gamma^\mu \zeta~.
\ee
The fact that $K^\mu$ is Killing follows directly from \eqref{KSE}.

\subsubsection{Curved-space supersymmetry algebra}
Given such a supersymmetric background, one can derive the curved-space supersymmetric Lagrangians and the supersymmetry variations as the ``rigid'' limit of the full supergravity; this is discussed thoroughly in Ref.\cite{Festuccia:2011ws, Closset:2012ru}. 

For instance, let us consider the chiral multiplet \eqref{chiral mult flat} coupled to a vector multiplet, with supersymmetry variations \eqref{chiral mult flat var}. Let $\fr$ denote the $U(1)_R$ charge of $\Phi$. This means that the scalar $\phi$ in $\Phi$ is valued in the space of sections:
\be
\Gamma\left( {\bf E}\otimes  {\bf L}^{\otimes r}_{R} \right)
\ee
where ${\bf E}$ is a vector bundle over $\CM_3$, corresponding to the gauge-group representation.~\footnote{Note that, depending on the global properties of the line bundle ${\bf L}_R$, there might be a restriction on the allowed $R$-charge $r$, so that the bundle $ {\bf L}_R^{\otimes r}$ is well-defined. We will come back to this point later on.}
Let us introduce the covariant derivative:
\be\label{def Dmu chiral}
D_\mu \varphi =\left(\nabla_\mu - i r_\varphi  (A_\mu^{(R)} +V_\mu) - i A_\mu\right)\varphi~,
\ee
with $r_\varphi= R[\varphi]$ the $R$-charge---that is, $r_\varphi= (r, r-1,r-2)$ for $\varphi=(\phi, \psi, F)$.
 Then, the curved-space version of the variations \eqref{chiral mult flat var}  reads:
\bea\label{susy chiral curved}
& \delta \phi = \sqrt 2 \zeta \psi~,\cr
& \delta \psi = \sqrt 2 \zeta F + \sqrt 2 i \left(\sigma + \fr H\right) \t \zeta \phi - \sqrt 2 i \gamma^\mu \t \zeta D_\mu \phi~,\cr
& \delta F = -\sqrt 2 i \left(\sigma + (\fr - 2)H\right) \t \zeta \psi + 2 i \phi  \t \zeta \lambda - \sqrt2 i D_\mu \big(\t \zeta \gamma^\mu \psi\big)~.
\eea
Here, $\zeta$ and/or $\t\zeta$ are non-trivial solutions to \eqref{KSE}. For any solutions $\zeta, \eta$ and $\t\zeta, \t \eta$ of the Killing spinor equations, the variations \eqref{susy chiral curved} realize the curved-space supersymmetry algebra:
\bea\label{susy alg curved space}
& \{\delta_\zeta, \delta_\eta\} \varphi=0~,\cr
& \{\delta_{\t\zeta}, \delta_{\t\eta}\} \varphi=0~,\cr
&\{\delta_\zeta, \delta_{\t \zeta}\} \varphi= -2 i \CL_K' \varphi +2 i \zeta \t\zeta (\sigma+ r_\varphi H)\varphi~,
\eea
with $K$ the Killing vector defined in \eqref{K def}, and $\CL'_K$ the covariant Lie derivative along $K$:
\be
\CL_K'=\CL_K- i r_\varphi K^\mu (A_\mu^{(R)}+V_\mu)- i A_\mu~,
\ee
for any two Killing spinors of opposite $R$-charges $\zeta$ and $\t \zeta$.

To write down the curved-space supersymmetric Lagrangian of $\Phi$ (for a canonical kinetic term), it is convenient to introduce a modified derivative:
\be
 {\mathscr D}_\mu = D_\mu + i  \fr_{0, \varphi} V_\mu,
 \ee
 with $D_\mu$ as in \eqref{def Dmu chiral}, and where $r_0$ denotes the superconformal $R$-charges of a {\it free} chiral multiplet, namely:
\be
r_0(\phi) = \half~, \qquad r_0(\psi)= -\half~.
\ee
Then, we have the supersymmetric Lagrangian:
\bea\label{Phi kin curved}
\SL_{\t\Phi\Phi}  =~&  { \mathscr D}^\mu \t \phi {\mathscr  D}_\mu \phi - i \t \psi \gamma^\mu {\mathscr D}_\mu \psi - \t F F  + \t \phi D \phi + 2 \big(\fr - \half\big) H \t \phi  \sigma \phi \cr
& + \t \phi\Big( \sigma^2  + {\fr \over 4} R + \half \big(\fr - \half\big ) V^\mu V_\mu  + \fr \big(\fr - \half\big) H^2 \Big) \phi \cr
&-i \t \psi \Big(\sigma + \big(\fr - \half\big) H\Big)\psi +  \sqrt 2 i  \big(\t \phi \lambda \psi +  \t \psi\t \lambda\phi \big)~.
\eea
Here, $R$ is the Ricci scalar. Note that, for $r=r_0=\half$, the Lagrangian becomes independent of $H$ and $V_\mu$, and we recover a conformally-coupled chiral multiplet. In general, however, those background fields are crucial to preserve supersymmetry. 

One can similarly write down the supersymmetric Lagrangians for the vector multiplet.~\cite{Closset:2012ru}

\subsubsection{A vector multiplet for the $R$-symmetry}
For later purpose, it is useful to point out that one can define a sub-multiplet of the full supergravity multiplet \eqref{sugra mult} which transforms as a vector multiplet, $\CV_R$. This is simply a vector multiplet for the $R$-symmetry, with the bosonic fields:
\be
\sigma^{(R)}= H~, \qquad \h A^{(R)}= A_\mu^{(R)}+ V_\mu~, \qquad  D^{(R)}= {1\ov 4} \big(R+2 H^2+2V_\mu V^\mu\big)~,
\ee
where $R$ is the Ricci scalar of $g_{\mu\nu}$.
A number of curved-space computations can be simplified when written in terms of $\CV_R$. Moreover, the curved-space supersymmetry algebra \eqref{susy alg curved space} can be understood as a coupling to $\CV_R$ in the WZ gauge, at least formally.~\footnote{The important difference with an ordinary vector multplet is that $\CV_R$ couples to the chiral multiplet fields through their $R$-charges. Moreover, we have to keep in mind that the ``supersymmetry parameters'' $\zeta$ and $\t\zeta$ are non-trivial solutions of the generalized Killing spinor equations on $\CM_3$.}

In the presence of any abelian global symmetry $U(1)_F$, one can consider mixing the $R$-symmetry with $U(1)_F$ to obtain a new $R$-symmetry. At the level of the conserved currents, we have:
\be\label{mixing jR}
j_\mu^{(R)} \rightarrow j_\mu^{(R)} + t_F \, j_\mu^{(F)}
\ee
with $t_F \in \R$ the mixing parameter. Due to the minimal coupling:
\be
 A_\mu^{(F)} j^\mu_{(F)}+\h A_\mu^{(R)} j^\mu_{(R)} + \cdots~,
\ee
where the ellipsis denotes terms required by supersymmetry, the mixing \eqref{mixing jR} is equivalent to mixing the background vector multiplet $\CV_F$ for $U(1)_F$ with  $\CV_R$ for~$U(1)_R$:
\be
\CV_F \rightarrow \CV_F+ t_F \CV_R~.
\ee
This simple fact is the key to understand the $R$-charge dependence of supersymmetric partition functions.~\cite{Closset:2014uda}

\subsection{Transversely holomorphic foliations and supersymmetry}
An oriented closed Riemannian three-manifold $\CM_3$ admits a generalized Killing spinor $\zeta$ if and only if it admits a transversely holomorphic foliation (THF) with an adapted metric.~\cite{Closset:2012ru, Closset:2013vra}  A THF is a one-dimensional foliation, defined by a nowhere-vanishing vector $\xi$. The orbits of $\xi$ are called the leaves of the foliation, and the space of leaves can be covered by complex coordinates with holomorphic transition functions. That is, there exists local coordinates $\tau, z, \b z$ on $\CM_3$, the ``adapted coordinates,''  such that:
\be
\xi = \d_{\tau}~, 
\ee
and such that changes of adapted coordinates are holomorphic: 
\be\label{change coord thf}
\tau'= \tau + \lambda(z, \bz)~,\qquad z'= f(z)~,\qquad  \bz'= \b f(\bz)~.
\ee
One can choose an adapted metric $g$ such that $\xi$ has unit norm. Let us denote the one-form dual to $\xi$ by $\eta$, with $\eta_\mu= g_{\mu\nu} \xi^\nu$. In adapted coordinates, we have:
\be
ds^2(\CM_3) = \eta^2 + 2 g_{z\bz }(\tau, z, \bz) dz d \bz~, \qquad \eta= d \tau + h(\tau, z,\bz) dz+\b h(\tau, z,\bz) d\b z~.
\ee
The THF with adapted metric can be defined entirely in terms of a one-form $\eta$ of unit norm, satisfying a certain integrability condition. Given $\eta$ on an oriented Riemanian three-manifold, we define the two-tensor:
\be
{\Phi_\mu}^\nu= - {\epsilon_\mu}^{\nu\rho}\eta_\rho~.
\ee
We thus have:
\be
\eta^\mu \eta_\mu=1~, \qquad  {\Phi_\mu}^\nu {\Phi_\nu}^\rho =- {\delta^\mu}_\nu+ \eta^\mu \eta_\rho~.
\ee
Here and in the following, we often write $\eta^\mu= \xi^\mu$, using the metric to raise and lower indices. 
Note that $\Phi$ behaves like a 3D analogue of an almost-complex structure.
The tensors $\eta$ and $\Phi$ define a THF if and only if the following integrability condition is satisfied:~\cite{Closset:2012ru}
\be
 {\Phi_\mu}^\nu (\CL_\xi  {\Phi_\nu}^\rho)=0~,
\ee
with $\CL_\xi$ the Lie derivative along $\xi$.

\paragraph{THF from Killing spinor.} The THF is related to the Killing spinor $\zeta$ by:
\be
\xi^\mu = {\zeta^\dagger \gamma^\mu \zeta\ov |\zeta|^2}~.
\ee
Note that $\xi$ is not a Killing vector, in general. The fact that this defines a THF is a non-trivial consequence of 
the Killing spinor equation \eqref{KSE}. 

\subsubsection{Holomorphic forms and the canonical line bundle of $\CM_3$}\label{subsec: holo form}
Given $(\CM_3; \eta, \Phi)$, a three-manifold equipped with a THF, there is a natural notion of an holomorphic one-form. Consider the projector:
\be\label{holo proj}
 {\Pi^\mu}_\nu =\half \left({\delta^\mu}_\nu - i {\Phi^\mu}_\nu- \eta^\mu \eta_\nu\right)~.
\ee
By definition, a {\it holomorphic one-form} on $\CM_3$ is such that:
\be\label{holo oneform}
  \omega_\mu {\Pi^\mu}_\nu= \omega_\nu~.
\ee
In adapted coordinates, such a one-form has a single component, $\omega= \omega_z dz$.  
The {\it canonical line bundle} of $\CM_3$, denoted by $\CK_{\CM_3}$, is the line bundle of holomorphic forms, with sections:
\be
\omega_z \in \Gamma[\CK_{\CM_3}]~.
\ee
Under a change of adapted coordinates \eqref{change coord thf}, we have $\omega_{z'} = (\d_z f)^{-1} \omega_z$.

\subsubsection{Supergravity background fields from the THF}
Given a THF with adapted metric on $\CM_3$, the other supergravity background fields are:~\cite{Closset:2012ru}\footnote{Our $\eta_\mu$ corresponds to $-\eta_\mu$ in Refs~\protect\cite{Closset:2012ru, Closset:2013vra}.}
\bea\label{gen sol KSE}
& H =\half \nabla_\mu \eta^\mu + {i\ov 2}\epsilon^{\mu\nu\rho} \eta_\mu\d_\nu \eta_\rho + i \kappa~,\cr
& V_\mu = - \epsilon_{\mu\nu\rho}\d^\nu \eta^\rho- \kappa \eta_\mu+ U_\mu~,\cr
& A_\mu^{(R)}=  \CA_\mu^{(R)} + \half \epsilon_{\mu\nu \rho} \d^\nu\eta^\rho+ {i\ov 4} \eta_\mu \nabla_\nu \eta^\nu - {i\ov 2} \eta^\nu \nabla_\nu \eta_\mu~,
\eea
with $\CA^{(R)}$ given by:
\be\label{def CAR ii}
\CA^{(R)}_\mu= {1\ov 4} {\Phi_\mu}^\nu \d_\nu \log\sqrt{g}+ \d_\mu s ~,
\ee
in the adapted coordinates $\psi, z, \bz$. Here, the function $\kappa$ and the vector $U_\mu$ are ambiguities in the determination of $H$ and $V_\mu$, which must satisfy:
\be
U_\mu{ \Phi^{\mu}}_\nu = - i U_\nu~, \qquad \nabla_\mu(U^\mu - \kappa \eta^\mu)=0~.
\ee
The function $s$ in \eqref{def CAR ii} is a $U(1)_R$ gauge parameter, which may satisfy constraints so that the background and the Killing spinor are globally defined. In the adapted frame:
\be\label{adapted frame def}
e^0 = \eta~, \qquad e^1= \sqrt{2 g_{z\bz}} dz~, \qquad e^{\b 1}= \sqrt{2 g_{z\bz}} d\bz~,
\ee
the Killing spinor $\zeta$ is simply given by:
\be
\zeta= e^{i s} \mat{0\\ 1}~.
\ee


\subsubsection{THF deformations and moduli}
Transversely holomorphic foliations sometimes come in continuous families, indexed by THF moduli, which are very similar to the complex structure moduli of complex manifolds. The deformation theory for THFs was discussed in Refs \cite{duchamp1979deformation,girbau1989deformations, Brunella1995, Brunella1996, Ghys1996}.
On $\CM_3$ equipped with a THF, one can define $(p,q)$-forms, $\omega^{(p,q)} \in \Lambda^{p,q}$, generalizing the case of the holomorphic form ($(1,0)$-form) above. Then, there exists a nilpotent Dolbeault-like operator:
\be
\t \d \; : \; \Lambda^{p,q} \rightarrow \Lambda^{p,q+1}~,
\ee
with its cohomology denoted by $H^{p,q}(\CM_3)$. At first order, non-trivial deformations of the THF are parameterized by elements of anti-holomorphic one-forms valued in the holomorphic tangent bundle:
\be
[\Theta^z] \in H^{0,1}(\CM_3, T^{1,0} \CM_3)~.
\ee
For an explicit description of these deformations, and a detailed discussion of some important examples, see Ref.~\citen{Closset:2013vra}.

\subsubsection{The classification of THFs}\label{subsubsec: THF class}
Transversely holomorphic foliations of three-manifolds have been well studied by mathematicians, and there exists a complete classification.~\cite{Brunella1995,Brunella1996,Ghys1996}  Topologically, any oriented $\CM_3$ that admit a THF is either a Seifert manifold or a torus bundle. 

\paragraph{A {\it Seifert manifold} $\CM_3$} is three-manifold that admits a locally-free $U(1)$ action.
Equivalently, it is a circle bundle over a two-dimensional orbifold:
\be\label{seifert fib}
S^1\longrightarrow \CM_3 \stackrel{\pi}\longrightarrow\h\Sigma_{g, n}~.
\ee
Here, $\h\Sigma_{g,n}$ denotes a closed Riemann surface of genus $g$ with $n$ orbifold points. An orbifold point (also known as a ramification point) is a marked point whose neighbourhood is modelled on $\C/\Z_q$ instead of $\C$, for some integer $q>1$; correspondingly, the metric on $\h\Sigma$ has a conical singularity at that point.  We will  discuss Seifert geometry in more detail momentarily.

\paragraph{A {\it torus bundle} over the circle,} on the other hand, can be obtained from a torus times an interval, $T^2 \times I$, by identifying the end points of the interval up to a large diffeomorphism of the torus. Given any $A\in SL(Z, \Z)$, we define an oriented three-manifold:
\be\label{M3A def}
\CM_3^A \cong  (T^2 \times I)/\sim_A~.
\ee
Theses two classes of manifold have a non-zero overlap. Recall that an $SL(2, \Z)$ transformation $A$ is called elliptic if $|\tr(A)| <2$, parabolic if  $|\tr(A)| =2$, or hyperbolic if $|\tr(A)| >2$. The torus bundle $\CM_3^A$ is also Seifert if and only if $A$ is either elliptic or parabolic. There is a small list of possibilities, up to conjugation:~\cite{Hatcher}
\be
A= \mat{1 & 0\\ p & 1}~, \quad 
\mat{-1 & 0\\ p & -1}~, \quad 
\mat{0 & -1\\ 1 & 0}~, \quad 
\mat{0 & -1\\ 1 & 1}~, \quad
\mat{-1 & -1\\ 1 & 0}~,
\ee
with $p\in \Z$.
The first two cases are parabolic, and the last three are elliptic (those  three correspond to $S$, $ST$ and $(ST)^2$, respectively, in the standard presentation of $SL(2, \Z)$.)

\paragraph{The full classification of THFs} goes as follows. Any THF on $\CM_3$ (a closed connected three-manifold) is contained in at least one of the following six cases:~\cite{Brunella1996,Ghys1996}  
\begin{itemize}
\item[(i)]  Seifert manifold $\CM_3$, with the THF given by the Seifert fibration. In that case, the leaf space is an orbifold $\h\Sigma$ as in \eqref{seifert fib}.
\item[(ii)] Linear foliation of $T^3$. Any constant vector $\xi$ of unit norm on a flat $T^3$ defines a THF.
\item[(iii)] ``Strong stable foliation'' of an hyperbolic torus bundle $\CM_3^A$---that is, with $A$ such that $|\tr(A)|>2$---, with the fibration defined as follows. Let $(\lambda, \lambda^{-1})$ be the eigenvalues of $A$, with $\lambda\in \R$ by the hyperbolic assumption. Then, the first eigenvector of $A$, $\xi_\lambda$, defines a  linear foliation of $T^2$. This foliation is invariant under the quotient \eqref{M3A def},  therefore it defines a THF on $\CM_3^A$.
\item[(iv)] Suspension of an holomorphic automorphism of $\mathbb{P}^1$. That is, the quotient of $S^2 \times \R$ by a M\"obius transformation:
\be
(z, \tau) \sim (M\cdot z, \tau+1)~,   \qquad M\cdot z\equiv {a z + b\ov c z +d}~,
\ee
with $(z, \tau)\in S^2\times \R$ and  $M\in PGL(2, \C)$. Here, the THF is the one corresponding to the adapted coordinates $(z, \tau)$.  The topology of this three-manifold is $S^2 \times S^1$.
\item[(v)]  Transversely $\C$-affine foliation of $S^2 \times S^1$. That is, the quotient:
\be\label{THF s2s1}
(z, \tau) \sim ({\bf q} z, \tau+ \log|{\bf q}|), \qquad {\bf q}\in \C~,\;\; |{\bf q}|< 1~,
\ee
with $(z, \tau) \in S^2 \times \R$ the adapted coordinates. 
 The topology of this three-manifold is  also $S^2 \times S^1$. The examples (iv) and (v) exhaust the possibilities for THFs on $S^2 \times S^1$.
\item[(vi)]  ``Poincar\'e foliation'' on a lens space $L(p,q)$, including $S^3= L(1,1)$.~\cite{Brunella1995,Brunella1996} Such foliations are indexed  by continuous parameters, the THF moduli, which are often called ``squashing parameters'' in the supersymmetric localization literature. We will come back to this case later on.  
\end{itemize}
The existence of a THF on $\CM_3$ is essentially equivalent to the existence of a complex structure on a four-manifold $\CM_4$ with a $\C^\ast$ action, such that $\CM_3= \CM_4/U(1)$ under that action; in fact, the above classification of 3D THFs relies on the Enriques-Kodaira classification of minimal complex surfaces.~\cite{Brunella1996} This fact has a simple physics counterpart: 3D $\CN=2$ theories on $\CM_3$ can be uplifted to 4D $\CN=1$ theories on $\CM_4$,~\cite{Closset:2012ru} and the presence of a generalized Killing spinor $\zeta$ in 4D is equivalent to the existence of a complex structure on the four-manifold.~\cite{Dumitrescu:2012ha, Klare:2012gn} In particular, the Poincar\'e foliations of $L(p,q)$ correspond to Hopf surfaces $\CM_4 \cong L(p,q) \times S^1$, as analysed in Ref.~\citen{Closset:2013vra, Aharony:2013dha}---the 3D squashing parameters are complex structure moduli of the Hopf surface.

Note that, topologically, all the three-manifolds of cases (i), (ii), (iv), (v) and (vi) are Seifert manifolds, while the hyperbolic torus bundles of case (iii) are not.~\footnote{A side note: Amongst Thurston's eight model geometries (see Ref.~\protect\cite{BLMS:BLMS0401} for a review), six admit a Seifert structure---the $S^3$ (spherical), $E^3$ (euclidean), $S^2 \times \R$, $H^2 \times \R$, $Sl(2, \R)$ and Nil geometries.  On the other hand, hyperbolic torus bundles, corresponding to the THF (iii) above, have a Solv geometry. The $H^3$ (hyperbolic) Thurston geometry is not compatible with a THF, on the other hand, and therefore one cannot couple 3D $\CN=2$ supersymmetric theories to hyperbolic three-manifolds in a supersymmetric manner.}

\subsection{Half-BPS geometries}\label{subsec: halfBPS geom}
All the exact results obtained by supersymmetric localization of 3D $\CN=2$ theories, so far, concern observables which are half-BPS---that is, they preserve at least {\it two} curved-space supercharges.

Let us consider supersymmetric backgrounds $\CM_3$ that admit two Killing spinors $\zeta$ and $\t\zeta$, of opposite $R$-charge. It follows that there exists a nowhere-vanishing Killing vector $K^\mu$ on $\CM_3$, bilinear in the Killing spinors, as defined in \eqref{K def}. There are two possibilities: 
\begin{itemize}
\item The Killing vector $K$ is {\it real} and generates a single isometry of $\CM_3$. The THF associated to $\zeta$ is given by:~\footnote{We also have a THF associated with $\t\zeta$, which only differs from $\eta$ by a sign in this case, $\t\eta= - \eta$.~\cite{Closset:2012ru}}
\be\label{xi from K}
\xi^\mu = {1\ov |K|}K^\mu~.
\ee
This give, by far, the largest class of half-BPS geometries, and all their supersymmetric partition functions have been computed. This will be the main focus of this review, from section \ref{sec: susy loc gen} to \ref{subsec: seifert operators}.
\item The Killing vector $K$ is {\it complex} and generates {\it two} independent isometries of $\CM_3$. 
Here, we must distinguish two subcases: Either one can make $K$ real by a continuous deformation of the THF moduli, bringing us to the previous case; or, $K$ is complex for any value of the THF moduli. This last condition is very restrictive, and corresponds to a THF of class (v) in subsection~\ref{subsubsec: THF class}, with $\CM_3$ of topology $S^2 \times S^1$. 
The corresponding partition function $Z_{S^2 \times S^1}$ computes the so-called 3D superconformal index. We will study that index in section~\ref{sec: S2S1 index}.
\end{itemize}

For backgrounds in the first case, and for their continuous deformations contained in the second case, the THF associated to the Killing spinor $\zeta$ corresponds to the classes (i), (ii), (iv) or (vi), in the classification of subsection~\ref{subsubsec: THF class}.

The generic case corresponds to class (i), where the foliation is a Seifert fibration---then, the Killing vector $K$ which appears in the curved-space supersymmetry algebra, as in \eqref{susy alg curved space}, generates the isometry along the $S^1$ fiber in \eqref{seifert fib}. However, if the two-dimensional base $\h\Sigma_{g, n}$ admits an isometry, we may also consider a deformation:
\be\label{K epsilon def}
K = K_{S^1}+ \epsilon  K_{\h\Sigma}~,
\ee
with the parameter $\epsilon \in \R$, and with $K_{S^1}$ and $K_{\h\Sigma}$ the Killing vectors along the fiber and the base, respectively. This setup gives rise to a one-dimensional family of supersymmetric backgrounds, for a given $\CM_3$. This is possible in a special but important case, when $g=0$ and $n \leq 2$. Then, the Seifert manifold has the topology of a lens space, $L(p,q)$, viewed as a circle bundle over $S^2$ with two marked points, and the parameter $\epsilon$ is related to a certain ``squashing parameter,'' usually denoted by $b \in \R$. This includes the case of the ``squashed three-sphere,''  $S^3_b$, which we review in detail in section~\ref{sec: lens space}. In those cases, one can naturally take the parameter $\epsilon$ (or, equivalently, $b$) to be complex. This covers many ``Poincar\'e foliations'' that fall in class (vi) above. It also covers the $S^2 \times S^1$ background of class (iv) with $M$ a diagonal matrix, giving rise to the so-called topologically-twisted index background.~\cite{Benini:2015noa}

Beyond lens spaces, the only other possibility, where we can consider the deformation \protect\eqref{K epsilon def} of a Seifert fibration, is for $g=1$ and $n=0$. This corresponds to a THF of class (ii).~\footnote{This case has not been studied systematically in the supersymmetric literature, as far as we know.}

Let us insist  on the obvious fact that the THF condition, which follows from the presence of a single curved-space supercharge with Killing spinor $\zeta$, is weaker than the half-BPS condition of having both $\zeta$ and $\t\zeta$. Nonetheless, the half-BPS background are actually generic, in the sense that Seifert manifolds (with a THF of example (i)) always admit a second supercharge. 
For the THFs of example (ii), we can take the flat metric on $T^3$ and preserve four supercharges. The cases (iv), (v) and (vi) all admit interesting one-parameter families of half-BPS background, which are usually denoted by:
\be
L(p,q)_b
\ee
with $b\in \C$ a ``squashing'' parameter of the lens space background. For $p \neq 0$, we have a Poincar\'e foliation of a lens space, falling in class (vi). For $p=0$, we denote by:
\be\label{susy top index}
L(0,1)_\epsilon \cong  S^2_\epsilon\times S^1~, \qquad \qquad
L(0,-1)_{\bf q} \cong  S^2_{\bf q}\times S^1~, 
\ee
the $S^2 \times S^1$ half-BPS backgrounds that give the twisted index (corresponding to a THF of class (iv)) and the superconformal index (corresponding to the THF of class (v)), respectively.

\subsection{Half-BPS Seifert geometry and the 3D A-twist}\label{subsec: 3DAtwist}
From now on, let us focus on half-BPS geometries that correspond to Seifert fibrations, or continuous deformations thereof. This covers every half-BPS background, with one exception. The exception is the  $S^2 \times S^1$ background (with the THF of example (v)) used to compute the superconformal index, which we will discuss in section~\ref{sec: S2S1 index}.

Unless otherwise stated, in the following and for simplicity, we choose the Killing vector $K^\mu= \t\zeta\gamma^\mu \zeta$ to be real.

\subsubsection{Seifert geometry: A primer}
Topologically, an oriented Seifert three-manifold  is fully specified by the data of its Seifert invariants, denoted by:
\be\label{seifert invariant 1}
\CM_3 \cong \big[\bd~; \, g~; \, (q_1, p_1)~, \cdots~, (q_n, p_n)\big]~.
\ee
Here, $\bd$ is the {\it degree} of the circle bundle and $g$ is the \emph{genus} of the base orbifold $\hat\Sigma_g$. The data at each orbifold points $x_n\in\hat\Sigma$ is determined by a pair of co-prime integers $(q_i, p_i)$, which corresponds to the insertion of an exceptional fiber at a $\mathbb{Z}_{q_i}$ orbifold point $x_i$ on the base.

A Seifert manifold $\CM_3^{(n+1)}$ with $n+1$ exceptional fibers can be obtained by Dehn surgery around a generic point on $x_{n+1}\in\hat \Sigma$, where $\hat\Sigma$ is a base of a Seifert manifold with $n$ exceptional fibers $\CM_3^{(n)}$ (at some fixed genus, $g$). After removing the infinitesimal tubular neighborhood around $x_{n+1}$, we glue back the boundary of a solid torus $\partial D^2\times S^1$ with $\partial\CM_3^{(n)} \cong T^2$ by an $SL(2,\mathbb{Z})$ transformation:
\be\label{dehn surgery}
\left(\begin{array}{c}\psi\\\varphi\end{array}\right) = \left(\begin{array}{cc}q & -t\\ p & s\end{array}\right)\left(\begin{array}{c}\psi'\\\varphi'\end{array}\right)\ ,~~~qs + pt=1\ ,
\ee
where $(\psi,\phi)$ and $(\psi',\phi')$ are coordinates on the boundary of $D^2\times S^1$ and $\partial\CM_3$ respectively. This operation introduces a $\mathbb{Z}_{q}$ orbifold point on $\hat\Sigma_g$. The Seifert fiber at $x_{n+1}$ is called the $(q,p)$-\emph{exceptional fiber}. 

The geometry of a Seifert manifold $\CM_3$ is invariant under the shift:
\be
\bd\rightarrow \bd-1,\qquad~p_i\rightarrow p_i + q_i~, 
\ee
for any $i$, as well as under the inversion $(q_i,p_i)\rightarrow (-q_i,-p_i)$. The Seifert invariant \eqref{seifert invariant 1} are usually normalised such that $q_i >0$ and $0 \leq p_i <q_i$. Note that inserting a $(q,p)=(1,1)$-exceptional fiber is equivalent to shifting the degree, $\bd\rightarrow \bd+1$.

\paragraph{Construction of Seifert manifolds.} From the above discussion, we see that any Seifert manifold $\CM_3$ can be constructed from the ``mother manifold" $S^2\times S^1$, by performing a combination of the following operations:
\begin{itemize}
\item[(i)] Add a handle to the base Riemann surface, shifting the genus as $g\rightarrow g+1$.
\item[(ii)] Shift the degree of the circle fibration, $\bd\rightarrow \bd+1$.
\item[(iii)] Add an exceptional $(q,p)$-fiber on the base Riemann surface.
\end{itemize}
As we will see in section~\ref{subsec: seifert operators}, half-BPS observables on any Seifert manifold $\CM_3$ can be constructed from the knowledlge of the $S^2 \times S^1$ observables, once we understand the above three geometric operations at the level of the path integral.

\paragraph{Examples of half-BPS Seifert geometries.} Supersymmetric partition functions on lens spaces  have been extensively studied in the literatures via various different localization techniques. The most widely studied example is that of the squashed three-sphere, $S^3_b$, which can be realized as a Seifert geometry when $b^2 \in\mathbb{Q}$:
\be\label{S3b as seifert}
S^3_b \cong [0~; \, 0~; \, (q_1, p_1)~, (q_2, p_2)]\ ,\quad~~q_1p_2+q_2p_1=1\ ,\quad~~b^2 = \frac{q_1}{q_2}\ .
\ee 
We will discuss the $S^3_b$ partition function (for any $b$) in section \ref{sec: lens space}. 
Another interesting class of half-BPS backgrounds is $L(0,1)\cong S^2_\epsilon \times S^1$ with a topological twist on the $S^2$. The corresponding partition function is known as the refined twisted index. This background can be realized as the Seifert fibration if and only if the deformation parameter $\epsilon$ is rational:
\be
\epsilon = \frac{t}{q}\ ,~~~\text{gcd}(q,t)=1\ .
\ee
The corresponding Seifert geometry can be written as
\be
S^2_{\epsilon}\times S^1 \cong [0~; \, 0~; \, (q, p)~, (q, -p)]\ ,~~qs+pt=1\ .
\ee 
We will discuss this example (for any $\epsilon$) in section~\ref{sec: S2 top index}.

Note that Seifert manifolds with $g=0$ and up to $2$ orbifold points are always {\it lens spaces}---including $S^3$ and $S^2 \times S^1$. In sections \ref{sec: bethe} and \ref{subsec: seifert operators}, we will review the approach recently studied in Refs.~\citen{Closset:2017zgf,Closset:2018ghr,Closset:2016arn,Benini:2016hjo,Benini:2015noa,Gukov:2015sna} that allows us to explicitly construct the half-BPS observables on \emph{any} Seifert manifold $\CM_3$.

\subsubsection{The 3D A-twist}
Consider $\CM_3$ a Seifert manifold. We choose a THF defined by an isometry $K$, as in~\eqref{xi from K}.
By a conformal rescaling of the metric, we can always take $|K|^2 =1$, which we will do from now on. We then have:
\be\label{standard met}
ds^2(\CM_3) = \beta^2 (d\h \psi + \CC_z dz +  \CC_\bz d\bz)^2 + 2 g_{z\bz} dz d\bz~,
\ee
in some adapted local coordinate ${\h\psi}, z, \bz$, with:
\be
K= {1\ov \beta} \d_{\h \psi}~, \qquad \eta=  \beta (d\h \psi + \CC_z dz +  \CC_\bz d\bz)
\ee
Note that the orbits of $K$ might not close, in general, which corresponds to the situation described around Eq.~\eqref{K epsilon def}.

The expressions  \eqref{gen sol KSE}   for the background supergravity fields simplify to:~\footnote{Here we must have $U_\mu=0$, and we choose $\kappa=0$. The supersymmetric observables are independent of that choice.~\protect\cite{Closset:2013vra}}
\be\label{HVA Seifert}
H= {i\ov 2}\epsilon^{\mu\nu\rho} \eta_\mu\d_\nu \eta_\rho \equiv i \h h~, \qquad
 V_\mu = - 2 \h h \eta_\mu~,\qquad
 A_\mu^{(R)}=  \CA_\mu^{(R)} + \h h \eta_\mu~,
\ee
where we use the fact that $\half \epsilon_{\mu\nu\rho}\d^\nu \eta^\rho=\h h \eta_\mu$.
Given the isometry and the transverse holomorphic structure, we can decompose any tensor in vertical, holomorphic and anti-homomorphic components. For instance, a one-form $\alpha$ is decomposed into:
\be\label{decompos alpha}
\alpha = \alpha_0 \eta + \alpha_z dz +\alpha_\bz d\bz~.
\ee
The projector onto the vertical component is  ${{\rm P_0}^\mu}_\nu= \eta^\mu \eta_\nu$, the projector onto the holomorphic component was given in \eqref{holo proj}, and the anti-holomorphic projector is its complex conjugate.
We discussed the case of a holomorphic one-form in  \eqref{holo oneform}.
Similarly, an holomorphic vector, $X$, is such that:
\be
 {\Pi^\mu}_\nu X^\nu = X^\mu~.
\ee
It has a single component, with $X= X^z(\d_z-\CC_z \d_\psi)$ in adapted coordinates. 

\paragraph{Adapted connection.}
The Levi-Civita connection $\nabla$ does not preserve the decomposition \eqref{decompos alpha}. It is very useful to introduce a modified connection, $\h \nabla$, adapted to the transversely holomorphic structure:~\cite{Closset:2012ru}
\be
\h\nabla_\mu g_{\nu\rho}=0~, \qquad\qquad \h\nabla_\mu \eta_{\nu}=0~.
\ee
The adapted spin connection is defined by:
\be\label{def omega hat}
\h \omega_{\mu\nu\rho}= \omega_{\mu\nu\rho} +\h h \left(\eta_\nu \Phi_{\mu\rho}-\eta_\rho \Phi_{\mu\nu}+\eta_\mu \Phi_{\nu\rho}\right)~,
\ee
with $\omega_{\mu\nu\rho}$ the standard spin connection, and $\h h$ as defined in \eqref{HVA Seifert}. Note that $\h\nabla$ has a non-zero torsion tensor:
\be
{T^{\nu}}_{\mu\rho}= -2 \h h \eta^\nu \Phi_{\mu\rho}~.
\ee

It is also useful to use the adapted frame coordinates with indices $(0, 1, \b 1)$, instead of local coordinates $(\h\psi, z, \bz)$, whenever possible; the adapted frame is as in \eqref{adapted frame def}. For instance, in the frame coordinates, the holomorphic one-form and the holomorphic vector read $\omega = \omega_{1} e^1$ and $X= X^1 \d_{1}$, respectively. In the frame basis, tensor components can be assigned a definite 2D spin, corresponding to frame rotations transverse to $\eta$. For instance, the components $\alpha_0$, $\alpha_1$ and $\alpha_{\b1}$ of a generic one-form \eqref{decompos alpha} have 2D spin $0$, $1$ and $-1$, respectively.

\paragraph{The 3D A-twist.} Using the adapted connection, the generalized Killing spinor equation can be written as:
\be\label{KSE twist}
\big(\h \nabla_\mu - i \CA^{(R)}_\mu\big)\zeta=0~, \qquad \qquad
\big(\h \nabla_\mu + i \CA^{(R)}_\mu\big)\t\zeta=0~,
\ee
with the $U(1)_R$ connection $\CA^{(R)}_\mu$ given in \eqref{def CAR ii}. The holonomy of  $\h\nabla_\mu$ is contained in $U(1)$, therefore it can be ``twisted'' away by a compensating $U(1)_R$ transformation. The generalized Killing spinors are given by:
\be\label{KS explicit}
\zeta=e^{i s}\mat{0\cr 1}~, \qquad\qquad \t\zeta=e^{-i s}\mat{1\cr 0}~,
\ee
in the adapted frame.
 This is the three-dimensional uplift of the topological A-twist of 2D $\CN=(2,2)$ supersymmetric field theories.~\cite{Witten:1988xj} Indeed, for $\CM_3 = \Sigma_g \times S^1$ with $K$ pointing along the $S^1$, $\h\nabla=\nabla$ and we obtain the 2D A-twist on $\Sigma_g$.

 Geometrically, the 3D A-twist corresponds to choosing the $U(1)_R$ line bundle over $\CM_3$, denoted by ${\bf L}_{R}$, such that:
\be\label{A twist as bundles}
({\bf L}_{R})^{\otimes 2}\cong \CK_{\CM_3}~,
\ee
with $\CK_{\CM_3}$ the canonical line bundle associated to the THF. This can be understood as follows. Given the nowhere-vanishing Killing spinors, we can construct the one-forms:
\be
p_\mu = \zeta\gamma_\mu \zeta~, \qquad \qquad \t p_\mu = \t\zeta\gamma_\mu\t \zeta~,
\ee
of $R$-charges $2$ and $-2$, respectively. We have:
\be\label{def p pb ii}
p_\mu dx^\mu =p_{\b1} e^{\b 1} = -e^{2 i s} e^{\b1}~, \qquad \qquad \t p_\mu dx^\mu =\t p_1 e^1= e^{-2 i s} e^{1}~.
\ee
In particular, $\t p_1$ is a nowhere-vanishing section of the line bundle $\CK_{\CM_3}\otimes ({\bf L}_{R})^{-2}$, which implies~\eqref{A twist as bundles}.

\subsubsection{Fields and Lagrangians in the A-twisted notation}\label{subsec: Atwist fields}
Let us again consider a 3D $\CN=2$ gauge theory. The various fields can be written in the ``A-twisted notation,'' in terms of forms of vanishing $R$-charge and definite 2D spin. These forms are constructed by contraction with the Killing spinors, in such a way as to absorb any $R$-charge carried by the fundamental fields. Given any field $\varphi$ of $R$-charge $r_\varphi$ and 2D spin $s_\varphi$ in the ``flat space'' notation, the associated twisted field has vanishing $R$-charge and A-twisted 2D spin:
\be\label{Atwisted spin}
s^{(A)}_\varphi = s_\varphi + \half r_\varphi~.
\ee
Indeed, in 2D notation---projecting onto the 2D frame transverse to $K$---, the non-zero Killing spinors are $\zeta_+$ and $\t\zeta_-$, which have $R$-charge $\pm 1$ and 2D spin $\mp \half$, respectively.

\paragraph{Vector multiplet.} Consider first the vector multiplet. The gauginos $\lambda, \t\lambda$ have $R$-charge $\pm1$, therefore they are affected by the twist. We define the new variables:
\be
\Lambda_\mu \equiv \t\zeta\gamma_\mu\lambda~, \qquad \qquad
\t\Lambda_\mu \equiv -\zeta\gamma_\mu\t\lambda~.
\ee
Expanding in vertical and horizontal components, we have:
\bea\label{gauginiA} 
&\Lambda_\mu dx^\mu =  \Lambda_0 e^0 + \Lambda_1 e^1~, \quad\qquad
&\t\Lambda_\mu dx^\mu =  \t \Lambda_0 e^0 + \t \Lambda_{\b 1} e^{\b 1}~,
\eea
where the vertical components $\Lambda_0$ and  $\t\Lambda_0$ are ``2D scalars'' (that is, scalars under changes of adapted coordinates),  and the horizontal components $\Lambda_1$ and $\t\Lambda_{\b1}$ are sections of $\CK_{\CM_3}$ and $\b\CK_{\CM_3}$, respectively.

Let us introduce the adapted covariant derivative:
\be
D_\mu = \h \nabla_\mu - i A_\mu~,
\ee
and let $\delta$ and $\t\delta$ denote the supersymmetry variations along $\zeta$ and $\t\zeta$, respectively. Then, the curved-space supersymmetry variations of the vector multiplet read:
\bea\label{susyVector twisted}
&\delta A_\mu = i \t\Lambda_\mu~,   
\quad &&  \t\delta A_\mu = - i \Lambda_\mu\cr 
&\delta\sigma =\t\Lambda_0~,
\quad &&  \t\delta \sigma = -\Lambda_0~,\cr  
& \delta \Lambda_0 =i \left(D -\sigma H - 2 i F_{1\b 1}\right)+ i D_0 \sigma~,
\quad\;  && \t\delta \Lambda_0 =0~,\cr
& \delta \Lambda_1 =2 F_{01} + 2 i D_1 \sigma~,
\quad &&   \t\delta \Lambda_1 =0~,  \cr
& \delta \t\Lambda_0 =0~,
\quad  &&\hspace{-0.8cm} \t\delta \t\Lambda_0 =i \left(D -\sigma H - 2 i F_{1\b 1}\right)- i D_0 \sigma~,\cr
&\delta \t\Lambda_{\b1} =0~,
\quad &&\hspace{-0.8cm}  \t\delta \t\Lambda_{\b1}= -2 F_{0\b 1} - 2 i D_{\b1}\sigma~, \cr
&\delta D= -D_0\t\Lambda_0 - 2 D_1 \t\Lambda_{\b 1}\quad
 &&\hspace{-0.8cm}  \t\delta D= -D_0\Lambda_0- 2D_{\b 1} \Lambda_1 \cr
 &\quad \quad -H \t\Lambda_0+ [\sigma, \t\Lambda_0]~,\quad 
 &&\hspace{-0.8cm}  \quad\qquad +H \Lambda_0 +[\sigma, \Lambda_0] 
\eea
The dependence of \eqref{susyVector twisted} on the geometric background is mostly implicit, through the covariant derivatives written in the frame basis (as well as through $H= i \h h$).~\footnote{To check the supersymmetry algebra, we use the fact that:
\be\nn
F_{01}= D_0 A_1- D_1 A_0~, \qquad
F_{0\b1}= D_0 A_{\b1}- D_{\b1} A_0~, \qquad
F_{1{\b1}}= D_1 A_{\b1}- D_{\b1} A_1 + H  A_0~.
\ee
Here, $\h h$ appears due to the non-zero torsion of the adapted covariant derivative.}
The curved-space Yang-Mills Lagrangian reads:
\bea\label{S YM full}
&  \SL_{\rm YM}&=&\;{1\ov e_0^2} \Big({1\ov 4} F_{\mu\nu}F^{\mu\nu} + \half D_\mu \sigma D^\mu \sigma - \half (D+ \sigma H)^2 + 4 i H \sigma F_{1\b 1} \cr 
&&&\qquad + 2 H^2 \sigma^2 
+ i \t\Lambda_0 D_0 \Lambda_0 + 2 i \t \Lambda_{\b1} D_1 \Lambda_0
 +2 i \t\Lambda_{0} D_{\b1} \Lambda_1  \cr
 &&&\qquad- i \t\Lambda_{\b1} D_0 \Lambda_1  - i \t\Lambda_0 [\sigma, \Lambda_0] - i \t\Lambda_{\b1} [\sigma, \Lambda_1] \Big)~. 
\eea
Importantly, this Lagrangian is $\CQ$-exact:
\be\label{SYM Qexact}
\SL_{\rm YM}={1\ov e_0^2} \delta \t\delta \left(\half \t\Lambda_0 \Lambda_0 - \half \t\Lambda_{\b1}\Lambda_1 + 2 \sigma F_{1\b1} - 2 i H \sigma^2 \right)~.
\ee

\paragraph{Chiral multiplet.}  
Consider a chiral multiplet $\Phi$ of $R$-charge $r \in \R$, transforming in a representation $\FR$ of the gauge group. The anti-chiral multiplet $\t\Phi$ has $R$-charge $-r$, and sits in the conjugate representation  $\b\FR$.
We introduce the A-twisted notation:
\be\label{compo Phi}
\Phi= \left(\CA~,\, \CB~,\, \CC~,\, \CF\right)~, \qquad\quad
\t\Phi= \left(\t\CA~,\, \t\CB~,\, \t\CC~,\, \t\CF\right)~,
\ee
with the A-twisted fields defined by:
\bea\label{Atwistvar}
&\CA = (\t p_1)^{r\over 2}\, \phi~, 
&\qquad&\t\CA = (p_{\b1})^{r\over 2} \t\phi~, \cr
& \CB =\sqrt2  (\t p_1)^{r\over 2} \zeta\psi~,\quad 
&& \t\CB =- \sqrt2  ( p_{\b1})^{r\over 2}  \t\zeta\t\psi~, \cr
&\CC =-{1\over \sqrt2} (\t p_1)^{r\over 2} p_{\b1} \, \t\zeta\psi~, \qquad
& &\t\CC ={1\over \sqrt2} (p_{\b1})^{r\over 2}\t p_1 \, \zeta \t\psi~, \cr
&\CF = (\t p_1)^{r\over 2} p_{\b1}\, F~, 
&&\t\CF = (p_{\b1})^{r\over 2}\t p_1 \,\t F~. \cr
\eea
Here $p_{\b 1}$ and $\t p_1$ are the sections of $\b\CK_{\CM_3} \otimes {\bf L}_R^{2}$ and $\CK_{\CM_3} \otimes {\bf L}_R^{-2}$, respectively, as defined in \eqref{def p pb ii}.  By constructions, all the A-twisted fields have $R$-charge zero and 2D spin \eqref{Atwisted spin}. In particular, $\CA, \CB$ have twisted spin ${r\ov 2}$ and $\CC, \CF$ have twisted spin ${r-2\ov 2}$.

The fields are valued in the canonical line bundle to the appropriate power. We have:
\bea
&\CA~,\, \CB \,\in \Gamma(\CK_{\CM_3}^{r\over 2}\otimes V_{\FR})~,\qquad && \CC~,\,\CF\, \in \Gamma(\CK_{\CM_3}^{r\over 2}\otimes \b\CK_{\CM_3} \otimes V_{\FR})~,\cr
&\t\CA~,\, \t\CB \,\in \Gamma(\b\CK_{\CM_3}^{r\over 2}\otimes \b V_{\b\FR})~,\qquad && \t\CC~,\,\t \CF\, \in \Gamma(\b\CK_{\CM_3}^{r\over 2}\otimes \CK_{\CM_3}\otimes \b V_{\b\FR})
\eea
where $V_{\FR}$, $\b V_{\b \FR}$ are the associated gauge vector bundles. In particular, $\CA, \CB$ have two-dimensional  spin ${r\ov 2}$, while $\CC, \CF$ have two-dimensional spin ${r-2\ov 2}$.
The supersymmetry transformations of the chiral multiplet read:
\bea\label{susytranfoPhitwistBis}
& \delta \CA = \CB~, \qquad\qquad  && \t\delta \CA=0~,\cr
& \delta \CB=0~, \qquad \qquad && \t\delta \CB= -2i\big(- \sigma +D_0\big)\CA~,\cr
& \delta \CC=\CF~, \qquad \qquad && \t\delta \CC= 2i D_{\b1}\CA~,\cr
& \delta \CF=0~, \qquad \qquad && \t\delta \CF=- 2i \big(-\sigma +D_0\big)\CC -2i D_{\b1}\CB -2i \t\Lambda_{\b1}\CA~,
\eea
where $D_\mu$ is the appropriately gauge-covariant adapted connection, and $\sigma$ and $\t\Lambda_{\b1}$ act in the representation $\FR$. We have:
\be
D_\mu \CA =\left( \d_\mu - i A_\mu - i r \CA_\mu^{(R)}\right)\CA~, \quad
D_\mu \CC =\left( \d_\mu - i A_\mu - i (r-2) \CA_\mu^{(R)}\right)\CC~,
\ee
with $\CA^{(R)}_\mu$ as defined in \eqref{def CAR ii}. Here, we used the $U(1)_R$ twist \eqref{A twist as bundles}.
For the antichiral multiplet, we similarly have:
\bea\label{susytranfotPhitwistBis}
& \delta \t\CA = 0~, \qquad\qquad\qquad\qquad\qquad\qquad\qquad
& \t\delta \t\CA=\t\CB~,\cr
& \delta \t\CB= -2i \big(\sigma +D_0 \big)\t\CA~, \qquad\qquad\qquad \qquad\quad\quad &\t\delta \t\CB=0~,\cr
& \delta \t\CC= -2i D_1 \t\CA~,
\qquad\qquad\qquad \qquad\quad 
&\t\delta \t\CC=\t\CF~,\cr
& \delta \t\CF= -2i \big(\sigma +D_0 \big)\t\CC +2i D_1  \t\CB + 2i \Lambda_1\t\CA~,
& \t\delta \t\CF=0~.
\eea
Using \eqref{susyVector twisted}, one can check that \eqref{susytranfoPhitwistBis} and \eqref{susytranfotPhitwistBis} realize the supersymmetry algebra:
\be\label{susy with gauge field i}
\delta^2=0~, \qquad\qquad \t\delta^2=0~, \qquad \qquad
\{\delta,\t\delta\}=- 2i  \left(- \sigma+ \CL_K^{(A)} \right)~,
\ee
where $\CL_K^{(A)}$ is the gauge-covariant Lie derivative, and $\sigma$ acts in the appropriate representation of the gauge group. 
The Lagrangian \eqref{Phi kin curved} for a chiral multiplet coupled to a vector multiplet reads:
\bea\label{kin chiral}
&\SL_{\t\Phi\Phi} &=&\; \t\CA \left(-D_0 D_0 - 4 D_1 D_{\b1} + \sigma^2 + D- \sigma H - 2 if_{1\b1}  \right)\CA - \t\CF \CF \cr
&&&  -{i\ov 2} \t\CB (\sigma+ D_0) \CB + 2 i \t\CC (\sigma-D_0)\CC + 2 i \t \CB D_1 \CC - 2 i \t\CC D_{\b1} \CB\cr
&&& - i \t\CB \t\Lambda_0 \CA + i \t\CA \Lambda_0 \CB - 2 i \t\CA \Lambda_1 \CC + 2 i \t\CC \t\Lambda_{\b1} \CA~.
\eea
This Lagrangian is also $\CQ$-exact:
\be\label{kinPhi Qexact}
\SL_{\t\Phi\Phi}  = \delta\t\delta \left({i\ov 2} \t\CA (\sigma+ D_0)\CA - \t\CC \CC\right)~.
\ee
Finally, we leave it as an exercise to write down the curved-space version of the superpotential term \eqref{Lag W}, and to check that it is $\CQ$-exact (and similarly so for the conjugate superpotential).

The only parts of the 3D $\CN=2$ gauge theory Lagrangian which are {\it not} $\CQ$-exact are the various Chern-Simons terms (including the FI terms).

\subsubsection{Global aspects (I): Orbifold line bundles and Seifert fibrations}\label{subsec:orbifold L}
In order to study partition functions on general half-BPS geometries, we first need a slightly more detailed understanding of Seifert manifolds.~\cite{orlik1972seifert,BLMS:BLMS0401} Consider the Seifert fibration:
\be\label{M3 gen fibration}
S^1 \longrightarrow  \CM_3 \longrightarrow \h \Sigma_{g, n}~,\qquad
\CM_3\cong  \big[\bd~; \, g~; \, (q_1, p_1)~, \cdots~, (q_n, p_n)\big]~,
\ee
with $n$ exceptional fibers. One can view the Seifert fibration as a special case of an orbifold circle bundle $\CL_0$ over the base:
\be
\CM_3 \cong S\Big[\CL_0 \rightarrow \h\Sigma\Big]~.
\ee
Let us explain this notion, which will be useful later on. (See Ref.~\citen{Closset:2018ghr} for a more detailed review, with examples.)

\paragraph{Two-dimensional orbifolds and line bundles.} Let $\h\Sigma= \hat\Sigma(q_1,\cdots,q_n)$ be a 2D orbifold with $n$ orbifold points.
Topologically, an orbifold holomorphic line bundle $L$ on $\h\Sigma$ is determined by a set of integers:
\be
\text{deg}(L) \in\mathbb{Z}\ ,~~~~~\qquad b_i(L)\in \mathbb{Z}_{q_i}\ ,~~i=1,\cdots\ ,n\ ,
\ee
where $\text{deg}(L)$ is the degree of the line bundle $L$. The integers $b_i(L)$ determines the local trivialisation at the $\mathbb{Z}_{q_i}$ orbifold point:
\be
(z_i,~s_i) \sim (e^{2\pi i/q_i}z_i,~ e^{2\pi ib_i(L)/q_i} s_i)\ ,
\ee
where $z_i$ and $s_i$ are local coordinates on the base and the fiber around the $i$-th orbifold point. 
Let us denote by $A$ the connection on the line bundle $L$. The first Chern number of the line bundle is then defined by:
\be
c_1(L) = \frac{1}{2\pi} \int_{\hat\Sigma_{g,n}} dA\ .
\ee
In terms of the topological data, it can be written as:~\cite{satake1957}
\be
c_1(L) = \text{deg}(L) + \sum_{i=1}^n \frac{b_i(L)}{q_i}\ .
\ee
It satisfies a simple relation under tensor product: 
\be
c_1(L_1\otimes L_2) = c_1(L_1) + c_1(L_2)\ ,
\ee
while the degree satisfies:
\be
\text{deg}(L_1\otimes L_2)= \text{deg}(L_1) + \text{deg}(L_2) + \sum_{i=1}^n \left\lfloor \frac{b_i(L_1)+ b_i(L_2)}{q_i}\right\rfloor\ ,
\ee
where $\lfloor ~\rfloor$ is the floor function.

\paragraph{The orbifold Picard group.} 
The 2D Picard group Pic$(\hat \Sigma)$---that is, the isomorphism class of the orbifold line bundles---, is generated by the line bundles:
\be
L_0~, \quad ~\text{ and }~\quad L_i~,\; \; i=1,\cdots,n\ ,
\ee
where $\text{deg}(L_0) =1,~b_i(L_0)=0$ and  $\text{deg}(L_j)=0,~b_i(L_j)=\delta_{ij}$. The 2D Picard group takes the form:
\be\label{pic orbifold}
\text{Pic}(\hat\Sigma) =\left\{ L_0,L_i~|~ L_i^{\otimes q_i}=L_0\ ,~ i=1,\cdots n\right\}\ .
\ee
Any line bundle $L$ on $\hat\Sigma$ can be represented as:
\be
L = L_0^{\n_0} \otimes L_1^{\n_1}\cdots \otimes L_n^{\n_n}\ ,
\ee
for a set of integers $\{\n_0,\cdots, \n_n\}$ (modulo the Picard group relations). The degree and the first Chern-class of $L$ are then:
\be
\text{deg}(L) = \n_0 + \sum_{i=1}^n \left\lfloor \frac{\n_i}{q_i} \right\rfloor\ ,\qquad ~~~ c_1(L) = \n_0 + \sum_{i=1}^n \frac{\n_i}{q_i}\ .
\ee
We call $\n_0$ and $\n_i$ the ordinary flux and the {\it fractional flux} for the U(1) gauge symmetry associated to the line bundle $L$, respectively. They correspond to the gauge fluxes localized on a set of points on $\hat\Sigma$:
\be
dA = 2\pi \n_0\delta^{2}(z-z_0) + 2\pi\sum_{i=1}^n \frac{\n_i}{q_i}\delta^2(z-z_i)\ .
\ee
In particular, $\n_i$ is the fractional flux localized at the $\mathbb{Z}_{q_i}$ orbifold point at $z=z_i$. The relation \eqref{pic orbifold} implies that, if we have $q_i$ units of the fractional fluxes, it can be moved away from the orbifold point and becomes one unit of the ``ordinary flux,'' corresponding to $L_0$.

\paragraph{The ``defining line bundle,'' $\CL_0$.}
The Seifert manifold \eqref{M3 gen fibration} can be viewed as the circle bundle $S[\CL_0]$ associated to a certain orbifold line bundle $\CL_0$ over $\hat\Sigma_{g,n}$, called the ``defining line bundle," with the topological data:
\be
\text{deg}(\CL_0) = \bd\ ,\qquad ~~ b_i(\CL_0) = p_i\ .
\ee
In order for the three-manifold $\CM_3$ to be non-singular, we need $q_i$ and $p_i$ to be mutually prime at each orbifold point.

\paragraph{$\pi_1(\CM_3)$ of $\CM_3.$} The fundamental group of $\CM_3$ can be represented as:
\bea
&\pi_1(\CM_3) = \left\langle a_l,b_l, g_i, h\ ,~l=1,\cdots,g\ , i=1,\cdots,n~\Big|\right.
\\
&\qquad\left.[a_l,h]=[b_l,h]=[g_i,h] = 0\ ,~ g_i^{q_i} h^{p_i}=1\ ,\prod_{l=1}^g[a_l,b_l]\prod_{i=1}^n g_i = h^d\right\rangle\ .
\eea
The first homology group is then
\be
H_1(\CM_3,\mathbb{Z}) \cong \widetilde{\text{Pic}}(\CM_3)\oplus \mathbb{Z}^{2g}\ ,
\ee
where $\widetilde{\text{Pic}}(\CM_3)$ is the pull-back of the orbifold Picard group Pic$(\hat\Sigma)$ through the Seifert fibration. We have:
\bea
\widetilde{\text{Pic}}(\CM_3) &= \text{Pic}(\hat\Sigma)/\langle\CL_0\rangle \\
& = \left\{\begin{array}{ll} \text{Tor}~H_1(\CM_3,\mathbb{Z}) \quad\qquad&{\rm if}~~ c_1(\CL_0)\neq 0~,\\
\text{Tor}~H_1(\CM_3,\mathbb{Z}) \oplus \mathbb{Z}\qquad & {\rm if}~~c_1(\CL_0)=0~.
\end{array}\right.
\eea
More explicitly, $\widetilde{\text{Pic}}(\CM_3)$ can be represented as
\be
\widetilde{\text{Pic}}(\CM_3) = \left\{[\gamma],[w_i]~\Big|~q_i[w_i]+p_i[\gamma]=0\ ,\forall i\ ,\sum_{i=1}^n [w_i] = \bd[\gamma] \right\}\ ,
\ee
where the generators can be obtained from pull-backs of the orbifold line bundles:
\be
\pi^*(L_0)\cong -[\gamma]\ ,~~\pi^*(L_i) \cong -s_i[\gamma] + t_i[w_i]\ ,
\ee
where $s_i, t_i$ are integers satisfying $s_iq_i+t_ip_i=1$.

\paragraph{Holomorphic line bundles on $\CM_3$.} 
On any three-manifold equipped with a THF, with adapted coordinates $(\tau, z, \bz)$, an holomorphic line bundle ${\bf L}$ is a complex line bundle with holomorphic transition functions, $f(z)$.

On a half-BPS Seifert manifolds, the (supersymmetry-preserving) holomorphic line bundles of interest to us are pull-back  of orbifold holomorphic line bundles on the base $\h\Sigma$, namely:
\be
[{\bf L}] \in \t\Pic(\CM_3)~.
\ee
The data of the orbifold line bundle $L$ such that ${\bf L} = \pi^\ast(L)$ provides us with a complete (although redundant)  topological description of the 3D line bundle ${\bf L}$. 
 In addition to the topological data, a 3D line bundle will also be specified by a complex line bundle {\it modulus,} denoted by $\nu\in \C$ for a ``flavor $U(1)$'' (or by $u\in \C$ for a dynamical gauge symmetry). It corresponds to an element of the Dolbeault-like cohomology $ H^{0,1}(\CM_3)$.
 To describe the line bundle modulus $\nu$ explicitly, consider the complexified gauge field:
 \be\label{CA def thf}
 \CA_\mu = A_\mu - i \eta_\mu \sigma~,
 \ee
 on ${\bf L}$, with $\sigma$ the scalar in the vector multiplet. The supersymmetry conditions include the condition:
\be
\CF_{\tau \bz}= 0~, \qquad \CF_{\mu\nu} = \d_\mu \CA_\nu- \d_\nu \CA_\mu~,
\ee
 which implies that the line bundle ${\bf L}$ is holomorphic.~\cite{Closset:2013vra}%
 ~\footnote{This discussion is valid for any $\CM_3$ with a THF. The new ingredient in the half-BPS case is that there is only a single continuous parameter, $\nu$, on which supersymmetric observables may depend.}
 On a Seifert manifold half-BPS background, the holomorphic modulus of ${\bf L}$, denoted by $\nu$, can be given explicitly in terms of the holonomy of \eqref{CA def thf} along the generic Seifert fiber, as:
\be\label{def nu from CA}
e^{2\pi i \nu} = e^{-i \int_\gamma \CA}~.
\ee

\paragraph{Large gauge transformations.}
From \eqref{def nu from CA}, it would seem that $\nu$ is not single valued.  Indeed, there is an equivalence $\nu \sim \nu+1$, corresponding to large gauge transformations along the Seifert fiber. 
Geometrically, on the base $\h \Sigma$, such a large gauge transformation corresponds to tensoring $L$ by the defining line bundle $\CL_0$:
\be\label{LGT top}
L \sim L\otimes \CL_0\ .
\ee
In terms of the complex modulus $\nu\in\mathbb{C}$ and of the (fractional) fluxes $\n_0\ ,\n_i$, the gauge equivalence take the form:
\be\label{lgt general shift}
(\nu~,\, \n_0~,\,\n_i)\sim (\nu+1~,\,\n_0+\bd~,\,\n_i +p_i)~,
\ee
in terms of the Seifert invariants \eqref{M3 gen fibration}. Thus, the natural range of $\nu$ depends on the topology of $\CM_3$. For instance, on the three-sphere, there are no topologically non-trivial line bundles, and thus it is natural to gauge fix the fluxes to zero; in that sense, $\nu$ of the line bundle ${\bf L}$ on $S^3$ is naturally valued on the full complex plane. (Such a line bundle is topologically but not holomorphically trivial, whenever $\nu\neq 0$.)

\subsubsection{Global aspects (II): $U(1)_R$ line bundle and spin structure}\label{subsec:LR details}
To fully define the half-BPS background on the Seifert manifold $\CM_3$ (and, in fact, on any supersymmetric three-manifold),  we must specify the $U(1)_R$ symmetry line bundle (which is also an holomorphic line bundle with respect to the THF defined by $\eta_\mu$), such that \eqref{A twist as bundles} holds; in other words, we must choose a ``square-root'' of the 3D canonical line bundle:
\be
{\bf L}_R \cong \sqrt{\CK_{\CM_3}}~.
\ee
Note that $\CK_{\CM_3}$ is the pull-back of the canonical line bundle $\CK$ on $\h\Sigma$ through the Seifert fibration. On $\Sigma_g$, is is well known that a choice of square root $\sqrt{\CK}$ is equivalent to a choice of {\it spin structure}. The same holds true in 3D, as explained in detail in Ref.~\citen{Closset:2018ghr}.

We choose some $L_R\in \Pic(\h\Sigma)$ which pulls back to ${\bf L}_R$. This is parameterized by
\be\label{LR rep in PicSigma}
L_R = L_0^{\t\n_0^R} \otimes \bigotimes_{i=1}^n L_i^{\n_i^R}\ ,
\ee
and by a complex line modulus as above, denoted by $\nu_R$. Unlike a ``flavor $U(1)$'' parameter $\nu$, the $U(1)_R$ parameter $\nu_R$ cannot be arbitrary while preserving supersymmetry. Indeed, the analogue of \eqref{CA def thf} for $U(1)_R$ reads:
\be
\CA_\mu^{(R)}= A_\mu +V_\mu - i \eta_\mu H~,
\ee
coinciding with \eqref{def CAR ii}.
For future reference, let us define another flux parameter:
\be\label{param LR}
\n_0^R=1-g+ \t\n_0^R~.
\ee  
We then introduce the parameterization:
\be\label{param LR ii}
\n_0^R \equiv { l_0^R\ov 2}+ \nu_R \bd~, \qquad\qquad \n_i^R \equiv {q_i-1\ov 2}+ {l_i^R q_i\ov 2}+  \nu_R p_i ~,
\ee
of $L_R$, with the constraints that $\n_0^R, \n_i^R$ are integers, and:
 \be\label{sum lR zero}
l_0^R\in \Z~, \qquad l_i^R\in \Z\; \;\;(i=1, \cdots, n)~, \qquad \text{such that:}\qquad \quad l_0^R + \sum_{i=1}^n l_i^R=0~.
\ee
In particular, $\nu_R$ is constrained to be integer or half-integer, $\nu_R\in \half \Z$. The freedom in choosing $L_R$ (or rather, ${\bf L}_R$) precisely corresponds to a choice of spin structure on $\CM_3$--see also Ref.~\citen{Closset:2018ghr} for a fuller explanation of this point.

\paragraph{R-charge quantization.} Fields charged under the $R$-symmetry, with $R$-charge $r$, must transform as well-defined sections of the line bundle ${\bf L}_R$. This gives the Dirac quantization:
\be
({\bf L}_R)^r \in \Pic(\h\Sigma)~.
\ee
In explicit computations of partition function, we will fix a representative $L_R \in \Pic(\h\Sigma)$, as in \eqref{LR rep in PicSigma}, in which case the Dirac quantization condition is simply:
\be
(g-1+\n_0^R) r \in \Z~, \qquad \n_i^R r\in \Z~, \; \forall i~.
\ee
In the generic case, the $R$-charge should be integer-quantized, $r\in \Z$. Some finer quantization conditions are often allowed in specific cases, however. The most interesting case is when ${\bf L}_R$ is topologically trivial. In that case, we can take $r\in \R$, and we should choose a representative $L_R$ which is also topologically trivial (that is, with zero fluxes, $\t \n_0^R=  \n_i^R=0$). For instance, this is the case on $S^3_b$ with $b\in \Q$, as in \eqref{S3b as seifert}, $\t\Pic(\CM_3)$ is trivial and we have:~\cite{Closset:2018ghr}
\be
\t\n_0^R=\n_1^R= \n_2^R=0~, \qquad  \nu_R = {q_1+ q_2\ov 2}~. 
\ee
Three-manifolds that admit a real $R$-charge, $r\in \R$, are particularly interesting from the point of view of SCFTs, since the 3D $\CN=2$ superconformal $R$-charges are generally irrational numbers.


\section{Supersymmetric localization on $\CM_3$: general aspects}\label{sec: susy loc gen}
In the last section, we reviewed in some detail the problem of coupling 3D $\CN=2$ theories to curved closed manifolds. Given those tools, let us consider the actual computation of supersymmetric curved-space observables. 

Consider our gauge theory on $\CM_3$ preserving two supercharges $\CQ$ and $\t\CQ$, which anti-commute to a Killing vector $K$. For simplicity, we take $K$ to be real---that is, the THF on $\CM_3$ is aligned with $K$. We can then use the A-twisted formalism of section~\ref{subsec: 3DAtwist}.

\subsection{The observables: half-BPS lines and partition functions}\label{subsec: 3D line obs}
Consider three-dimensional flat space, written as a product $\R^3 \cong \C \times \R$, with some coordinates $(z, \bz) \in \C$ and $x^3 \in \R$.  
It is convenient to write the 3D  $\CN=2$ supersymmetry algebra in two-dimensional notation, from the point of view of the $\C$-plane:
\bea\nn
&  \{ Q_{-}, \t Q_{-}\} = 4P_{z}~, \qquad &&
 \{Q_{+},  \t Q_{+}\} =- 4P_{\bz}~, \cr
& \{Q_{-}, \t Q_{+}\} =   -2i (Z+i P_3) ~,\qquad && \{Q_{+}, \t Q_{-}\} =2i(Z- i P_3)~,
\eea
This takes the form of the 2D $\CN=(2,2)$ supersymmetry algebra, with a complex central charge:
\be
Z_{\rm 2d} = Z+ i P_3~.
\ee
We are interested in observables that preserve the two supercharges $Q_-$ and $\t Q_+$. The corresponding A-twisted supercharges are:
\be
\CQ= \zeta_+ Q_-~, \qquad 
\t\CQ= \zeta_- \t Q_+~, 
\ee
in 2D notation. By construction, they are 2D scalar nilpotent operators. This construction uplift to 3D on $\CM_3$ with a real Killing vector $K^\mu$, which determines the $\C \times \R$ splitting of the tangent bundle. 
The curved-space supersymmetry algebra takes the form:
\be
\CQ^2=0~, \qquad 
\t\CQ^2=0~, \qquad 
\{\CQ, \t\CQ\}= -2i (Z+ \CL_K)~, 
\ee
with $Z$ the flat-space central charge. This is obviously the 2D A-twist on the plane $\C$ transverse to $K$, in the adapted coordinates $(z, \bz) \in \C$.

We are interested in operators, denoted by $\SL$, that commute with these two supercharges:
\be\label{def SL op}
[\CQ, \SL]=0~, \qquad \qquad
[\t\CQ, \SL]=0~.
\ee
In two-dimensions, the local operators, $\omega$, that commute with $Q_-$ and $\t Q_+$ are called the {\it twisted chiral operators.}~\footnote{Recall that this use of the word ``twisted'' in ``twisted chiral'' is not directly related to the ``twist'' in ``A-twist.'' This is standard terminology. The 2D twisted chiral ring is compatible with the A-twist, while the 2D chiral ring is compatible with the B-twist.} Since the condition $[\CQ, \omega]=[\t\CQ, \omega]=0$ is not Lorentz-covariant in 3D, the 3D uplift of a twisted chiral operator cannot be local. Instead, it is a line, wrapped along the $\R\cong \{ x^3\}$ direction. The operators $\SL$ in \eqref{def SL op} are {\it half-BPS line operators,} wrapped along an orbit $\gamma$ of the Killing vector~$K$. The observables that we will compute are of the form:
\be
\langle \SL_i \SL_j \cdots \rangle_{\CM_3}~,
\ee
with the various half-BPS lines inserted along paths $\gamma_i, \gamma_j, \cdots$ parallel to $K$.
The 3D A-twist guarantees that these correlators are topological in the transverse directions---that is, they are independent of the insertion points along the $\C$ plane. 
A simple example of such a line is a supersymmetric Wilson loop in a representation $\FR$ of $\GG$. In adapted coordinates, it reads:
\be
 W_{\FR} = \Tr_{\FR} {\rm Pexp}{\left( -i \int_{\gamma} \left(A_\mu dx^\mu - i  \sigma d\tau\right)\right)}~.
\ee
This simplest example of a half-BPS line is of course the trivial one:
\be
\SL= {\bf 1}~,
\ee
whose insertion computes the {\it supersymmetry partition function} on $\CM_3$:
\be
Z_{\CM_3} = \langle  {\bf 1} \rangle_{\CM_3}~.
\ee
Once we understand how to compute the supersymmetric partition function itself, the insertion of other lines will turn out to be straightforward.

\subsection{Supersymmetric localization: the general procedure}
Let us explain the gist of the supersymmetric localization argument, as far as we need it for the following discussion. A much more detailed reviews of supersymmetric localization techniques can be found in Ref,~\cite{Pestun:2016zxk} and especially in Ref.~\citen{Willett:2016adv} for the 3D case.

\subsubsection{Half-BPS loci}
Consider any path integral whose integrand is invariant under some nilpotent supercharge---$\CQ$ such that $\CQ^2=0$. On general grounds, such a supersymmetric path integrals should ``localize'' onto the fixed points of $\CQ$.~\cite{Witten:1991zz} That is, the full path integral only receives contributions from a subspace of field space, $\CM_{\rm BPS}$, on which $\CQ$ vanishes:
\be
 \int [D\varphi] \quad \rightsquigarrow\quad  \int_{\CM_{\rm BPS}} d\varphi_0~,
 \ee
In favourable cases, including all the examples that we will consider, the right-hand-side is an {\it ordinary} integral (or super-integral), with coordinates $\varphi_0$ on $\CM_{\rm BPS}$. We will often call $\varphi_0$ the ``zero-modes.''

For a gauge theory on a half-BPS three-manifold $\CM_3$, with two supercharges and a real Killing vector $K^\mu= \eta^\mu$, the equations $\CQ \varphi= \t\CQ \varphi=0$ for the vector multiplet give:
\be\label{BPS equs gauge field 0}
D_\mu \sigma+i  F_{\mu\nu}K^\nu=0~, \qquad D =\sigma H + \half\epsilon^{\mu\nu\rho}Â K_\mu F_{\nu\rho}~,
\ee
with all the fermionic fields vanishing. For the chiral multiplet, we have:
\be
\big(- \sigma +K^\mu D_\mu\big)\CA=0~,  \quad D_{\b 1} \CA=D_{\b 1}\CB=0~, \quad \big(- \sigma +K^\mu D_\mu\big)\CC=0~,\quad \CF=0~,
\ee
and similarly for the anti-chiral multiplet.

\subsubsection{Localizing action: the semi-classical approximation is exact}
The localization argument is based on an important physical ``lemma'':  in a supersymmetric theory with a supercharge $\CQ$, the expectation value of any $\CQ$-exact operator vanish, 
$\langle \CQ\Psi \rangle=0$. This is easily understood in the Hamiltonian formalism---given any supersymmetric state $|0\rangle$, we have: $\langle 0 | [\CQ, \Psi]| 0\rangle=0$, since $\CQ |0 \rangle =0$ by assumption. It directly follows that we have:
\be
\langle \CQ(\Psi) \,  \CO \rangle=0~.
\ee
for any $\CQ$-closed operator $\CO$. Supersymmetric localization is then based on a simple trick. Let us add to the action a $\CQ$-exact term:
\be
t S_{\rm loc}[\varphi]=t  \CQ(\Psi_{\rm loc})~,
\ee
with a coefficient $t\geq 0$. We choose $S_{\rm loc}$ such that the path integral remains ``well defined''---in particular, it should not introduce any divergence at infinity in field space. 
Supersymmetric observables are independent of that deformation. At first order:
\be
{d\ov d t} \langle  \CO \rangle_t \Big|_{t=0}= - \langle \CQ(\Psi_{\rm loc})  \,  \CO \rangle=0~,
\ee
and this generalizes to any order. The {\it localization limit} corresponds to taking $t$ to infinity, assuming that:
\be\label{loc assumption}
 \langle  \CO \rangle_{t \rightarrow \infty} =  \langle  \CO \rangle_{t=0}~,
\ee
by the above argument. 
At arbitrarily large $t$, the semi-classical approximation becomes exact, and we ``localize onto'' the solutions to the equations of motion for the localizing action:
\be
\CM_{\rm EOM} = \big\{ \varphi= \varphi_0 \; \big|\; {\delta S_{\rm loc} \ov \delta \varphi} =0\big\}~.
\ee
We then have:
\be
 \langle  \CO \rangle= \int_{\CM_{\rm EOM}} d \varphi_0 \, \CO(\varphi_0) \, Z^{\oneloop}(\varphi_0) \,e^{-S[\varphi_0]}~,
\ee
schematically, with $Z^\oneloop(\varphi_0)$ the one-loop contribution from the fluctuations of all the fields around the localization locus at $\varphi=\varphi_0$.

Note that, in Euclidean signature, there is no natural ``reality conditions'' on the fields; we should think of all the fields as {\it a priori} complexified---for instance, the chiral multiplet scalars $\CA$ and $\t\CA$ need not be complex conjugate of each other. We must then prescribe a middle-dimensional ``contour'' for the path integral, as part of the definition of the theory.

It is up to us to choose a ``good'' localizing action that leads to a maximally simplified problem.
For the 3D gauge theories of interest here, we noted in \eqref{SYM Qexact} and \eqref{kinPhi Qexact} that the kinetic terms for the gauge and matter fields are $\CQ$-exact. We may then choose:
\be
\SL_{\rm loc}= {1\ov e^2} \SL_{\rm YM} +{1\ov g^2} \SL_{\t\Phi\Phi}~,
\ee
and consider the limit $e, g \rightarrow 0$. This is the UV limit of the 3D $\CN=2$ gauge theory. Other choices are possibles, which may lead to different-looking answers---see for instance Ref.\cite{Benini:2013yva} (Of course, the final answer should always be the same if \eqref{loc assumption} indeed holds, but different localization ``schemes'' can lead to rather different presentations of the same answer.)

\subsection{One-loop determinants on the BPS locus}
For some purposes, it is useful to introduce a complexified gauge field $\CA_\mu \equiv A_\mu - i \eta_\mu \sigma$, as in \eqref{CA def thf}. 
Let us denote by $\CF= d\CA - i [\CA, \CA]$ the curvature of $\CA$. Then,  the BPS equations \eqref{BPS equs gauge field 0} for the vector multiplet can be written as:
\be\label{BPS eqs vec 2}
D_0 \sigma= 0~, \qquad \CF_{01}=0~, \qquad \CF_{0\b 1}= 0~, \qquad  D = 2 i \CF_{1\b 1}-\sigma H~,
\ee
in the adapted frame. The second and third equations imply that $\CA$ is the connection of an holomorphic vector bundle---in particular, for a $U(1)$ gauge group, it is a connection over an holomorphic line bundle; see Ref.~\citen{Closset:2013vra} and references therein.

The gauge-fixing can be done in various ways. For a Seifert-manifold background with $c_1(\CL_0)\neq 0$, the simplest choice is the temporal gauge, defined by $A_0=0$.\footnote{When $c_1(\CL_0)=0$, for example, for $\Sigma_g\times S^1$, the holonomy along $S^1$ cannot be gauged away and the temporal gauge cannot be chosen.} Then the BPS equations together with the reality conditions on the fields imply that the vector multiplet localizes to the configuration with
\be\label{sigma constant}
\sigma =\text{diag}(\sigma_a)\in \mathbb{R}^\rk\ ,~~a=1,\cdots, \rk\ ,
\ee
which gives the $\rk$-dimensional integral contour along the product of the real lines as studied in the literatures, {\it e.g.} in Ref.~\citen{Kapustin:2009kz}. The contour in the $\sigma$-plane may be deformed in a way that the integral converges.

Alternatively, one can also consider the following partial gauge fixing
\be
\eta^\mu(\CL_K a_\mu)=0\ ,
\ee
which leads to a more natural and unifying description in the 2D $A$-model point of view.\cite{Blau:2006gh,Closset:2017zgf} This equation implies $\partial_0 A_0=0$, which allows us to
define the ``two-dimensional'' field:
\be
u \equiv -\beta \CA_0 = i \beta (\sigma + i A_0)~.
\ee
We introduced a factor of $\beta$, which enters the adapted metric \eqref{standard met}, to render $u$ dimensionless. (Or, more precisely, we can define $u$ as in \eqref{def nu from CA}.)
 From \eqref{susyVector twisted}, it is clear that $\CA_0= A_0 -i \sigma$ preserves both supersymetries:
 \be
\delta u= \t\delta u =0~.
\ee
In the language of 2D $\CN=(2,2)$ supersymmetry, the complex scalar field $u$ is the lowest component of a twisted-chiral multiplet, with components:~\cite{Closset:2015rna}
\be\label{cbback}
\CU = \big(u~,\, \Lambda_1~,\,  -\t\Lambda_{\b 1}~, \, -4 F_{1\b 1}\big)~.
\ee
Moreover, the other half of the 3D vector multiplet can then be organised into a 2D twisted-anti-chiral multiplet, with bottom component:
\be
\t u = -i \beta (\sigma - i A_0)~.
\ee
Localization with the Yang-Mills action forces $u, \t u$ to be constant, in the localization limit, $e\rightarrow 0$. Since $[u, \t u]=0$, we can diagonalise $u$ to:
\be
u = {\rm diag}(u_a)~, \qquad a=1, \cdots, \rk~.
\ee
Finally, we also need to sum over the magnetic fluxes valued in the 2D Picard group Pic$(\hat\Sigma)$ associated to the gauge group $\GG$, subject to the equivalence relation \eqref{lgt general shift}. This determines the domain of $u$-integral. We then expect, very schematically:
\be\label{ZM3 very schem}
Z_{\CM_3} = \sum_{\text{top. sectors}, \m } \int_\CC du\;  e^{-S_0(u)} \, Z^{\rm 1-loop}(u)_\m~,
\ee
with a sum over topological sector, and a ``Coulomb branch integral'' in each sector.

We will see how this works out in examples, momentarily. In the rest of this section, let us only  consider some ``Coulomb branch background'' \eqref{cbback} for the 3D vector multiplet, consisting of the constant modes $u, \t u$ for the 2D scalars, and of some as-yet unspecified 3D bundle with connection $A_\mu$, such that \eqref{BPS eqs vec 2} is satisfied. We can then easily compute the relevant one-loop determinants, which are the contributions from small fluctuations around this background, and which will enter the integrand of \eqref{ZM3 very schem}.

\paragraph{Chiral-multiplet one-loop determinant.}
One can easily present general results for all the one-loop determinants on the Coulomb branch background. Consider a chiral multiplet coupled to a $U(1)$ vector multiplet with charge $Q=1$, and with R-charge $r\in \R$. The one-loop determinant is simply the Gaussian path integral:
\be\label{Zphi path int}
Z^\Phi_{\CM_3} = \int D\Phi D\t\Phi  \, e^{-\int_{\CM_3} d^3 x \sqrt{g} \SL_{\t\Phi\Phi}}~,
\ee
with the Lagrangian \eqref{kin chiral}, in the A-twisted notation of section \ref{subsec: Atwist fields}:
\be
\SL_{\t\Phi\Phi} = \t\CA \Delta_\phi \CA + (\t \CB, \t \CC) \,\Delta_\psi \mat{\CB\cr \CC}- \t\CF \CF~.
\ee
Here, the kinetic operators, evaluated on the BPS locus, are given by:
\be
\Delta_\phi = - D_0 D_0 - 4 D_1 D_{\b 1} + \sigma^2~, \qquad \Delta_\psi  = -2 i \mat{ {1\ov 4} (\sigma + D_0) & -D_1 \cr D_{\b 1} & - \sigma + D_0}~.
\ee
The superdeterminant \eqref{Zphi path int} simplifies due to supersymmetry. Indeed, we note that the bosonic and fermionic operators can be related as:
\be
\Delta_\psi \cdot \mat{2i (\sigma -D_0) & 0 \cr 2 i D_{\b 1} & 1} = \mat{\Delta_\phi  &  2i D_1  \cr 0 & 2i (\sigma-D_0)}~.
\ee
We then find:
\be\label{oneloop Phi inter i}
Z^\Phi_{\CM_3} =  {\det \Delta_\psi \ov \det \Delta_\phi} = {\det \left[2i(\sigma-D_0)_{r-2}\right] \ov \det \left[2i(\sigma-D_0)_r\right]}~.
\ee
Here, the operators $\sigma-D_0$ in the numerators and denominators acts on the fields $\CC$ and $\CA$, of R-charges $r-2$ and $r$, respectively. Let $\CH_{r-2}$ and $\CH_r$ denote the two Hilbert spaces spanned spanned by such modes. The operators $-2i D_{1}$ and $2i D_{\b 1}$ are mutually adjoint operators which intertwine between them:
\be
-2 i D_1 \; : \CH_{r-2} \rightarrow \CH_{r}~, \qquad 
2 i D_{\b1} \; : \CH_{r} \rightarrow \CH_{r-2}~.
\ee
There is a one-to-one correspondence between the modes of $\CH_{r}$ with non-zero eigenvalue for $D_1 D_{\b 1}: \CH_{r} \rightarrow \CH_{r}$ and the modes of $\CH_{r-2}$ with non-zero eigenvalue for $D_{\b1} D_{1}: \CH_{r-2} \rightarrow \CH_{r-2}$. Therefore, the only non-trivial contribution to \eqref{oneloop Phi inter i} comes from the kernels of $D_1$ and $D_{\b 1}$, respectively, and we find:~\cite{Pestun:2007rz, Hama:2011ea, Closset:2013sxa}
\be\label{oneloop Phi inter ii}
Z^\Phi_{\CM_3}(u; r) = {\det_{{\rm coker}(D_{\b 1})} \left[2i(\sigma-D_0)_{r-2}\right] \ov \det_{{\rm ker}(D_{\b 1})}  \left[2i(\sigma-D_0)_r\right]}~.
\ee
This is a completely general result, for any half-BPS $\CM_3$ on a Seifert manifold. We will evaluate \eqref{oneloop Phi inter ii} explicitly, in examples, below. An important property of \eqref{oneloop Phi inter ii}, which is emphasised by our notation, is that it is locally holomorphic in $u$, simply because $u$ enters the eigenvalue equations holomorphically, through:
\be\label{operator smD0}
i (\sigma- D_0) = {1\ov \beta} \left(-i \d_{\h\psi} + u+ \nu_R r\right)~.
\ee
Note also the simple dependence on the $R$-charge. 

The generalization to any gauge group and matter content is straightforward. Consider $\Phi$ in some (generally reducible) representation $\FR$ of $\GG$, with weights $\rho$. It also transforms in some representation $\FR_F$ of the flavor group, with flavor weights $\omega$. We may choose a different $R$-charge, $r_\omega$, for each $\omega$.\footnote{In general that might break the flavor group explicitly. This is a choice we are allowed to make.}
We then have:
\be\label{Z mat gen}
Z^{\rm matter}_{\CM_3}(u) =\prod_{\omega \in \FR_F} \prod_{\rho\in \FR} Z^\Phi_{\CM_3}(\rho(u) + \omega(\nu); r_\omega)~,
\ee
with $Z^\Phi_{\CM_3}(u; r)$ given by \eqref{oneloop Phi inter ii}.

\paragraph{Vector multiplet.} The one-loop contribution of the vector multiplet on the Coulomb branch can be computed by a simple trick: each W-boson $W_\alpha$, for a non-zero root $\alpha \in \Fg^\ast$, contributes as a chiral multiplet of weight $\rho= \alpha$ and $R$-charge $r=2$. This can be seen by a direct computation in the A-twisted theory, wherein the gauge fields components along the holomorphic plane, $A_{1}, A_{\b 1}$, become the A-twisted chiral multiplet scalars~\cite{Closset:2015rna}. This can also be understood in terms of the Higgs mechanism \cite{Benini:2015noa}. 
We thus find:
\be\label{Z vec gen}
Z^{\rm vector}_{\CM_3}(u) =\left(Z_{\CM_3}^{(0)}\right)^{{\rm dim}(\GG)} \prod_{\alpha \in \Delta}   Z^\Phi_{\CM_3}(\alpha(u); r=2)~,
\ee
where $\Delta$ is the set of non-zero roots. The first factor is a contribution from the gaugino, including the ones in the Cartan of $\Fg$, corresponding to our choice of quantization in section \ref{subsec: quantize susy mult}. We will discuss it in various examples below.

\section{The squashed-sphere partition function}\label{sec: lens space}

The supersymmetric partition function on $S^3_b$,  the three-sphere with squashing parameter $b$, is probably the most studied example of a 3D supersymmetric partition function.~\cite{Kapustin:2009kz, Jafferis:2010un, Hama:2010av, Hama:2011ea, Imamura:2011wg, Alday:2013lba, Tanaka:2013dca, Nian:2013qwa, Benini:2013yva} In this section, we review the derivation of the sphere partition functions $Z_{S^3_b}$, and some applications of the result. 

\subsection{The squashed three-sphere}

Consider the three-sphere $S^3$, viewed as torus fibered over an interval, with $\chi \in [0, 2\pi)$ and $\varphi \in [0, \pi)$ the torus coordinate, and $\theta \in [0, \pi]$ the interval. We take the real Killing vector:
\be
K=  {1\ov R_0}\left(b \d_\varphi + b^{-1} \d_\chi\right)~,
\ee  
with $b \in \R_{>0}$. We then choose an adapted metric:
\be\label{S3b hhl metric}
ds^2(S^3_b) = R_0^2 \left( {1\ov 4} h(\theta)^2  d\theta^2+ b^{-2} \cos^2{\theta\ov 2} d\varphi^2 + b^2  \sin^2{\theta\ov 2} d\chi^2\right)~.
\ee
Here, the function $h(\theta)$ is a smooth positive function which behaves as
$h(\theta) \sim b + \CO(\theta^2)$ at the north pole, $\theta\sim 0$, and as
$h(\theta) \sim b^{-1} + \CO((\pi-\theta)^2)$ at the south pole, $\theta\sim \pi$.  Our THF is simply given by:
\be
\eta = K_\mu dx^\mu= R_0 \left(b \sin^2{\theta\ov 2} d\chi + b^{-1} \cos^2{\theta\ov 2} d\varphi\right)~.
\ee
To display the transversely holomorphic structure explicitly, it is convenient to introduce the coordinate change:
\be
\mat{\h\phi\cr \h\psi} = M \mat{\varphi\cr \chi}~, \qquad M\equiv \mat{\alpha_0 b ^{-1} &- \alpha_0 b \cr c_1 & c_2} \in SL(2, \R)~.
\ee
The local coordinates adapted to the THF are $\h\psi$ and $z= f(\theta) e^{i\h\phi}$, with:
\be
K= {1\ov \beta} \d_\psi~, \quad \eta= \beta (d\h\psi + \CC)~, \quad \CC= - {b c_1 - b^{-1} c_2 - (b c_1 + b^{-1} c_2) \cos\theta\ov 2 \alpha_0} d\h\phi~,
\ee
where $\beta = R_0 \alpha_0$. The metric \eqref{S3b hhl metric} then takes the canonical form \eqref{standard met}, with:
\be
2 g_{z\bz} dz d\bz = {R_0^2 \ov 4} \left(h(\theta)^2 d\theta^2 + (b c_1 + b^{-1} c_2)^2 \sin^2{\theta}\, d\h\phi^2\right)~.
\ee

\subsubsection{Coulomb branch localization and one-loop determinants}
One can localize 3D $\CN=2$ gauge theories on $S^3_b$ as sketched above. On $S^3$, the only non-trivial solutions to the BPS equations \eqref{BPS equs gauge field 0} are:
\be
F_{\mu\nu}=0~, \qquad D= \sigma H~,
\ee
for a constant $\sigma$. In this section, we choose the temporal gauge $A_0=0$ discussed around \eqref{sigma constant}. Let us also introduce the dimensionless parameters:
\be
\h \sigma_a \equiv R_0 \sigma_a~, \qquad \h m_\alpha = R_0 m_\alpha~,\qquad \h\sigma_R \equiv -{i\ov 2}(b+ b^{-1})~.
\ee
Here, $\h m_\alpha$ is the real mass for the flavor symmetry $U(1)_\alpha$, and $\h\sigma_R$ can similarly be viewed as a ``real mass'' for the R-symmetry \cite{Closset:2014uda}. Then, the $S^3_b$ path integral takes the form \cite{Kapustin:2009kz, Jafferis:2010un, Hama:2010av, Hama:2011ea}:
\be
Z_{S^3_b}(\h m) = \int \prod_a d\h\sigma_a\;  Z^{\rm CS}_{S^3_b}(\h\sigma, \h m) \,  Z^{\oneloop}_{S^3_b}(\h\sigma, \h m)~,
\ee
where the contour is given by the real line integral, or an appropriate deformation thereof.~\cite{Closset:2017zgf}
One then simply needs to compute the one-loop determinants on $S^3_b$ in the background of a constant $\sigma$ (and with $A_\mu=0$).

Let us apply the general formula \eqref{oneloop Phi inter ii} for $Z^\Phi$ to this case \cite{Hama:2011ea}.  
On general grounds, a well-defined $\CA$- or $\CC$-type chiral-multiplet mode will be of the form:
\be\label{A gen modes S3b}
\CA_{n, m}(\theta, \chi, \varphi) = \CA_0(\theta) e^{i n \chi + i m \varphi}~, \qquad n, m\in \Z~.
\ee
Let us note that such modes are eigenmodes of the operator \eqref{operator smD0}, with:
\be
i (\sigma-D_0) \CA^{(r)}_{n, m}={1\ov R_0} (i \h \sigma+ i \h\sigma_R r + b^{-1} n + b m)\CA^{(r)}_{n, m}~.
\ee
Here, we included the dependence on the $R$-charge, for a mode of twisted spin ${r\ov 2}$. 
The modes that actually contribute to  \eqref{oneloop Phi inter ii} are the $\CA$- and $\CC$-modes in the kernel and cokernel of $D_{\b 1}$, respectively:
\be
D_{\b1} \CA_{n, m}=0~, \qquad D_1 \CC_{n, m}=0~.
\ee
This gives:
\bea
&\left[{2\ov h(\theta)} \d_\theta + i \left(\cot{\theta\ov 2} b^{-1} \d_\chi - \tan{\theta\ov 2} b\, \d_\varphi\right)\right]\CA_{n, m}=0~,\cr
&\left[{2\ov h(\theta)} \d_\theta - i \left(\cot{\theta\ov 2} b^{-1} \d_\chi - \tan{\theta\ov 2} b\, \d_\varphi\right)\right]\CC_{n, m}=0~,\cr
\eea
which determine the profiles $\CA_0(\theta)$ of the modes \eqref{A gen modes S3b}. Using the asymptotics of $h(\theta)$ at the poles, it is easy to check that the modes $\CA_{n, m}$ are normalizable if and only if $n, m \geq 0$, while the modes $\CC_{n, m}$ are normalizable if and only if $n, m \leq 0$. We thus obtain the one-loop determinant \cite{Hama:2011ea}:
\be\label{ZS3bPhi formal}
Z_{S^3_b}^\Phi(\h\sigma; r) = \prod_{n =0}^\infty \prod_{m=0}^\infty {n b ^{-1}+m  b- i\h\sigma + {b+b^{-1}\ov 2}(2-r) \ov n b^{-1} + m b +i \h\sigma +{b+b^{-1}\ov 2} r }~,
\ee
where the $\CA$- and $\CC$-modes contribute to the denominator and numerator, respectively.

The formal answer \eqref{ZS3bPhi formal} needs to be regularized carefully. While we derived it for $b\in \R$,  there exists a natural analytic continuation of the answer to $b\in \C$ (except on the negative real axis). An elegant regularization of \eqref{ZS3bPhi formal} is given in terms of the so-called {\it quantum dilogarithm} $\t\Phi_b$ \cite{Faddeev:1993rs, Garoufalidis:2014ifa}, which is defined by:
\be\label{def Phib quantum dilog}
\t\Phi_b(\h\sigma) \equiv \left(e^{-{2\pi\ov b} \h\sigma}\, e^{-\pi i \left({1\ov b^2}+1\right)}; e^{-2\pi i b^{-2}}\right)_\infty  \; \left(e^{-2\pi b\h\sigma}\, e^{\pi i \left(b^2+1\right)}; e^{2\pi i b^{2}}\right)^{-1}_\infty~,
\ee
if we take $b$ such that ${\rm Im}(b^2)>0$.
Here, $(a; q)_\infty$ denotes the $q$-Pochhammer symbol:
\be
(a; q)_\infty \equiv \prod_{k=0}^\infty (1- a q^k)~.
\ee
The quantum dilogarithm admits an analytic continuation to $b\in \C-\{\R_{\leq 0}\}$, and in particular to the case $b \in \R_{>0}$ that we started with.
We claim that the correct regularization of \eqref{ZS3bPhi formal} is given by:
\be\label{ZS3bPhi final}
Z_{S^3_b}^\Phi(\h\sigma; r)=Z_{S^3_b}^\Phi(\h\sigma+ \h\sigma_R r)\equiv \t\Phi_b\big(\h\sigma+ \h\sigma_R(r-1)\big)~,
\ee
corresponding to the $U(1)_{-\half}$ quantization of the chiral multiplet.~\cite{Closset:2018ghr} In particular, we have the identity:
\be
 \t\Phi_b(\h\sigma) \t\Phi_b(-\h\sigma)= e^{-\pi i \h \sigma^2}e^{-{\pi i \ov 12}\left(b^2+ b^{-2}\right)}~,
\ee
which corresponds to integrating out a pair of chiral multiplets, of $R$-charges $r=1$ and gauge charges $\pm 1$ in the $U(1)_{-\half}$ quantization, with a superpotential mass term, which shifts the CS levels by $\Delta k_{GG}=-1$ and $\Delta k_g=-2$.
We then have  \eqref{Z mat gen} for the full chiral-multiplet contribution, namely:
\be\label{S3b matter}
Z_{S^3_b}^{\rm matter}(\h \sigma, \h m)=\prod_{\omega \in \FR_F} \prod_{\rho\in \FR} Z^\Phi_{S^3_b}(\rho(\h\sigma) + \omega(\h m)+ \h\sigma_R r_\omega)~.
\ee

Similarly, the vector multiplet contribution \eqref{Z vec gen} is given by:
\be
Z_{S^3_b}^{\rm vector}(\h\sigma)= \left(Z^{(0)}_{S^3_b} \right)^{{\rm dim}(\GG)} \; \prod_{\alpha \in \Delta_+}  4 \sinh\left({\pi b \alpha(\h\sigma)}\right)\, \sinh\left({\pi b^{-1} \alpha(\h\sigma)}\right)~,
 \ee
 where $\Delta_+$ is the set of positive roots, and we defined:
\be
Z^{(0)}_{S^3_b}=  \big(Z_{S^3_b}^{\rm RR}\big)^{\half}\,  Z_{S^3_b}^{\rm grav}= e^{-{\pi i \ov 12} \left(b^2+ b^{-2} +3\right)}~.
\ee
This is the contribution of each gaugino to the $U(1)_R$ and gravitational contact terms, corresponding to the UV CS contact terms \eqref{kappaRR App vec}.

\subsubsection{The $\sigma$-integral formula}
Putting it all together, including the classical contributions, one finds:
\be\label{ZSb loc formula}
Z_{S^3_b}(\h m; r)= {1\ov |W_\GG|} \int  d^\rk \h\sigma \; Z_{S^3_b}^{\rm CS}(\h\sigma, \h m)\, Z_{S^3_b}^{\rm vector}(\h\sigma)\, Z_{S^3_b}^{\rm matter}(\h\sigma, \h m)~.
\ee
Here, $\h m_\alpha$ are the real-mass parameters associated to the flavor symmetry.
The classical piece $Z^{\rm CS}(\h\sigma,\h m)$ comes from CS terms, and takes the general form:
\be\label{ZS3b cl full}
Z_{S^3_b}^{\rm CS} = Z_{S^3_b}^{\rm GG}(\h\sigma)^{k_{GG}}\; Z_{S^3_b}^{\rm G_1G_2}(\h\sigma_1,\h\sigma_2)^{k_{G_1 G_2}}\; Z_{S^3_b}^{\rm GR}(\h\sigma)^{k_{GR}}\; (Z_{S^3_b}^{\rm RR})^{k_{RR}}\; (Z_{S^3_b}^{\rm grav})^{k_g}~.
\ee
The various supersymmetric CS actions evaluated on the $S^3_b$ background are given by:~\cite{Hama:2011ea, Imamura:2011wg, Closset:2012vp}
\bea\label{CS terms S3b}
&  Z_{S^3_b}^{\rm GG}(\h\sigma) = e^{\pi i \h\sigma^2}~, \quad && Z_{S^3_b}^{\rm G_1G_2}(\h\sigma_1,\h\sigma_2)= e^{2\pi i \h\sigma_1 \h\sigma_2}~, \cr
& Z_{S^3_b}^{\rm GR}(\h\sigma)= e^{2\pi i \h\sigma_R \h\sigma}= e^{\pi \left(b+ b^{-1}\right) \h\sigma}~, \quad 
&& Z_{S^3_b}^{\rm RR}= e^{\pi i \h\sigma_R^2}= e^{-{\pi i \ov 4} \left(b^2 + b^{-2} +2\right)}~, \cr
& Z_{S^3_b}^{\rm grav}= e^{{\pi i \ov 24}\left(b^2+ b^{-2}\right)}~.
\eea
The generalization to any non-abelian CS term is straightforward.
One can similarly write down CS contact terms for the flavor symmetries, by replacing the gauge parameters $\h\sigma_a$ by the flavor parameters $\h m_\alpha$ appropriately.

\subsection{The round three-sphere and $F$-maximization}
It is interesting to consider the special case of the ``round $S^3$''  in some detail \cite{Jafferis:2010un, Hama:2010av}.  This corresponds to setting the squashing parameter to $b=1$. The supersymmetric partition function \eqref{ZS3b cl full} becomes:
\be\label{ZS0 loc formula}
Z_{S^3}(\h m; r)= {1\ov |W_\GG|} \int  d^\rk \h\sigma \; Z_{S^3}^{\rm CS}(\h\sigma, \h m)\, Z_{S^3}^{\rm vector}(\h\sigma)\, Z_{S^3}^{\rm matter}(\h\sigma, \h m)~,
\ee
with the classical contribution obtained by setting $b=1$ in \eqref{CS terms S3b}. The vector multiplet contribution is given by:
\be
Z_{S^3}^{\rm vector}(\h\sigma)= e^{-{5 \pi i \ov 12}{{\rm dim}(\GG)}}  \prod_{\alpha \in \Delta_+}  4 \sinh^2\left({\pi \alpha(\h\sigma)}\right)~.
\ee
The matter contribution takes the form:~\cite{Jafferis:2010un} 
\be
Z_{S^3}^{\rm matter}(\h \sigma, \h m)=\prod_{\omega \in \FR_F} \prod_{\rho\in \FR} \CF(i \rho(\h \sigma)+ i \omega(\h m) + r_\omega-1)~,
\ee
in terms of the function:
\be\label{CF S3 def}
\CF^\Phi(u) \equiv \exp\left({1\ov 2 \pi i} \dilog(e^{2\pi i u}) + u \log\left(1- e^{2\pi i u}\right)\right)~,
\ee
which is holomorphic in $u$.\footnote{While this is not immediately obvious from this definition, the branch-cut ambiguities cancel out between the dilog and the log in \protect\eqref{CF S3 def}. See Ref.~\protect\citen{Closset:2017zgf} for further discussion. Up to some entire function related to our choice of fermion quantization (and up to sending $\sigma$ to $-\sigma$), $\CF(u)$ is equivalent to the function $e^{f(u)}$ introduced in Ref.~\protect\citen{Jafferis:2010un}. }

\paragraph{$F$-theorem and $F$-maximization.}
For any 3D CFT, one can define a quantity of intrinsic intererest, called $F_{S^3}$, as the finite part of the logarithm of the $S^3$ partition function. It plays the role of an ``$c$-function'' under RG flow, in the sense that, for any non-trivial RG flow from a UV to an IR CFT, we must have:
\be
F_{S^3, UV}  > F_{S^3, IR}~.
\ee
This is known as the ``$F$-theorem.''~\cite{Jafferis:2011zi, Klebanov:2011gs, Casini:2012ei}
Here, $F_{\rm CFT}$ is defined as the real number:
\be
F_{S^3} = - {\rm Re} \left[ \log Z_{S^3}\right]~,
\ee
obtained upon renormalization of the sphere partition function; this is scheme-independent, because there are no dimensionless local terms that could modify the answer. (The imaginary part of $\log Z_{S^3}$, on the other hand, can be affected by background Chern-Simons terms.~\cite{Closset:2012vg})

For an $\CN=2$ SCFT that arises in the IR of an asymptotically-free gauge theory, we may define:
\be
F_{S^3}(\h m; r) \equiv - {\rm Re}\left[ \log Z_{S^3}(\h m; r)\right]~,
\ee
with the supersymmetric partition function given by \eqref{ZS0 loc formula}. The CFT answer is obtained by setting the real masses to zero, $\h m=0$, and by choosing $r=r^\ast$ the {\it superconformal} R-charge:
\be
F_{S^3} =  F_{S^3}(0; r^\ast)~.
\ee
The superconformal $R$-charge is the one that {\it maximizes} $F_{S^3}(0; r)$; namely, if we consider a generic $R$-charge:
\be
R= R_0 + t^\alpha F_\alpha~,
\ee
where $t^\alpha$ are mixing parameters between some reference $R$-charge $R_0$ and any abelian flavor-symmetry charges $F_\alpha$, we must have:~\cite{Jafferis:2010un, Closset:2012vg} 
\be
\d_{t^\alpha} F_{S^3}(0; t_\ast)=0~, \qquad \d_{t^\alpha} \d_{t^\beta}  F_{S^3}(0; t_\ast) = - {\pi^2 \ov 2} \tau_{\alpha\beta}~,
\ee
for $t=t_\ast$ defining the superconformal $R$-charge. Here, $\tau_{\alpha\beta}$ is a positive-definite matrix, which encodes the two-point functions of the $U(1)_\alpha$ conserved currents.~\cite{Closset:2012vg}.

\subsection{Computing $Z_{S^3_b}$: Examples}
Let us now discuss the three-sphere partition function of the simple gauge theories introduced in section~\ref{subsec: 3 examples}.

\subsubsection{Example (1): Supersymmetric $U(N)_k$ CS theory.} In this case, we have:
\be\label{ZS3b exp(1)}
Z_{S^3_b}^{U(N)_k} = {1\ov N!} \int_{\CC_{\h\sigma}} d^N\h\sigma \, e^{\pi i k \sum_{a=1}^N \h\sigma_a^2}\, e^{2\pi i \h\xi \sum_{a=1}^N \h\sigma_a} \, Z_{S^3_b}^{\rm vector}(\h\sigma)~,
\ee
with $\h\xi$ the FI term, and:
\be
Z_{S^3_b}^{\rm vector}(\h\sigma)=e^{-{\pi i N^2\ov 12}(b^2 +b^{-2} +3)} \prod_{\substack{a, b=1\\ a>b}}^N 4 \sinh(\pi b(\h\sigma_a-\h\sigma_b))\sinh(\pi b^{-1}(\h\sigma_a-\h\sigma_b))~.
\ee
Here, we can either use the naive contour $\h\sigma_a \in \R$, or equivalently we can consider a contour $\CC_{\h\sigma}$ obtained by rotating the real axis for each $\h\sigma_a$ slightly clockwise if $k>0$, or anti-clockwise if $k<0$, so that the term $e^{\pi i k \h\sigma^2}$ provides an exponential damping factor as ${\rm Re}(\h\sigma) \rightarrow \pm \infty$.

An interesting (and particularly simple) special case is the $U(1)_k$ theory. We have:
\be
Z_{S^3_b}^{U(1)_k} =e^{-{\pi i \ov 12}(b^2 +b^{-2} +3)} \int_{\CC_{\hat\sigma}} d\h\sigma  e^{\pi i  k \h\sigma^2 + 2\pi i \h\xi \h\sigma}= \alpha(k)\, e^{-{\pi i \ov 12}\left(b^2+ b^{-2}\right)}\, {e^{-{\pi i \h\xi^2\ov k}}\ov \sqrt{|k|}}~,
\ee
in the conventions of section~\ref{subsec: review UNk levelrk}, with the phase $\alpha(k)=1$ if $k>0$ and $\alpha(k)=-i$ if $k<0$. 
Consider the special case $k=1$. Indeed, we have:
\be\label{ZS3b U1 explicit}
Z_{S^3_b}^{U(1)_1}= e^{-{\pi i \ov 12}\left(b^2+ b^{-2}\right)}\, e^{-\pi i \h\xi^2}~.
\ee
This agrees with the fact that $U(1)_1$ is equivalent to an empty theory with non-zero $U(1)_T$ and gravitational contact terms. By comparing with \eqref{CS terms S3b}, we see that this matches the contributions from the CS local terms with levels $\Delta k_{TT}=-1$ and $\Delta k_g=-2$.

 Another simple duality relation is between $U(1)_2$ and $U(1)_{-2}$, namely:
\be
Z_{S^3_b}^{U(1)_2} =i \, e^{-\pi i \h\xi^2}\,Z_{S^3_b}^{U(1)_{-2}}~,
\ee
also in perfect agreement with the duality \eqref{levelrk susy}-\eqref{rel level susy levelrk}, since $\Delta k_{TT}=-1$, $\Delta k_{RR}=-1$ and $\Delta k_g =-6$ in this case.

\subsubsection{Example (2): $U(1)_{\half}{+}\Phi$ theory.}
In this case, we simply have:
\be
Z_{S^3_b}^{U(1)_{\half}{+}\Phi} = e^{-{\pi i \ov 12}\left(b^2+ b^{-2}+3\right)}\, \int_{\CC_{\hat\sigma}} d\h\sigma e^{\pi i \h\sigma^2+ 2\pi i \h\xi \h\sigma}\,  \t \Phi_b(\h\sigma+ \h\sigma_R(r-1)) 
\ee
for the ``tetrahedron theory,'' in the presence of an FI term $\h \xi$, and with an arbitrary $R$-charge $r\in \R$ for the chiral multiplet.
The duality between this theory and a chiral multiplet $T^+$ implies a ``Fourier transform'' identity for the quantum dilogarithm.~\cite{Faddeev:2000if} Without loss of generality, consider the case $r=1$. Then, we have:
\be\label{dual rel ZS3b EMS}
 e^{-{\pi i \ov 12}\left(b^2+ b^{-2}+3\right)}\, \int_{\CC_{\hat\sigma}} d\h\sigma e^{\pi i \h\sigma^2+ 2\pi i \h\xi \h\sigma}\,  \t \Phi_b(\h\sigma) = e^{\pi i \sigma_R^2 - 2 \pi i \h\xi \h\sigma_R} \, \t\Phi_b(\h\xi - \h \sigma_R)~,
\ee
in agreement with \eqref{EMS duality}-\eqref{kTR kRR rel mirrsym 1}. 

It is interesting to consider the special case of the round sphere, $b=1$, as well. Then, the duality relation \eqref{dual rel ZS3b EMS} takes the form:
\be
\int_{\CC_{\hat\sigma}} d\h\sigma e^{\pi i \h\sigma^2+ 2\pi i \h\xi \h\sigma}\, \CF^\Phi(i \h \sigma) = e^{-{7\pi i \ov12}} e^{-2 \pi \h\xi} \CF^\Phi(i\h \xi-1)~,
\ee
with the holomorphic function $\CF^\Phi(u)$ defined in \eqref{CF S3 def}. One can also follow the RG flow to the $U(1)_1$ theory, as described in section~\ref{subsec: EMS}, at the level of the supersymmetric partition function. 
Indeed, we have:
\be
\t\Phi_b(\h\sigma)\sim e^{-\pi i \h\sigma^2} e^{-{\pi i\ov 12} \left(b^2+ b^{-2}\right)} \quad \text{as}\;\; \h\sigma \rightarrow -\infty~, \qquad 
\t\Phi_b(\h\sigma)\sim 1 \quad \text{as} \;\;\h\sigma \rightarrow \infty~.
\ee
Taking the limit $\h\xi \rightarrow - \infty$, the equality \eqref{dual rel ZS3b EMS} reduces to \eqref{ZS3b U1 explicit}.

\subsubsection{Example (3): 3D SQCD with $U(N)$ gauge group.}
For 3D SQCD, we have the integral formula:
\be
Z_{S^3_b}^{U(N), N_f}(\h m, \h{\t m}, \h m_A,  \h\xi) = {1\ov N!}  \int_{\CC_{\hat\sigma}} d^N \h\sigma \, Z^{\rm CS} Z^{\rm vec} \, Z^{\rm mat}~,
\ee
with:
\bea\nn
& Z^{\rm CS} = \prod_{a=1}^{N} e^{2\pi i \h \xi \h \sigma_a}~,\cr
&Z^{\rm vec}= e^{-{\pi i\ov 12}(b^2 + b^{-2} +3) N^2} \prod_{1\leq a < b \leq N} 4 \sinh\left({\pi b (\h\sigma_b- \h\sigma_a)}\right)\, \sinh\left({\pi b^{-1}(\h\sigma_b- \h\sigma_a)}\right)~, \cr
 &Z^{\rm mat}= \prod_{a=1}^N \left[ e^{\pi i N_f \h\sigma_a^2}\, \prod_{i=1}^{N_f} Z_{S^3_b}^\Phi(\h\sigma_a- \h m_i +\h m_A+ r \h\sigma_R)\, \prod_{j=1}^{N_f} Z_{S^3_b}^\Phi(-\h\sigma_a+ \h{\t m_i} +\h m_A+ r \h\sigma_R)\right]
\eea
Here, $\h m_i$ and $\h {\t m_j}$ such that $\sum_{i=1}^{N_f} \h m_i=0$ and $\sum_{j=1}^{N_f}\h {\t m_j}=0$ denote the $SU(N_f) \times SU(N_f)$ real masses, while $\h m_A$ is the $U(1)_A$ real mass, and $\h \xi$ the FI parameter.
Note that we included the necessary bare CS level in the definition of $Z^{\rm mat}$, here.
Aharony duality takes the form of complicated integral identities:
\be
Z_{S^3_b}^{U(N), N_f}(\h m, \h{\t m}, \h m_A,  \h\xi)=Z_{\rm ct}(\h m, \h{\t m}, \h m_A, \h\xi)\;  Z_{S^3_b}^{U(N_f-N), N_f}(\h{\t m}, \h m,- \h m_A, -\h\xi)~,
\ee
with $Z_{\rm ct}(\h m, \h{\t m},  \h m_A, \h\xi)$ denoting the contribution from the CS contact terms \eqref{AD CS ct1} for the global symmetry.
These identities between so-called hyperbolic hypergeometric functions were first discovered by mathematicians~\cite{vdbult_thesis}, and the connection to infrared dualities was discovered soon after~\cite{Willett:2011gp}. We will give an alternative description of these duality relations in later sections, in the special case of rational squashing, $b^2 \in \Q$.


\section{The topologically twisted index}\label{sec: S2 top index}

In later sections, we will see that supersymmetric partition function on general half-BPS Seifert manifolds can be constructed from the so-called genus-zero twisted index~\cite{Benini:2015noa,Nekrasov:2014xaa, Gukov:2015sna}  defined on $S_A^2 \times S^1$; namely $S^2$ with a topological twist.

In this section, we study the $S^2$ twisted index in detail. More generally, we consider a one-parameter family of half-BPS backgrounds labelled by a parameter $\epsilon \in \mathbb{R}$, which deforms the A-twist on $S^2$ by a so-called $\Omega$ background.~\cite{Dimofte:2010tz, Closset:2014pda} This parameter corresponds to  a chemical potential for the azimuthal rotation, $J_\varphi$, on $S^2$, in the index picture. The $S^2 \times S^1$ metric can be chosen as: 
\be
ds^2 = \beta d\tau^2 + \frac{R^2}{4}(d\theta^2 + \sin^2\theta (d\varphi+\epsilon d\tau)^2)\ .
\ee
The theory is topologically twisted on $S^2$ using the $U(1)_R$ symmetry, with the $U(1)_R$ background flux:
\be
\frac{1}{2\pi}\int_{\Sigma_g}dA^{(R)} = -1~.
\ee
It follows that the R-charge must be quantized, $r\in \mathbb{Z}$. We can also introduce the holonomy for the R-symmetry along $S^1$ direction:
\be
v_R = -\frac{1}{2\pi}\int_{S^1} A^{(R)}~, \qquad v_R \in \half \Z~,
\ee
which is correlated with a choice of spin structure along the $S^1$---for $v_R= 0$ mod $1$, we must choose the periodic boundary condition for fermions, while for $v_R= \half$ mod $1$, we have the anti-periodic spin structure.~\cite{Closset:2018ghr}
The Killing vector $K$ reads:
\be
K = \frac{1}{\beta}(\partial_\tau  - \epsilon \partial_{\varphi})\ ,
\ee
with $\eta = \beta d\tau$. The generic orbit of $K$ closes only when $\epsilon\in \mathbb{Q}$. The half-BPS background with $\epsilon=0$ corresponds to the standard A-twist on $S^2$.  The other supergravity background fields can be chosen to vanish on $S^2_\epsilon \times S^1$, namely $V_\mu=H=0$.

We will now review the path integral derivation of the $\epsilon$-refined twisted index, defined as:
\be
\CI_{S^2_\epsilon\times S^1}(y)_\n = \text{Tr}_{\CH_{S^2}(\n)}\left[ (-1)^F e^{2\pi i v_R (F+R)} e^{2\pi i \epsilon J_\varphi}\prod_{\alpha} y_\alpha^{Q_F^\alpha}\right]~,
\ee
via Coulomb branch localization \cite{Benini:2015noa,Benini:2016hjo,Closset:2016arn}. Here, the trace is over the Hilbert spaces on $S^2$ at fixed background fluxes, $\n_\alpha$, for the flavor symmetry, twisted by the complexified fugacities $y_\alpha$, and with a choice of spin structure---when $v_R=0$ (mod 1), we have the usual insertion $(-1)^F$ in the index, while for $v_R= \half$ (mod 1), we have $(-1)^R$ instead, with $R$ the (integer-quantized) $R$-charge.

\subsection{Localization to the Coulomb-branch BPS locus}

Using the $Q$-exactness of the SYM action, as in equation \eqref{SYM Qexact}, we can take the limit $e_0\rightarrow 0$ so that the path integral localizes to the zero locus of $S_{YM}$. The BPS conditions on the bosonic fields read:
\be
D = 2i F_{1\bar 1}~,\qquad D_0 \sigma = 0~,\qquad F_{0\bar 1} = -i D_{\bar 1}\sigma~,\qquad  F_{01} = -i D_{1}\sigma~.
\ee
Imposing that the fields $a_\mu$ and $\sigma$ be real, we then have:~\footnote{The field $D$ need not be real. In fact, its natural contour in Euclidean signature would be $D \in i \R$.}
\be
D = 2if_{1\bar 1}~, \qquad D_\mu \sigma = 0~, \qquad F_{0\bar 1} = F_{01}=0~.
\ee
If we impose the gauge-fixing condition $\partial_0A_0 = 0$, we can define:
\be
u = i\beta (\sigma + i a_0)\ ,
\ee
where $a_0$ is the holonomy of the gauge field along $S^1$
\be
a_0 = \frac{1}{2\pi \beta}\int_{S^1_{\beta}}A\ .
\ee
We can diagonalize $a_0$ using the residual gauge symmetry:
\be
a_0 = \text{diag}(a_{0,a})\ ,\qquad ~~a=1,\cdots, \rk\ ,
\ee
which leaves the maximal torus times the Weyl symmetry of $\GG$ as a residual gauge symmetry:
\be
\GG\rightarrow \mathbf{H} \rtimes W_\GG\ ,~~~\qquad \mathbf{H} \cong \prod_{a=1}^{\rk} U(1)_a\ .
\ee
On the BPS locus, $\sigma$ can also be diagonalised as $\sigma = \text{diag}(\sigma_a)$ using the relation $D_\mu\sigma = 0$. We define complex parameters:
\be
u_a = i\beta (\sigma_a + ia_0) \ ,~~ \qquad u_a = 1,\cdots, \rk
\ee
which parameterizes the classical Coulomb branch of the 3D theory compactified on a circle. For a later purpose, it is convenient to consider ``small fluctuations'' along the BPS locus, which include the fields $u$, some fermionic zero-modes $\Lambda_0, \t \Lambda_0$, as well as a constant auxiliary field $\hat D \in \mathbb{R}$ defined as:
\be
D = 2i f_{1\bar 1} + i\hat D~.
\ee
These ``zero-modes'' sit in a supersymmetry multiplet:
\be
\CV_0 = (u, \t u, \Lambda_0,\t\Lambda_0,\hat D)\ ,
\ee
with the two supersymmetry transformations:
\bea\label{zero mode alg}
&\delta u= 0\ ,\quad &&\delta \t u =-2i\beta\t\Lambda_0 \ ,\quad && \delta \Lambda_0 = -\hat D\ ,\quad && \delta\t\Lambda_0 =0 \ ,\quad &&\delta\hat D = 0\ ,\\
&\t\delta u= 0 \ ,~~ &&\t\delta\t u = 2i\beta\Lambda_0\ ,~~&&\t\delta \Lambda_0 = 0\ ,~~&&\t\delta\t\Lambda_0 = -\hat D\ ,~~&&\t\delta\hat D = 0\ .
\eea
In the $e_0\rightarrow 0$ limit, the path integral localizes to the finite dimensional integral over the zero modes $\CV_0$. In addition, we must sum over the non-trivial $\mathbf{H}$-bundles on $S^2$,~\cite{Blau:1993tv, Blau:1994rk, Benini:2015noa} correspondinh to GNO-quantized fluxes on the sphere:
\be
\m = \frac{1}{2\pi} \int_{S^2} dA \in \Gamma_{\GG^\vee}\ ,
\ee
which are valued in the co-character lattice $\Gamma_{\GG^\vee}$ defined by:
\be\label{cocharacter}
\Gamma_{\GG^\vee} = \{\m ~|~ \rho(\m) \in \mathbb{Z}\ ,\forall \rho \in \Gamma_{\GG}\} \cong \mathbb{Z}^{\rk}~,
\ee
with $\Gamma_{\GG}$ the weight lattice of $\GG$.
Then the path integral can be written as a localized expression:
\be\label{localized}
Z = \lim_{e_0\rightarrow 0}\sum_{\m \in \Gamma_{\GG^\vee}}\int [d\CV_0]~ e^{-S_{\text{classical}}(\CV_0)}Z_{\text{1-loop}}(\CV_0)\ ,
\ee
where $S_{\text{classical}}$ is the contribution from the classical action and $Z_{\text{1-loop}}(\CV_0)$ is the one-loop determinant around the BPS locus. At the same time, one localizes the matter fields in chiral multiplets by using the fact that the canonical kinetic terms are themselves $Q$-exact. Then, the matter fields contribute non-trivially to $Z_{\text{1-loop}}(\CV_0)$ in \eqref{localized}.

\subsection{Classical and one-loop contributions}\label{subsec: twisted index Z1loop}
Before discussing the integral \eqref{localized}, let us first discuss the integrand in more detail. Here, we focus on the classical and one-loop contribution in the strict BPS limit, with $\h D=0$ and vanishing gauginos. The final answer will be written in terms of these building blocks only.

\subsubsection{Classical action contribution}
The only non-trivial contribution from the classical action is the Chern-Simons action for the gauge and global symmetries. For each flux sector $\m$, we have:
\be\label{classical D=0}
e^{-S_{\text{classical}}(u,\m)} = Z_{GG}(x,\m)^{k_{GG}} Z_{G_1 G_2}(x,\m)^{k_{G_1 G_2}} Z_{GR}(x,\m)^{k_{GR}} Z_{RR}^{k_{RR}}\ ,
\ee
where:
\be
Z_{GG} =  (-1)^{\m(1+2v_R)} e^{2\pi i u\m}\ ,\qquad Z_{G_1 G_2} = e^{2\pi i (u_1\m_2+ u_2\m_1)}~.
\ee
 The second term is the contribution from the mixed abelian CS terms with levels $k_{G_1 G_2}$. This includes the contribution of the FI parameters:
\be
\prod_{I} q_I^{\m_I} (x_I)^{\n_{T_I}}\ ,
\ee
where $q_I = e^{2\pi i \xi_I}$ and $x_I= e^{2\pi i u_I}$ for the $U(1)_I$ factors in $\GG$. The contribution from gauge-$R$ and $RR$ Chern-Simons level is
\be
Z_{GR} = (-1)^{2v_R\m}e^{-2\pi i u}\ ,\qquad ~~ Z_{RR} = -1\ .
\ee

\subsubsection{One-loop determinant at $\hat D =0$}

The one-loop contribution on the SUSY locus with $\hat D=0$ can be written as
\be\label{1-loop D=0}
Z_{\text{1-loop}}(x,\m) = Z_{\text{1-loop}}^{\text{vector}}(x,\m)Z_{\text{1-loop}}^{\text{chiral}}(x,\m) \ ,
\ee
for each flux sector $\m$. The contribution from the chiral multiplet in a representation $\mathfrak R$ can be easily computed by expanding the mode along $S^1$ and summing over the KK tower:
\be\label{Z1l chiral S2epS1}
Z_{\text{1-loop}}^{\text{chiral}}(x,\m) = \prod_{\rho \in {\mathfrak R}} \left(e^{2\pi i\left(\rho(u)+ r v_R - \frac{\epsilon}{2}(\rho(\m)-r)\right)};e^{2\pi i \epsilon}\right)^{-1}_{\rho(\m) -r +1}\ ,
\ee
where $x\equiv e^{2\pi i u}$, and $(x;q)_n$ is the $q$-Pochhammer symbol, defined by:
\be\label{qPochdef}
(x; q)_n \equiv \begin{cases}  \prod_{l=0}^{n-1} (1 -q^l x),  \qquad & {\rm if}\quad n >0~, \\
 \prod_{l=1}^{|n|} (1- q^{-l} x)  \qquad & {\rm if}\quad n \leq 0~.  \end{cases}
\ee
The easiest way to compute the contribution from the vector multiplet is to use the fact that the contribution from each W-boson from a non-zero root $\alpha \in \mathfrak g$ is same as the one from a chiral multiplet of charge $\alpha$ and R-charge 2.~\cite{Benini:2015noa, Closset:2015rna} If we choose the symmetric quantization for a pair $\alpha$ and $-\alpha$, we find:
\be
Z_{\text{1-loop}}^{\text{vector}} = \prod_{\alpha \in \Delta^+}(-1)^{2v_R \alpha(\m)}\sin \left(\pi \left(\alpha(u) - \frac{\epsilon \alpha (\m)}{2}\right)\right)\sin \left(\pi \left(\alpha(u) + \frac{\epsilon \alpha (\m)}{2}\right)\right)\ ,
\ee
where $\Delta^+$ is the space of the positive roots.

\subsection{The integration contour}

Let us come back to the path integral localized onto the Coulomb branch:
\be\label{localized 2}
Z = \lim_{e\rightarrow 0}\sum_{\m \in \Gamma_{\GG^\vee}}\int_\Gamma d\hat D\int_{\fM} du d\t u\int d\Lambda_0d\t\Lambda_0~ e^{-S_{\text{classical}}(\CV_0)}Z_{\text{1-loop}}(\CV_0)~.
\ee
Here, $\Gamma$ is the contour defined by the real line $\mathbb{R}^\rk$ and $\fM$ can be identified with the (universal cover of the) classical Coulomb branch, $\fM\cong {\frak h}_{\mathbb{C}} \cong (\mathbb{C^*})^{\rk}$. Note that the integrand of \eqref{localized 2}, when evaluated at $\Lambda_0 = \t\Lambda_0 = \hat D=0$, has various singularities in the domain $\fM$.

\paragraph{Chiral multiplet singularities:}
The first type of singularities are defined by the loci where the chiral multiplets become massless. The chiral multiplet with gauge charge $\rho$ and R-charge $r$ defines a singular hyperplanes $H_\rho$ given by
\be\label{chiral sing}
H_{\rho,r, j} = \{u \in {\frak h}_{\mathbb{C}} ~|~ \rho(u)+ r v_R + j \epsilon+ k = 0\ , k\in \mathbb{Z}\}~,  
\ee
assuming that $\n_{\rho, r} \equiv \rho(\m) - r+1  \geq 0$, and with $j = -{\n_{\rho, r}-1\ov 2}, -{\n_{\rho, r}-1\ov 2}+1, \cdots, {\n_{\rho, r}-1\ov 2}$. They correspond to the poles of \eqref{Z1l chiral S2epS1}.

\paragraph{Monopole singularities}
The second type of the singularities are at the asymptotic infinities where Im$(u^a)\rightarrow \pm \infty$. The singular hyperplanes are defined by
\be\label{monopole sing}
H_{a\pm} = \{u \in \mathfrak{h}_{\mathbb{C}}~|~\text{Im}(u_a)\rightarrow \mp \infty\}\ .
\ee
We can extract the behavior of the integrand at asymptotic infinities from the one-loop determinant. We find
\be\label{asymptotic}
\lim_{\text{Im}(u)\rightarrow \mp\infty} e^{-S_{\text{classical}(u)}}Z_{\text{1-loop}}(u) \rightarrow x_a^{\pm \left(Q_{a\pm}(\m) + Q^F_{a\pm}(\n) -r_{a\pm}\right)}\ ,
\ee
where $Q_{a\pm}$, $Q^F_{a\pm}$ and $r_{a\pm}$ are the induced charge of the monopole operators~\cite{Aharony:1997bx, Benini:2011cma} $T^\pm_a$ defined by:
\bea
{Q_{a\pm}}^b &= Q^b[T_a^\pm] = \pm k^{ab} - \frac12 \sum_{i}\sum_{\rho_i \in {\mathfrak R}_i} |\rho_i^a|\rho_i^b~, \\
{Q^F_{a\pm}}^b &= Q^F[T_a^{\pm}] = \pm k_{GF}^{aF} - \frac12 \sum_i \sum_{\rho_i \in {\mathfrak R}_i} |\rho_i^a|Q_i^F\ ,
\eea 
and:
\be
r_{a\pm} = R[T_a^{\pm}] = \pm k_R^a - \frac12 \sum_i \sum_{\rho_i\in {\mathfrak R}_i}|\rho_i^a|(r_i-1)-\frac12 \sum_{\alpha \in {\mathfrak g}} |\alpha^a|\ .
\ee
These singularities must be regulated, which can be done as in Ref.\cite{Benini:2013nda,Benini:2013xpa,Hori:2014tda} We divide $u$-integral into two pieces:
\be
\int_{\fM} du d\t u = \int_{\fM \backslash \Delta_\varepsilon} dud\t u ~+~ \int_{\Delta_\varepsilon}dud\t u\ , 
\ee 
where $\Delta_\varepsilon$ is the $\varepsilon$-neighborhood of the singular hyperplanes. One can show that the integrand in the second contribution is bounded as long as we keep $e$ finite, and therefore it vanishes in the limit $\varepsilon \rightarrow 0$ sufficiently faster then taking $e\rightarrow 0$. 

The integration over the gaugino zero modes can be most easily done by using the residual supersymmetry of the zero mode multiplet $\CV_0$. The algebra \eqref{zero mode alg} implies
\be
\partial_{\Lambda_0}\partial_{\t\Lambda_{0}} Z_{\m}\big|_{\Lambda_0 = \t\Lambda_0 = 0} = \frac{1}{\hat D} 2i\beta\partial_{\t u} Z_\m  \big|_{\Lambda_0 = \t\Lambda_0 = 0}\ ,
\ee
where $Z_\m$ is the integrand of \eqref{localized 2}. Then the Coulomb branch integral becomes:
\be
Z = \lim_{\varepsilon, e\rightarrow 0}\sum_{\m \in \Gamma_{\GG}^\vee}\int_\Gamma\frac{d\hat D}{\hat D}\oint_{\partial (\fM\backslash \Delta_\varepsilon)} du ~ e^{-S_{\text{classical}}(u,\hat D)}Z_{\text{1-loop}}(u,\hat D)\ ,
\ee
where the contour $\Gamma$ is taken to be a $\rk$-dimensional real line, slightly shifted to the imaginary axis with the vector $\delta \in {\mathfrak h}$. The integral gets contribution from the various disconnected components of the contour $\partial(\fM\backslash \Delta_\varepsilon)$, which reduces to the $\rk$-dimensional residue integrals of the poles constructed from the intersection of the hyperplanes \eqref{chiral sing} and \eqref{monopole sing} in $\fM$.
The contribution from various components of the contours in $\fM$ can be decomposed into
\be
Z = Z_{\text{bulk}} + Z_{\text{boundary}}\ ,
\ee
where $Z_{\text{bulk}}$ is the contribution from the contour around the chiral multiplet singularities \eqref{chiral sing}, while $Z_{\text{boundary}}$ is the contribution from the contours at $|u|\rightarrow \infty$, around the monopole singularities \eqref{monopole sing}. The contribution from each contours in $Z_{\text{bulk}}$ is determined by carefully performing the $\hat D$ integrals.~\cite{Benini:2013xpa,Hori:2014tda} The result can be written as:
\be
Z_{\text{bulk}} = \frac{1}{|W_\GG|}\sum_{\m\in\Gamma_{\GG}^\vee} \sum_{\{u_*\}}\underset{u = u_*}{\text{JK-Res}} (Q_{u_*},\eta)~d^\rk u~e^{-S_{\text{classical}}(u)} Z_{\text{1-loop}}(u)\ ,
\ee
where the integrand is the one-loop and classical contribution evaluated at $\hat D = 0$, as given in subsection~\ref{subsec: twisted index Z1loop}. The so-called Jeffrey-Kirwan (JK) residue~\cite{JK1995, 1999math......3178B} selects a particular subset of the poles from the set of all charge vectors $\{Q_{u^*}\}$, which define singular hyperplanes of type \eqref{chiral sing}, whose intersection is a co-dimension $r=\rk$ singularity at a point $u=u_*$ in $\fM$. This operation is defined by the property that, for a rank-$r$ charge vector $(Q_1,\cdots, Q_r)$,
\be
\underset{u=u_*}{\text{JK-Res}}(Q_*(u),\eta) \frac{d^r u}{Q_1(u)\cdots Q_r(u)} = \begin{cases} 
\displaystyle\frac{1}{|\text{det}(Q_1,\cdots, Q_{r})|}\ , & \eta\in \text{Cone}{(Q_1,\cdots, Q_{r})}\\ \displaystyle 0\ ,&\text{else} \end{cases}
\ee
where $\text{Cone}(Q_1,\cdots, Q_r)$ is the positive cone spanned by the charge vectors $(Q_1,\cdots, Q_r) \in \R^r$. 
Evaluating $Z_{\text{boundary}}$ for general gauge theories can be complicated due to the geometry of the contours at $|u|\rightarrow \infty$. For rank one theories, one can explicitly show that the contribution of the poles from the hyperplane of type \eqref{monopole sing} are determined by the gauge charge of the monopole operators $Q_{\pm}$. For the theories with non-zero monopole charges $Q_{\pm}\neq 0$, the $\hat D$ integral at $|u|\rightarrow \infty$ gives:
\be
Z_{\text{boundary}} = \frac{1}{|W_{\mathbf{G}}|} \sum_{\m\in \Gamma_{\mathbf{G}}^\vee} \underset{u=\pm i \infty}{\text{JK-Res}} (Q_\pm, \eta)~ d u~ e^{-S_{\text{classical}(u)}}Z_{\text{1-loop}}(u)\ ,
\ee
and therefore the index can be written uniformly as:
\be\label{integral formula rk1}
Z = \frac{1}{|W_\GG|}\sum_{\m\in\Gamma_{\GG}^\vee} \sum_{\{u_{0} = u_*, \pm i\infty\}}\underset{u = u_0}{\text{JK-res}} (Q_{u_0},\eta)~du~e^{-S_{\text{classical}}(u)} Z_{\text{1-loop}}(u)\ .
\ee
The monopole singularity with $Q_+ =0$ or $Q_-=0$ needs a careful treatment. 
Suppose that the residue of such a pole is non-zero. Since $Q_\pm =0$, the asymptotic behaviour of the integrand \eqref{asymptotic} implies that the residue is non-zero for all integers $\m\in \mathbb{Z}$ and therefore the sum of the contribution from this pole does not converge. In fact, one can choose a regularization scheme which allows us to determine the contribution of this pole at each flux sector and gives a convergent formula.~\cite{Bullimore:2018jlp}

For a non-abelian gauge group $\GG$, we conjecture that the final formula can still be written as the Jeffrey-Kirwan residue:
\be\label{integral formula S2S1eps}
Z = \frac{1}{|W_\GG|}\sum_{\m\in\Gamma_{\GG}^\vee} \sum_{\{u_{0}\in \fM_{\text{sing}}\}}\underset{u = u_0}{\text{JK-Res}} (Q_{u_0},\eta)~d^{\rk}u~e^{-S_{\text{classical}}(u)} Z_{\text{1-loop}}(u)\ .
\ee
where $\fM_{\text{sing}}$ includes all the singularities from the hyperplanes \eqref{chiral sing} and \eqref{monopole sing}.

In explicit examples, the JK-residue prescription for the Coulomb-branch integral can be hard to deal with. In later sections, we will explore a different approach, which can be argued to be equivalent to the Coulomb branch integral while avoiding its technical difficulties.

\subsection{Examples}
Let us briefly discuss the $S^2_\epsilon$ twisted index for the first two examples introduced in section~\ref{subsec: 3 examples}.

\subsubsection{Example (1): Supersymmetric $U(N)_k$ CS theory}

For concreteness, let us choose $k>0$. The twisted index can be written as
\bea
Z_{S^2_\epsilon\times S^1}^{U(N)_k} &= \frac{1}{N!}\sum_{\substack{(\m_1,\cdots, \m_N)\\\in \mathbb{Z}^N
}}\oint_{\text{JK}(\eta)} \left(\prod_{a=1}^N du_a\right)~ q^{\sum_{a=1}^N \m_a} e^{2\pi i k u_a \m_a}Z_{\text{1-loop}}^{\text{vector}}(u,\m)~,
\eea
If we choose $\eta_a<0$, $\forall a$, then the Jeffrey-Kirwan residue integral picks up a pole from $u_a \rightarrow i\infty$. In this limit, the partition function receives the contribution from the topological sector $\m_a=0$ only. We can easily see that:
\be
Z_{S^2_\epsilon\times S^1}^{U(N)_k} = 1\ ,
\ee
and in particular, it is independent of the parameters $\epsilon, q$ and $v_R$. This agrees with the fact that the Hilbert space of the pure CS theory on $S^2$ is one-dimensional.~\cite{Witten:1988hf}

\subsubsection{Example (2): $U(1)_{\frac12}+\Phi$ theory}
Here, let us set $v_R=0$ to avoid clutter, thus considering the periodic spin structure on $S^2 \times S^1$. 
The integral formula for the twisted index of this theory reads:
\bea
&Z_{S^2_\epsilon\times S^1}^{U(1)_{\frac12}+\Phi}& =&\;\sum_{\m\in \mathbb{Z}} \oint_{\text{JK}(\eta)} du~ e^{2\pi i \xi \m}e^{2\pi i u\n_T}  (-1)^{(1+ 2 v_R) \m} e^{2\pi i u\m} \\
&&&\qquad\qquad\qquad \times \left(e^{2\pi i\left(u - \frac{\epsilon}{2}(\m-r)\right)};e^{2\pi i \epsilon}\right)^{-1}_{\m -r +1}\ .
\eea
We again choose $\eta<0$ and pick up the residue from the pole at $u \rightarrow i\infty$. The contour integral at each $\m$ can be performed explicitly after using the $q$-binomial theorem:
\be\nn
\left(z; e^{2\pi i \epsilon} \right)^{-1}_{\n} = \sum_{n=0}^\infty \frac{(e^{2\pi i \epsilon \n};e^{2\pi i \epsilon})_n}{(e^{2\pi i \epsilon};e^{2\pi i \epsilon})_n} z^n~.
\ee
The residue integral is non-vanishing only for $-\n_T\geq \m$ and, for each $\m$, only the term $n =-\m-\n_T$ contributes. We then find:
\bea
&Z_{S^2_\epsilon\times S^1}^{U(1)_{\frac12}+\Phi} = \sum_{n=0}^\infty   (-1)^{(\n_T+n) (1+2 v_R)} e^{- 2\pi i \xi (\n_T+n)}\\
&\qquad \qquad\qquad\quad  \times \frac{\left(e^{2\pi i (-n-\n_T-r+1)\epsilon}; e^{2\pi i \epsilon}\right)_n}{(e^{2\pi i \epsilon};e^{2\pi i \epsilon})_n}
~e^{2\pi i n\left(r v_R+ \frac{\epsilon}{2}(\n_T+n+r)\right)}\ .
\eea
Using the identity:
\be
(z;e^{2\pi i \epsilon})_n = (-z)^{n}e^{\pi i\epsilon n(n-1)} (z^{-1}e^{2\pi i\epsilon(1-n)};e^{2\pi i \epsilon})_n~,
\ee
one finds:
\bea\nn
&Z_{S^2_\epsilon\times S^1}^{U(1)_{\frac12}+\Phi} =(-1)^{\n_T(1+2 v_R)} e^{-2 \pi i \xi \n_T}  \sum_{n=0}^\infty \frac{\left(e^{2\pi i (\n_T+r)\epsilon};e^{2\pi i\epsilon}\right)_n}{(e^{2\pi i \epsilon};e^{2\pi i \epsilon})_n}~e^{2\pi i n(-\xi+ (1-r)v_R-\frac{\epsilon}{2}(\n_T+r-1))} \\
& =(-1)^{\n_T(1+2 v_R)} e^{-2 \pi i \xi \n_T}  \left(e^{2\pi i (-\xi +(1-r)v_R- \frac{\epsilon}{2}(\n_T+r-1))};e^{2\pi i \epsilon}\right)_{\n_T+r}^{-1} \\
&=(-1)^r \left[(-1)^{2 v_R \n_T} e^{-2\pi i \xi} \right]^{-r}\,  \left(e^{2\pi i (\xi +(1-r)v_R- \frac{\epsilon}{2}(\n_T+r-1))};e^{2\pi i \epsilon}\right)_{\n_T+r}^{-1}~.
\eea
As expected from the elementary mirror symmetry, we recovered the refined index of the theory of a free chiral multiplet of unit charge under $U(1)_T$, $R$-charge $1-r$, and with the relative CS contact terms \eqref{kTR kRR rel mirrsym 1}.

\section{Bethe vacua and half-BPS line-operator algebras}\label{sec: bethe}
As explained in section~\ref{sec:3}, in a precise sense, most half-BPS geometries are Seifert geometries:
\be
S^1 \rightarrow \CM_3 \rightarrow \h\Sigma~.
\ee
It turns out that one can understand the partition functions on such manifolds by ``reducing'' along the Seifert fiber, $S^1$, and considering the corresponding A-twisted 2D theory on $\h \Sigma$ (with all KK modes taken into account).~\cite{Closset:2017zgf, Closset:2018ghr} 
This will be explained in section~\ref{subsec: seifert operators} below. In preparation for this discussion, let us first consider the 3D theory on $\R^2 \times S^1$, as well as on the closed product manifold $\Sigma_g \times S^1$. We will also discuss the algebra of half-BPS Wilson line operators in 3D $\CN=2$ gauge theories.

\subsection{The Coulomb branch vacua and the 2D $A$-model}
Let us first review some general facts about 2D $\CN=(2,2)$ gauge theories and the topological A-twist thereof.

\subsubsection{The effective twisted superpotential in 2D}
The low-energy dynamics of the 2D gauge theory on its Coulomb branch is governed by the \emph{effective twisted superpotential}:
\be
\CW(\sigma,m)~,
\ee
which can be obtained by integrating out the contributions from massive chiral multiplets and W-bosons, a generic vacuum expectation values for the scalar $\sigma$ in the 2D vector multiplet. The twisted superpotential depends holomorphically on $\sigma$ and on the complex masses $m$, which arise as background vector multiplets, $\sigma_F=m$. The one-loop result reads:~\cite{Witten:1993yc}
\be\nn
\CW(\sigma,m) = \xi(\sigma) - \frac{1}{2\pi i }\sum_i\sum_{\rho_i \in R_i} (\rho_i(\sigma)+m_i)\left(\log \rho_i(\sigma)+m_i -1 \right) -\frac12 \sum_{\alpha\in \Delta^+} \alpha(\sigma)~,
\ee
where $\xi$ denotes the 2D complexified FI parameters.
This is also the exact result, due to a non-renormalization theorem. Note that the twisted superpotential suffers from the branch cut ambiguity:
\be\label{ambiguity 2d}
\CW_{\text{eff}}(\sigma,m) \rightarrow \CW_{\text{eff}}(\sigma,m) + n^a \sigma_a + n^\alpha m_{\alpha} \ ,
\ee
for $n^a,m^\alpha \in \mathbb{Z}$.

The ``2D $A$-model'' associated to the 2D $\CN=(2,2)$ field theory is obtained by passing to the simultaneous cohomology of the flat-space supercharges $Q_-$ and $\t Q_+$---in other words, by definition, the $A$-model observables are correlation functions of $Q$-closed operators, modulo $Q$-exact insertions. These operators include the gauge invariants polynomials in the the twisted-chiral scalars $\sigma$. In the simplest cases, these operators, schematically given by:
\be
\CO_n(\sigma) = \Tr(\sigma^n)~,
\ee
furnish a complete set of generators of the twisted chiral ring. 
We will also consider ``quasi-local'' twisted-chiral operators consisting of arbitrary (gauge-invariant) functions of $\sigma$.

Taking advantage of the topological A-twist,~\cite{Witten:1988xj} one can define the $A$-model on any genus-$g$ closed Riemann surface, $\Sigma_g$. The low-energy effective theory  on $\Sigma_g$ has contributions from an infinite numbers of topological sectors, labelled by the GNO-quantized fluxes:
\be\label{GNO lattice}
\frac{1}{2\pi}\int_{\Sigma_g} d^2 x \sqrt g~ (-2if_{1\bar 1})  = \m \in \Gamma_{\mathbf{G}^\vee}\ ,
\ee
with $\Gamma_{\mathbf{G}^\vee}$ defined in \eqref{cocharacter}. (The flux sectors are the gauge-theory equivalent to the holomorphic instantons in the more familiar 2D A-twisted NLSM.) The low-energy effective Lagrangian can then be written as:
\be\label{2d effective action}
S_{\text{eff}} = \int_{\Sigma_g}d^2x \sqrt{g} \left[-2f_{1\bar 1a}\frac{\partial \CW}{\partial\sigma_a} + \t\Lambda_{\bar 1}^a \Lambda_1^b \frac{\partial^2\CW}{\partial\sigma_a\partial\sigma_b}\right]  + \frac{i}{2}\int_{\Sigma_g}d^2x \sqrt{g} ~\Omega R~,
\ee
with $R$ the Ricci scalar of $\Sigma_g$, and modulo $Q$-exact terms. 
The twisted effective Lagrangian \eqref{2d effective action} depends only on the twisted superpotential, $\CW$, together with the \emph{effective dilaton}, $\Omega$, which determines the coupling of the $A$-model to the Riemann surface $\Sigma_g$.~\cite{Witten:1993xi, Nekrasov:2014xaa} Both $\CW(\sigma, m)$ and $\Omega(\sigma, m)$ are locally holomorphic functions of $\sigma$ and $m$.

\subsubsection{The 2D vacua, a.k.a. the Bethe vacua}

\begin{figure}[t]
\centerline{\includegraphics[width=2.5cm]{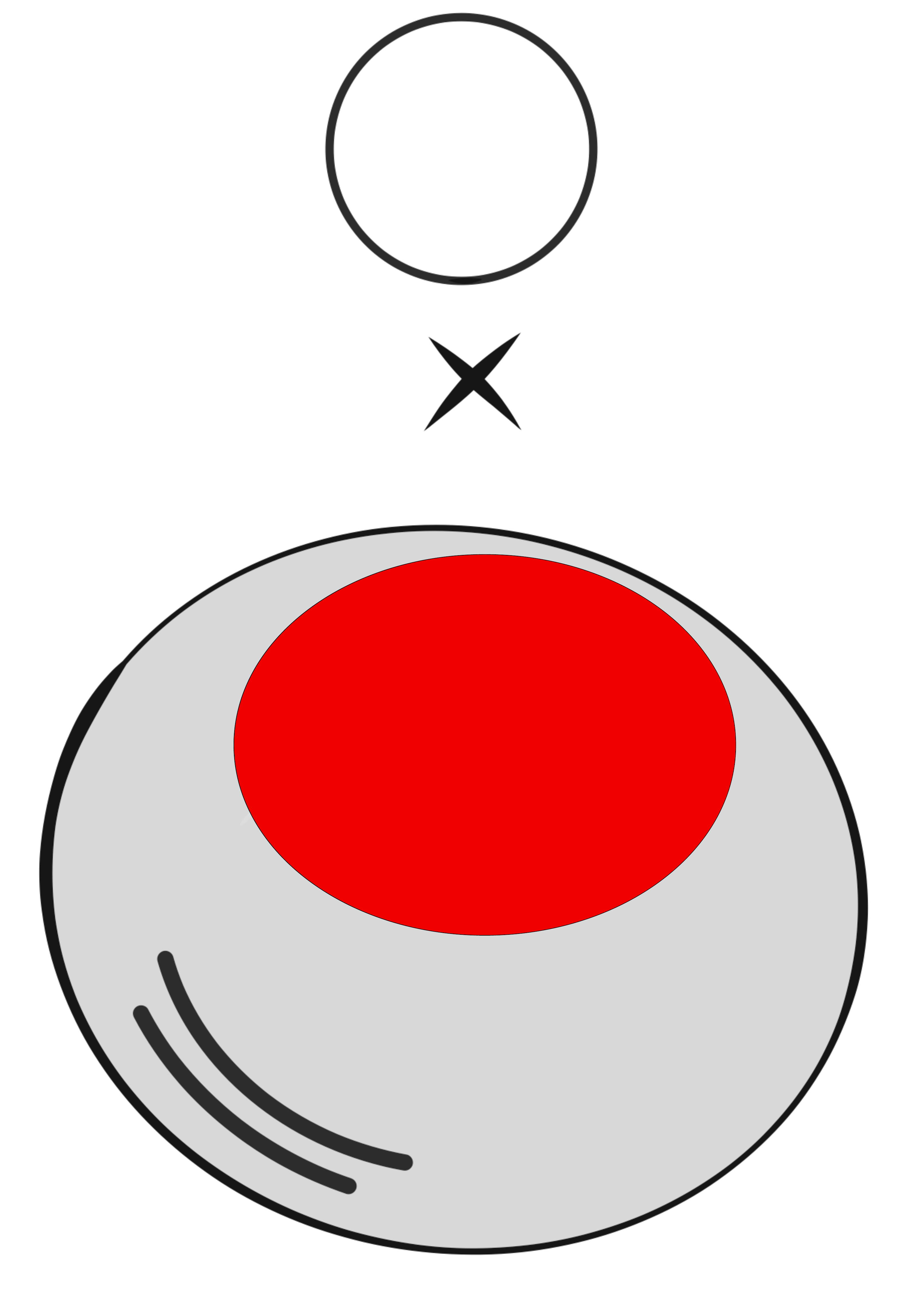}
\qquad\qquad
\includegraphics[width=2.5cm]{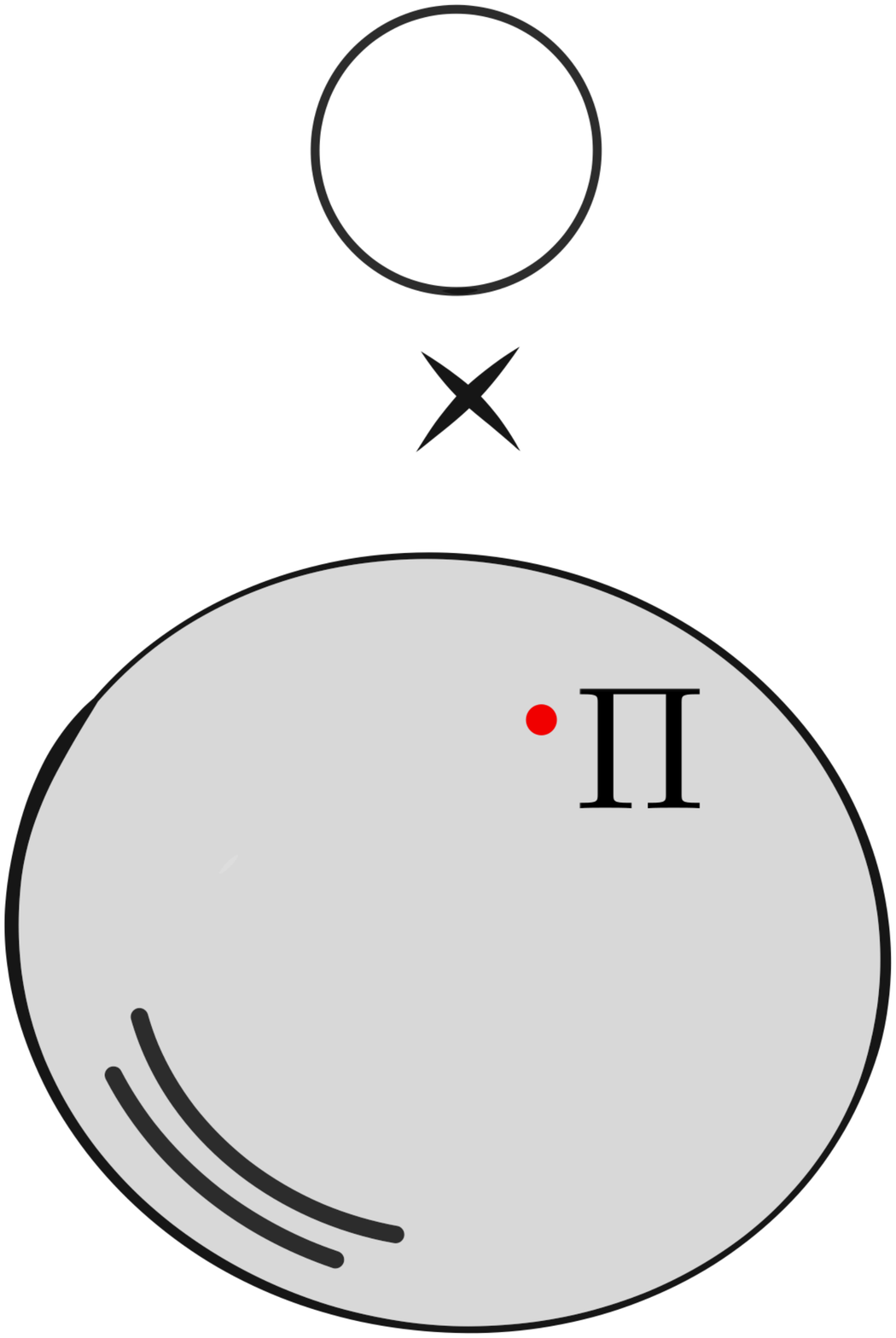}
}
\caption{Any gauge flux (dynamical or flavor) through $\Sigma_g$ (as depicted on the left-hand-side figure, for some random flux profile) can be concentrated to a point (as shown on the right), which can then be interpreted as a 2D local operator. For the 3D theory on $\Sigma_g \times S^1$, the flux operator, $\Pi$, is a line operator wrapped over the $S^1$ (with this circle shown here as well). \label{fig:flux}}
\end{figure}

The Coulomb branch vacua are fully determined by the twisted superpotential of the theory. For later convenience, let us define the \emph{flux operators}~\cite{Closset:2017zgf} for the gauge and flavor symmetry, respectively, as:
\be\label{flux op}
\Pi_a (\sigma,m) = \exp\left(2\pi i \frac{\partial \CW}{\partial \sigma_a}\right)~, \qquad 
\Pi_\alpha (\sigma,m) = \exp\left(2\pi i \frac{\partial \CW}{\partial m_a}\right)~,
\ee
in terms of the effective twisted superpotential.
As is clear from the effective action \eqref{2d effective action}, the insertion of a flux operator increases the total flux on the Riemann surface by a unit element in the co-character lattice of the symmetry group. Using the topological invariance, the supersymmetry-preserving flux can be continuously localized to a point on the Riemann surface, which can then be identified with the quasi-local operators \eqref{flux op}. See Figure~\ref{fig:flux}.

Summing over the gauge fluxes \eqref{GNO lattice}, we obtain a set of polynomial equations in $\sigma$ that determines the Coulomb branch vacua:
\be
\Pi_a(\sigma,m) = 1\ , \qquad ~~a=1,\cdots, \rk\ .
\ee 
These equations are known as the \emph{Bethe equations}. In certain supersymmetric theories, they coincide with the Bethe equations for a class of the integrable systems, as studied in the context of the  Nekrasov-Shatashvili  Bethe/Gauge correspondence \cite{Nekrasov:2009uh, Nekrasov:2009ui}. For our purpose, the term ``Bethe equations'' and ``Bethe vacua'' is just a convenient name to denote 2D Coulomb-branch vacuum equations and vacua, respectively. Note that the flux operators and the vacuum equations are free from the branch-cut ambiguity \eqref{ambiguity 2d}.

\begin{figure}[t]
\centerline{\includegraphics[width=2.5cm]{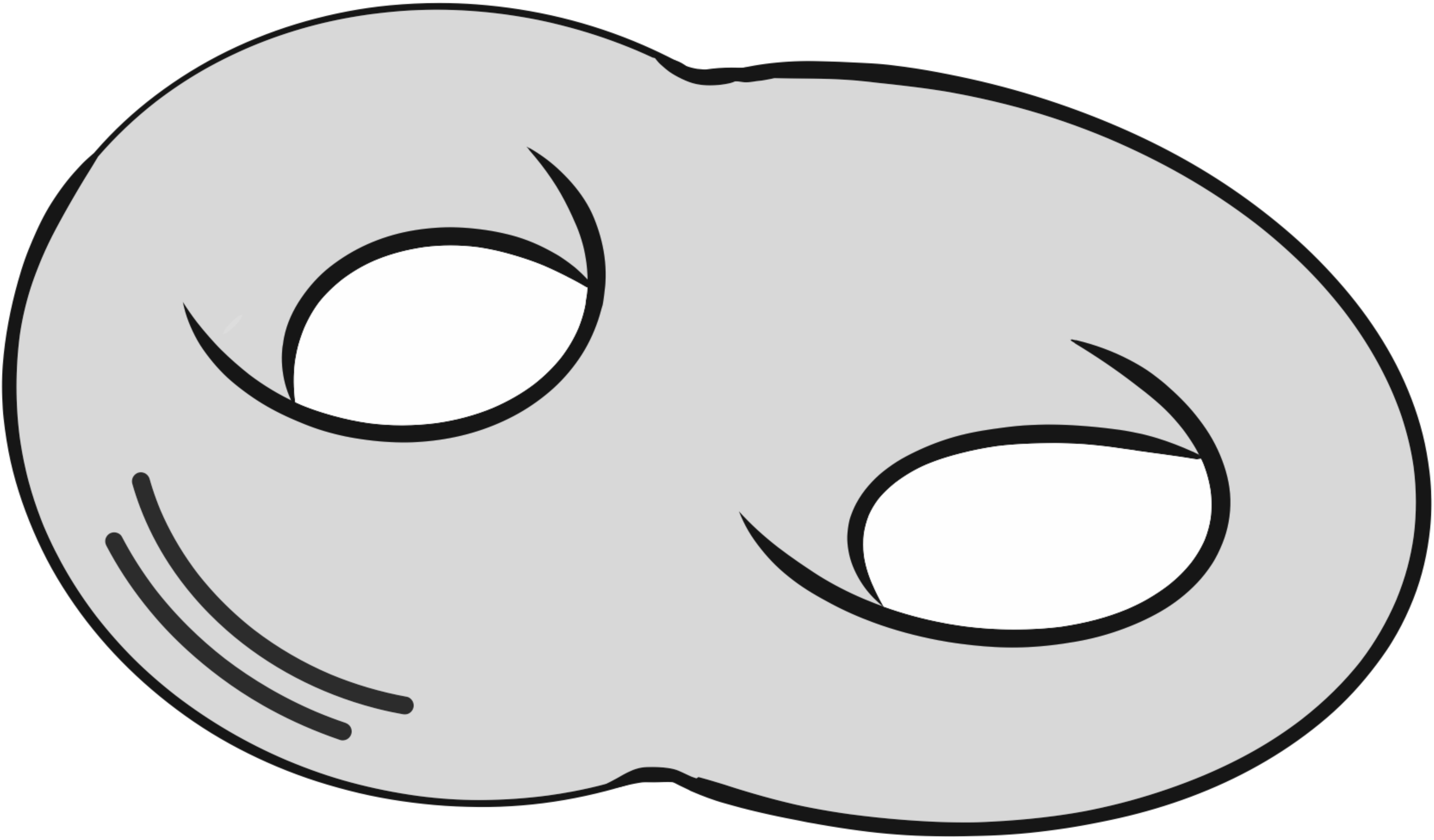}
\qquad \qquad
\includegraphics[width=2.5cm]{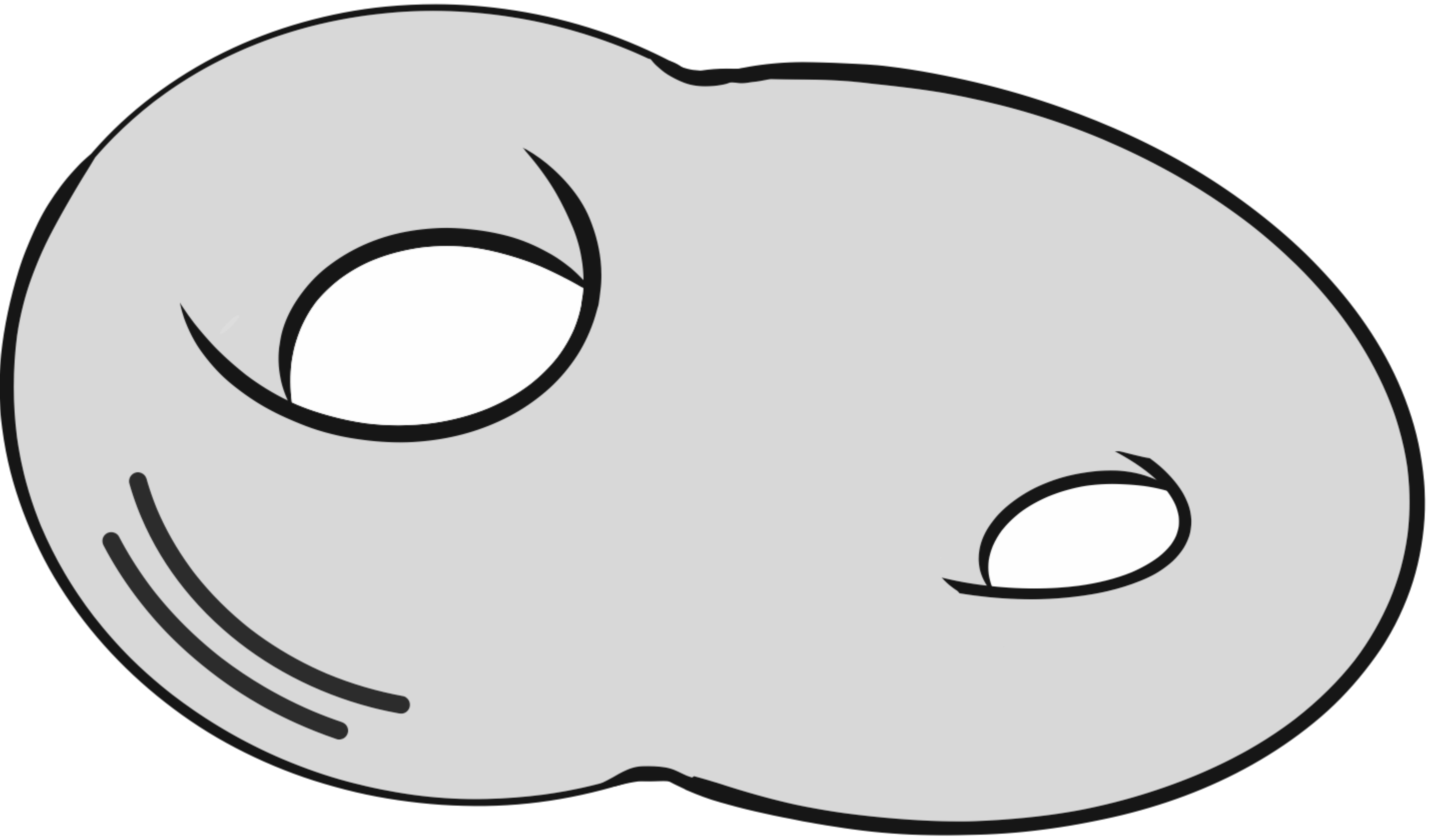}
\qquad \qquad
\includegraphics[width=2.5cm]{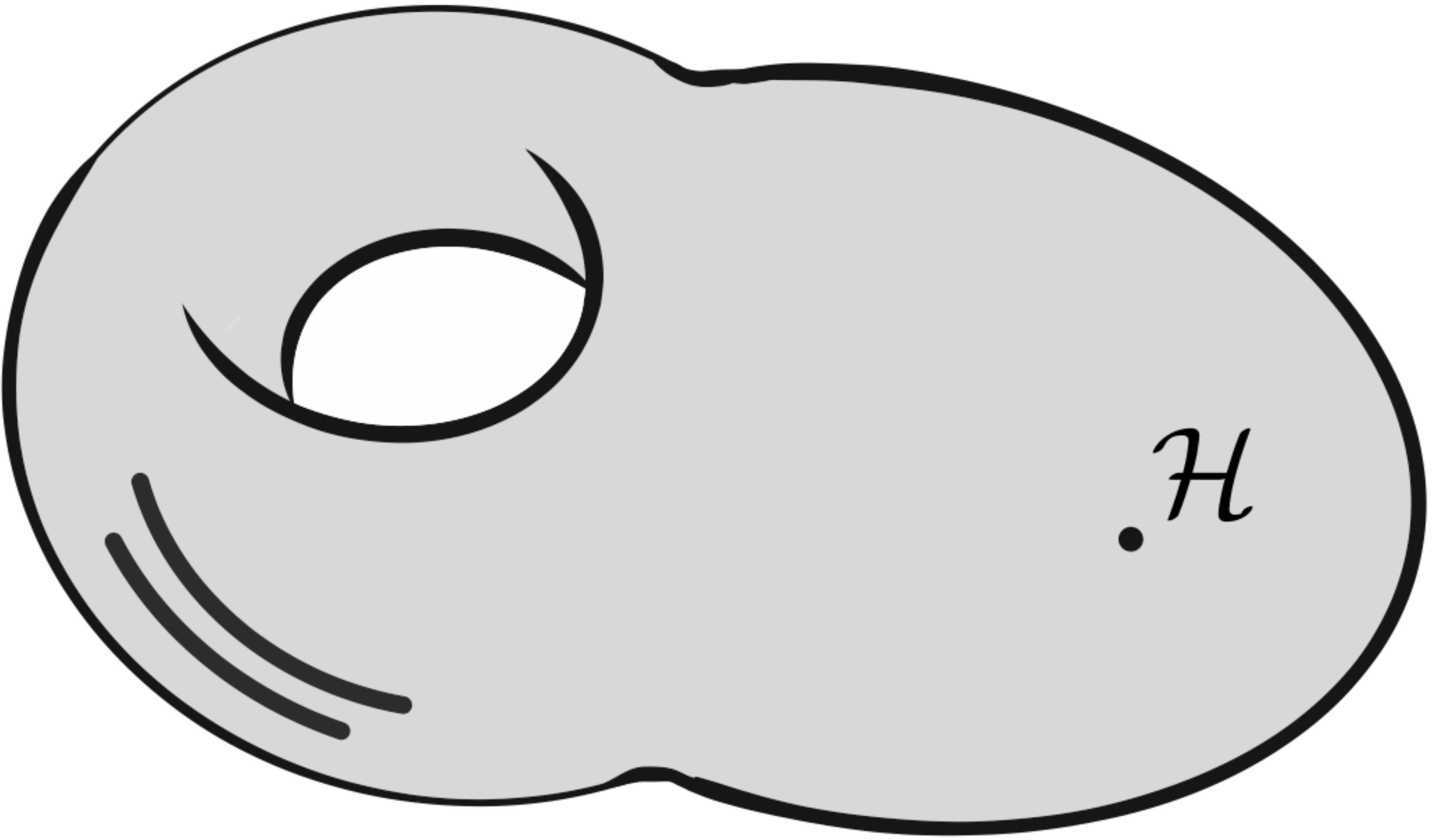}
}
\caption{Shrinking an handle to a point, we obtain the handle-gluing operator, $\CH$, as a quasi-local operator of the $A$-model. \label{fig:hgo}}
\end{figure}

When the gauge group $\GG$ is non-abelian, the Coulomb branch vacua are in one-to-one correspondence with the solution to the Bethe equations modulo the unbroken Weyl symmetry, $W_{\GG}$. In general, there exist solutions which are fixed under the action of $W_{\GG}$, which corresponds to the solutions with unbroken non-abelian symmetry. We will refer to such vacua as the \emph{degenerate vacua}. Following the standard lore,~\cite{Hori:2006dk} we claim that they do not correspond to physical vacua and we therefore discard them.~\footnote{For the 3D generalization to be discussed below, and in the special case of a Chern-Simons theory coupled to an adjoint chiral multiplet, this claim has been rigorously proven in Ref~\protect\citen{2003math.....12154T}, in the language of moduli stacks of $\GG$-bundles on $\Sigma_g$.} We then define the set of \emph{Bethe vacua} as:
\be
\CS_{\text{BE}} = \left\{\hat \sigma_a\Big|~\Pi_a = 1\ ,a=1,\cdots, \rk\ ,~ w(\sigma) \neq \sigma\ , w\in W_{\GG} \right\}/W_{\GG}\ .
\ee
The quotient by $W_\GG$ corresponds to the fact that the allowed solutions to the Bethe equations fill out complete Weyl-group orbits, and we count each such orbit as one Bethe vacuum. 
Let us assume that there are finitely many such solutions, corresponding to isolated massive vacua. In the absence of flavor background flux on $\Sigma_g$, the expectation value of any twisted chiral ring operator, $\CO(\sigma)$, can be written as:
\be
\langle \CO(\sigma)\rangle_{\Sigma_g} = \sum_{\hat\sigma \in \CS_{\text{BE}}} \CO(\hat\sigma) \,\CH(\hat\sigma)^{g-1}\ ,
\ee
where:
\be\label{handle gluing}
\CH(\hat\sigma) = e^{2\pi i \Omega (\sigma)}\det_{ab} \frac{\partial^2 \CW(\sigma)}{\partial\sigma_{a}\partial\sigma_b}\ ,
\ee
is the so-called \emph{handle-gluing operators}. 
This formula was first derived by Vafa for topological Landau-Ginzburg models.~\cite{Vafa:1990mu}  More recently, it was generalized to gauge theories by Nekrasov and Shatashvili.~\cite{Nekrasov:2014xaa} For any Riemann surface $\Sigma_g$, with $g>0$, we can use the topological invariance of $\Sigma_g$ to shrink the volume of any ``handle" so that it becomes a point-like singularity on $\Sigma_{g-1}$, as shown in Figure~\ref{fig:hgo}. In this limit, the insertion of a handle can be thought of as a quasi-local operator in the (extended) twisted chiral ring, given by the expression \eqref{handle gluing}. (The contribution from the effective dilaton to \eqref{handle gluing} follows from the curvature coupling in \eqref{2d effective action}, while the Hessian of the superpotential is a contribution from fermionic zero-modes.)

The twisted chiral ring operators on the classical Coulomb branch are constructed from the Weyl invariant polynomials $\CO(\sigma)$ in $\sigma_a$, with $a=1,\cdots, \rk$. Quantum mechanically, the $A$-model correlation functions are subject to the chiral ring relation, which can be written as:
\be
\langle\CO(\sigma) P_a(\sigma)\rangle_{\Sigma_g} = 0\ ,~~\qquad a=1,\cdots, \rk
\ee
where we denote by $P_a(\sigma)=0$ the Bethe equations written as a set of the polynomial equations.

\subsection{The 3D $A$-model}\label{subsec: 3D Amod}
Let us move on to the 3D $\CN=2$ theories defined on $\mathbb{R}^2\times S^1$, by viewing the 3D theory as a 2D theory with an infinite number of fields. Integrating out all massive fields, we again obtain an effective field theory on the Coulomb branch, parameterized by 2D fields denoted by:
\be
u= {\rm diag}(u_a)~, \qquad a=1, \cdots, \rk~,
\ee 
and by flavor parameters (complexified 3D real masses) denoted by $\nu=(\nu_\alpha)$. 
For the 3D theory on $\R^2 \times S^1$, we can also specify a non-trivial holonomy $\nu_R$ along the $S^1$, with:
\be\label{yR def}
y_R \equiv e^{2\pi i \nu_R} = (-1)^{2\nu_R}~.
\ee
corresponding to the periodic spin structure for $\nu_R=0$ mod $1$, or the anti-periodic spin structure for $\nu_R= \half$ mod $1$, respectively. Let us first discuss the case $\nu_R=0$, before giving the generalization to $\nu_R$ arbitrary.~\cite{Closset:2018ghr}

\subsubsection{The effective twisted superpotential in 3D (with $\nu_R=0$)}
The Coulomb branch low-energy dynamics is now determined by the 3D twisted superpotential:
\be
\CW(u,\nu)\ ,
\ee 
which can be obtained by summing over all the massive fluctuations at a generic point on the classical Coulomb branch, including the contribution from the Kaluza-Klein modes along $S^1$. The twisted superpotential gets contribution from the classical Chern-Simons action and from the chiral multiplet at one-loop:
\be
\CW(u,\nu) = \CW_{\text{CS}} + \CW_{\text{1-loop}}\ .
\ee
The classical contribution from the gauge, flavor and gravitational CS terms reads:~\cite{Closset:2017zgf}
\bea
\CW_{\text{CS}}(u,\nu) =& \sum_{a}\frac12 k_{aa}u_a(u_a+1)+\sum_{a>b} k_{ab}u_a u_b + \sum_{\alpha}\frac12 k_{\alpha\alpha}\nu_\alpha(\nu_\alpha+1)\\
&+\sum_{\alpha>\beta} k_{\alpha\beta}\nu_\alpha \nu_\beta+ \sum_{a,\alpha} k_{a\alpha}u_a\nu_\beta +\frac{1}{24}k_g\ .
\eea
where, for each simple factor $\GG_\gamma$ of the the gauge group, it is understood that $k_{ab} = h_{ab}k_\gamma$ with the Killing form $h$. The one-loop contribution $\CW_{\text{1-loop}}$ can be written as a sum over contributions from the chiral multiplets of the theory. A chiral multiplet $\Phi$ of unit charge under some $U(1)$ contributes:
\be
\CW^{\Phi}(u)  \equiv \frac{1}{(2\pi i )^2}\text{Li}_2 (x)~,
\ee
where $x = e^{2\pi i u}$ as usual. Then, the one-loop contribution comes entirely from the chiral multiplets, in the representation $\frak R$ of the gauge group:
\be\label{1loop W 3D}
 \CW_{\text{1-loop}} = \sum_{\rho\in {\frak R}}\CW^{\Phi}(\rho(u)+ \nu)~, 
\ee
with the dependence on the flavor parameters given schematically. Here, we used the ``$U(1)_{-1/2}$ quantization'' for  the chiral multiplets.
The 3D W-bosons do not contribute to $\CW$ (with $\nu_R=0$), due to our choice of symmetric quantization between $\CW_\alpha$ and $\CW_{-\alpha}$.

The 3D twisted potential suffers from the branch-cut ambiguity:
\be\label{ambiguity 3d}
\CW_{\text{eff}} \rightarrow \CW_{\text{eff}}+n^a u_a + n^\alpha \nu_\alpha + n^0\ ,~~ n^a, n^\alpha, n^0 \in \mathbb{Z}\ ,
\ee
as is apparent from the multi-valuedness of the dilogarithm. Nonetheless, as in 2D, the physical observables are  well-defined and single-valued. 

Once we couple the theory to the curved background $\Sigma_g\times S^1$ with the topological A-twist on $\Sigma$, the effective dilaton $\Omega$ in \eqref{2d effective action} can be similarly computed by summing over the KK modes along $S^1$. We have:
\be
\Omega(u,\nu) = \Omega_{\text{CS}} + \Omega_{\text{1-loop}}\ ,
\ee 
where the classical contribution is determined by the CS action for the background R-symmetry:
\be
\Omega_{\text{CS}} = \sum_{a}k_{aR}u_a + \sum_{\alpha}k_{\alpha R}\nu_\alpha + \frac12 k_{RR}\ .
\ee
The one-loop term gets contributions from both the chiral multiplets $\Phi$, and the $W$-bosons:
\be\label{Omega 1loop}
 \Omega_{\text{1-loop}}= -\frac{1}{2\pi i }\sum_{\rho\in {\mathfrak R}} (r-1)\log(1-x^\rho y)-\frac{1}{2\pi i }\sum_{\alpha\in {\mathfrak g}}\log(1-x^\alpha)\ ,
\ee
with $y= e^{2\pi \nu}$ denoting the flavor dependence, again schematically.

\subsubsection{The Bethe vacua}
As in 2D, we can define the 3D flux operators for gauge and flavor symmetry from the 3D twisted twisted potential $\CW$:
\be\label{flux op 3D}
\Pi_a (u,\nu)= \exp\left(2\pi i \frac{\partial \CW}{\partial u_a}\right)\ ,\qquad ~~\Pi_\alpha (u,\nu)=\exp\left(2\pi i \frac{\partial \CW}{\partial m_a}\right)\ .
\ee
Then, the 3D ``Bethe vacua'' are the Coulomb branch vacua for the 3D theory on a circle, which are solutions to the Bethe equations:
\be\label{bethe vacua 3D formula}
\CS_{\text{BE}} = \left\{\hat u_a\Big|~\Pi_a(u,\nu) = 1\ ,a=1,\cdots, \rk\ ,~ w(u) \neq u\ , w\in W_{\GG} \right\}/W_{\GG}\ .
\ee
Note that the flux operators and the vacuum equations are again free from the branch-cut ambiguity \eqref{ambiguity 3d}. The partition function on $\Sigma_g\times S^1$ can be immediately written as:
\be\label{bethe sum 1}
Z_{\Sigma_g\times S^1} = \sum_{\hat u_a \in \CS_{\text{BE}}} \CH(\hat u_a)^{g-1} \prod_{\alpha \in F}\Pi_\alpha(\hat u_a, \nu_\alpha)^{\n_\alpha}\ ,
\ee
where we also consider the insertion of flavor flux operators, which introduce background flavor fluxes $\n_\alpha$. Here, we also defined the 3D handle-gluing operators:
\be
\CH( u_a) = e^{2\pi i \Omega(u,\nu)} \det_{ab}\frac{\partial^2\CW}{\partial u_a \partial u_b}~,
\ee
similarly to the 2D case. Note that, as a special case, we have the ``flavored Witten index:''
\be
Z_{T^2 \times S^1} = Z_{T^3}  =  \sum_{\hat u_a \in \CS_{\text{BE}}} 1= |\CS_{\text{BE}}|~,
\ee
which is the Witten index of the 3D theory regularized with generic flavor real masses and holonomies.~\cite{Intriligator:2013lca, Closset:2016arn}

\subsubsection{Spin structure dependence of the 3D $A$-model}
We can consider $\R^2 \times S^1$, or $\Sigma_g \times S^1$, with a non-trivial $U(1)_R$ holonomy $\nu_R$ along the $S^1$, corresponding to a change of spin structure as explained above. Having $\nu_R \neq 0$ affects the above discussion as follows. The CS contributions to the twisted superpotential takes the form:
\be
\CW_{\text{CS}} = \half k_{GG} (u^2+ (1+ 2 \nu_R)u) + k_{GR} \nu_R u~,
\ee
for the gauge sector, and similarly for the flavor sector (in terms of $\nu_\alpha$ instead of $u_a$). Note that the mixed gauge-$U(1)_R$ CS level appears non-trivially here, when $\nu_R \neq 0$ (contributing an additional sign to the gauge flux operator). For a $U(1)$ gauge group with CS level $k\in \Z$,  the classical contribution to the gauge flux operator (with flux $\m\in \Z$)  on $\Sigma_g \times S^1$ reads:~\cite{Closset:2017zgf}
\be
\Pi_{\rm CS}^\m = (-1)^{(1+2\nu_R)\m k} x^{\m k}~,
\ee
and therefore depends on the choice of spin structure if $k$ is odd. The one-loop contribution to the twisted superpotential, on the other hand, is given by:~\cite{Closset:2018ghr}
\be\label{1loop W 3D nuR}
 \CW_{\text{1-loop}} = \sum_{\rho\in {\frak R}}\CW^{\Phi}(\rho(u)+\nu+ \nu_R r)+ 2 \nu_R\, \rho_W(u)~, 
\ee
with $r \in \Z$ denoting the $R$-charge(s) of the chiral multiplet(s). Note also the contribution from the $W$-bosons, given in terms of the Weyl vector  $\rho_W=\half \sum_{\alpha\in \Delta^+}\alpha$ of $\GG$. 
Finally, $\nu_R$ only enters the effective dilaton $\Omega$ through the one-loop term \eqref{Omega 1loop}, where the argument $x^\rho y$ in the matter contribution should be replaced by $y_R^r x^\rho y$.

\subsection{Contour-integral formulas and Bethe equations for the twisted index}

The Bethe sum formula \eqref{bethe sum 1} for the partition function at genus zero provides an alternative expression for the integral representation of the twisted index in section~\ref{sec: S2 top index}, in the unrefined limit $\epsilon = 0$. 

Let us sketch how we can recover the expression \eqref{bethe sum 1} starting from the contour-integral formula \eqref{integral formula rk1}, for the theories with $\rk=1$.
Let us first define a new contour $\CC^\eta_0 \in \fM\backslash \Delta_\varepsilon$ which depends on the choice $\eta\in {\mathfrak t}^*$:
\be
\CC_0^\eta = \left\{u \in \partial(\fM\backslash \Delta_\varepsilon)\; |\; ~\text{sign}(\text{Im}(\partial_u \CW)) = -\text{sign}(\eta)
\right\}\ .
\ee
One can show that this contour can be continuously deformed to the Jeffrey-Kirwan residue integral \eqref{integral formula rk1}.~\cite{Closset:2018ghr} Then, the integral formula can be written as:
\be
Z_{S^2\times S^1} =  \frac{1}{|W_{\GG}|}\sum_{\m \in \mathbb{Z}} \int_{\CC_0^\eta} du~  e^{-2\pi i \Omega(u)}\Pi(u,\nu)^\m \prod_{\alpha}\Pi_\alpha(u,\nu)^{\n_\alpha}\ ,
\ee
where we wrote the integrand in terms of the effective dilaton and of the flux operators.  (Note that $v_R= \nu_R$ when $\epsilon=0$.) The contour $\CC_0^\eta$ has a property that, at any point $u\in\CC_0^\eta$, we have $|\Pi(u)|<1$ if $\eta<0$, and $|\Pi(u)|>1$ if $\eta>0$. Defining:
\be
I(u) =e^{-2\pi i \Omega(u)}\prod_{\alpha}\Pi_\alpha(u,\nu)^{\n_\alpha}\ ,
\ee
the following choices of $\eta$ for each $\m$ allows us to perform the geometric sum over $\m$, which converges at every point on the contour:
\bea\label{contour integr manip}
Z_{\Sigma\times S^1} &= \sum_{\m = -\infty}^{-1}\int_{\CC_0^{\eta>0}} du~I(u)\Pi(u)^\m + \sum_{\m=0}^\infty \int_{\CC_0^{\eta<0}} du~ I(u)\Pi(u)^\m \\
&= \int_{\CC_0^{\eta>0}} du~\frac{I(u)}{\Pi(u)-1} +  \int_{\CC_0^{\eta<0}} du~\frac{I(u)}{1-\Pi(u)} \\
&= \oint_{\CC_{\text{BE}}} du~ \frac{I(u)}{1-\Pi(u)}\ .
\eea
Here, we introduced the new contour $\CC_{\text{BE}}$ defined as:
\be
\CC_{\text{BE}} = -\CC_0^{\eta>0} + \CC_0^{\eta<0}\ .
\ee
Under a mild technical assumption on the R-charges and flavor parameters of the chiral multiplets, we can show that the only poles surrounded by the new contour $\CC_{\text{BE}}$ are at $\Pi(u)=1$. One can then easily derive  the expression \eqref{bethe sum 1} from the last line in \eqref{contour integr manip}.

For higher-genus Riemann surfaces, there exist subtleties in relating the contour-integral and the Bethe-vacua expressions directly, mainly due to integrand singularities coming from the W-bosons. The Bethe-sum formula \eqref{bethe sum 1} always hold,~\cite{Closset:2018ghr} however, while the proper definition of the contour integral is not fully understood in general. (See Refs.~\citen{Benini:2016hjo, Closset:2016arn} for partial discussions.)

\subsection{Bethe equations and infrared dualities}
The Bethe-vacua  formula \eqref{bethe sum 1} for the $\Sigma_g \times S^1$ partition functions provides a very simple tool for studying various infrared dualities of 3D $\CN=2$ gauge theories. Any infrared duality between two theories, $\CT$ and $\CT^D$, implies the existence of the one-to-one mapping among the solutions to the Bethe equations:
\be\label{solution pair}
\{\hat x_a\}\quad ~~\leftrightarrow\quad ~~\{\hat x^D_{\bar a}\}\ ,
\ee
where $\hat x_a$'s and $\hat x^D_{\bar a}$'s are the solution for the Bethe equations for the two theories $\CT$ and $\CT^D$, respectively (with $a$ and $\b a$ running over the Cartan subalgebra of the two dual gauge groups):
\be
\Pi_a(\h x)=1\quad ~~\leftrightarrow \quad ~~\Pi^D_\ba(\h x^D) = 1\ .
\ee
More generally, for any twisted chiral ring operator $\SL(x)$ in $\CT$, we can find a dual element $\SL_D(x)$ in $\CT^D$ such that, for any dual pair of Bethe vacua \eqref{solution pair}, the dual operators must agree ``on-shell,'' namely:
\be
\SL(\hat x) = \SL_D(\h x^D)\ .
\ee
In particular, this must be true for the handle-gluing operator and the flavor flux operators of the dual theories. The infrared dualities thus implies the non-trivial identities:
\be\label{duality}
\CH(\h x) = \CH^D(\h x^D)~, \qquad \qquad 
\Pi_\alpha(\h x)  = \Pi^D_\alpha(\h x^D)\ .
\ee
To illustrate this point, let us study these relations in a couple of examples.

\subsubsection{Elementary mirror symmetry}
Consider the elementary 3D $\CN=2$ mirror symmetry introduced in section~\ref{subsec: EMS}. Here, to illustrate the general case, we consider the possibility of a non-trivial $\nu_R$ (and the associated choice of spin structure) on $\Sigma_g \times S^1$. The twisted superpotential for the theory $\CT$ reads:
\be
\CW(u, \xi) = {1\ov (2\pi i )^2} \dilog(y_R^r x)+ \half (u^ 2+ (1+ 2\nu_R)u) + \xi u~,
\ee
with $\xi= \nu_T$  the complexified FI parameter---that is, the flavor parameter for the topological symmetry, $U(1)_T$. Here, $x= e^{2\pi i u}$, $y_R$ is defined as in \eqref{yR def}, and we will also use the notation $q= e^{2\pi i \xi}$ for the $U(1)_T$ fugacity.
The Bethe equation has a single solution:
\be\label{xhat EMS}
\Pi(x) = -\frac{y_R q x}{1-y_R^r x} = 1~, \qquad \h x= {1\ov y_R^r - y_R q}~.
\ee
corresponding to the trivial vacuum of the dual theory $\CT^D$, the free chiral multiplet $T^+$. The handle gluing operator for the theory $\CT$ is given by:
\be
\CH(x) = \left(\frac{1}{1- y_R^r x}\right)^{r-1} \d_u^2 \CW = \left(\frac{1}{1-y_R^r x}\right)^{r}\ .
\ee
Plugging in the Bethe vacuum solution \eqref{xhat EMS}, we obtain
\be
\CH(\hat x) = (-1)^{r^2} q^{-r}  \left(\frac{1}{1-y_R^{1-r}q^{-1}}\right)^{-r}\ ,
\ee
which precisely agrees with handle-gluing operator of the dual theory $\CT^D$, with the relative CS contact terms $\Delta k_{RT}=-r$ and $\Delta k_{RR}= r^2$.  
Similarly,  the flavor flux operator for $U(1)_T$ is simply $\Pi_T(x) = x$ in the gauge theory, which gives:
\be
\Pi_T(\hat x) = {y_R^{-r}\ov 1 - y_R^{1-r} q}~,
\ee
which agrees with the $U(1)_T$ flux operator $\Pi^D$ in the dual description, including the sign $y_R^{-r}$ in the numerator from the mixed $U(1)_R$-$U(1)_R$ CS level $\Delta k_{RT}$. 

\subsubsection{Bethe equations and the Aharony duality}

Let us also consider the Aharony duality between the gauge theories $\CT$ and $\CT_D$ introduced in section~\ref{subsec: Aharony duality}. 
Here, for simplicity, we will only consider the case $\nu_R=0$. (One can easily generalize the discussion below to $\nu_R \neq 0$.)
The twisted superpotential of the $U(N_c)$ gauge theory $\CT$ takes the form:
\be
\CW= \sum_{a=1}^{N_c} \left[ {1\ov (2\pi i)^2} \sum_{i=1}^{N_f}\left(\dilog(x_a y_i^{-1})+ \dilog(x_a^{-1} \t y_i)\right) + {N_f \ov 2 }u_a(u_a+1) + \xi u_a \right]~.
\ee
Here, we have the flavor fugacities $y_i, \t y_j$ such that $\prod_{i=1}^{N_f} y_i^{-1}=\prod_{j=1}^{N_f} \t y_j= y_A^{N_f}$, with $y_A$ the $U(1)_A$ fugacity, and we have $q=e^{2\pi i \xi}$ for $U(1)_T$. 
 The Bethe equation for the theory $\CT$ can be written as:
\be
P(x_a) = 0\ ,\quad ~~a=1,\cdots, N_c\ ,~~x_a\neq x_b\ ,~~\forall a\neq b\ , 
\ee
in terms of the single-variable polynomial:
\be\label{bethe aharony}
P(x) \equiv \prod_{i=1}^{N_f} (x-y_i) - q y_A^{-N_f}\prod_{i=1}^{N_f}(x-\t y_i)~,
\ee
of degree $N_f$. If we denote by $\{\hat x_\beta\}_{\beta=1}^{N_f}$ the set of $N_f$ roots of $P(x)$, the ``Bethe roots,'' then the Bethe vacua are in one-to-one correspondence with choices of $N_c$ distinct Bethe roots $\{\hat x_a\}_{a=1}^{N_c} \subset \{\hat x_\beta\}_{\beta=1}^{N_f}$ . In particular, we have:
\be\label{ZT3 UN SQCD}
Z_{T^3}= |\CS_{\text{BE}}| = {N_f \choose N_c}~,
\ee
for the flavored Witten index of 3D $\CN=2$ $U(N_c)$ SQCD.

One can easily check that the Bethe equation of the dual, $U(N_f-N_c)$ gauge-theory description, take the form:
\be
P(x_\ba) = 0\ ,\quad ~~\ba=1,\cdots, N_f-N_c\ ,~~x_\ba\neq x_{\b b}\ ,~~\forall \ba\neq \b b\ , 
\ee
in terms of the {\it same} polynomial \eqref{bethe aharony}. A Bethe vacua in the theory $\CT^D$ is simply a choice of $N_f-N_c$ distinct Bethe roots; in particular, the number of Bethe vacua, \eqref{ZT3 UN SQCD}, is the same in both descriptions. The duality maps a Bethe vacuum $\{\hat x_a\}_{a=1}^{N_c} \subset \{\hat x_\beta\}_{\beta=1}^{N_f}$ to its complement in the set of Bethe roots:
\be\label{AD duality map}
\{ \h x\} \equiv \{\hat x_a\}_{a=1}^{N_c} \subset \{\hat x_\beta\}_{\beta=1}^{N_f} \qquad \leftrightarrow \qquad \{ \h x^D\} \equiv \{\hat x_\ba\}_{a=1}^{N_f- N_c} = \{ \h x\}^c~.
\ee
This simple duality map between Bethe vacua makes it rather easy to check the duality relations \eqref{duality}, as we now show.

The handle-gluing operator of the theory $\CT$ takes the form:
\be\nn\label{CH sqcd 1}
\CH = \prod_{a=1}^{N_c} \left[ H(x_a) \prod_{i=1}^{N_f} \Big[(1- x_a y_i^{-1})^{1-r} (1- x_a^{-1} \t y_i)^{1-r} \Big]\right] \prod_{\substack{a, b=1\\ a \neq b}}^{N_c} {1\ov 1- x_a x_b^{-1}}~,
\ee
where we used the fact that the Hessian determinant takes the simple form:
\be\label{Hx sqcd def}
\det_{ab}\d_{u_a} \d_{u_b} \CW= \prod_{a=1}^{N_c} H(x_a)~, \qquad \quad H(x) \equiv  \sum_{i=1}^{N_f} {x (\t y_i -y_i)\ov (x-y_i)(x-\t y_j)}~.
\ee
On the other hand, the handle-gluing operator of the dual gauge theory takes the form:
\be
\CH^D(x) = \CH_{\rm ct} \, \CH_{\rm singlets}\,  \CH_{\rm gauge}^D(x)~,
\ee
where the three factors are contributions from the relative CS terms \eqref{AD CS ct1}:
\be\label{CH ct sqcd}
 \CH_{\rm ct}= (-1)^{\Delta k_{RR}} y_A^{\Delta k_{AR}}= (-1)^{N_f+N_c} y_A^{2 N_f^2 +(4 N_f^2 - 2 N_f N_c)(r-1)}~,
\ee
the contribution from the gauge singlets:
\be
\CH_{\rm singlets}=  \prod_{i, j=1}^{N_f} \left({1\ov 1- y_i^{-1} \t y_j}\right)^{2 r-1} 
  \left({1\ov 1- q y_A^{-N_f}} \right)^{r_T -1} \left({1\ov 1- q^{-1} y_A^{-N_f}} \right)^{r_T -1}~,
\ee
where $r_T\equiv -N_f(r-1)-N_c+1$  denotes the $R$-charge gauge singlets $T^\pm$,
and the contribution from the  $U(N_f-N_c)$ gauge-theory sector:
\be\nn
 \CH_{\rm gauge}^D = \prod_{\ba=1}^{N_f-N_c} \left[-H(x_\ba) \prod_{i=1}^{N_f} \Big[(1- x_\ba^{-1} y_i)^{r} (1- x_\ba \t y_i^{-1})^{r} \Big]   \right] \prod_{\substack{\ba,\b b=1\\ \ba \neq \b b}}^{N_f-N_c} {1\ov 1- x_\ba x_{\b b}^{-1}}~,
\ee
in terms of the same function $H(x)$ as in \eqref{Hx sqcd def}.

\paragraph{Matching the handle-gluing operators across the duality.}  To show the equality of the handle-gluing operators, \eqref{duality}, using the duality map \eqref{AD duality map}, let us specialize to the case of $r=1$. (The general case is easily obtained by mixing this $R$-charge with $U(1)_A$.)
We have the useful identity:
 \be
\partial_x P(x) |_{x=\hat x_\beta} = -\hat x_\beta^{-1} H(\hat x_\beta)\prod_{i=1}^{N_f}(\hat x_\beta -y_i)~,
\ee
on any Bethe root $x= \h x_\beta$. Then, the handle-gluing operator \eqref{CH sqcd 1} evaluated on any Bethe vacuum $\h x= \{\h x_a\}_{a=1}^{N_c}$ reads:
\be
\CH(\h x) =  \prod_{a=1}^{N_c} \Big[ -\h x_a^{N_c} \d_x P(\h x_a)  \prod_{i=1}^{N_f} (\h x_a- y_i)^{-1} \Big] \, \prod_{\substack{a, b=1\\ a \neq b}}^{N_c} (\h x_a- \h x_b)^{-1}~.
\ee
while in the dual gauge theory, we have:
\bea
&\CH^D_{\rm gauge}(\h x^D) &=&\; q^{-(N_f-N_c)}\prod_{\ba =1}^{N_f-N_c} \Big[ (-1)^{N_f} x_\ba^{-N_c}  \d_x P(\h x_\ba)  \prod_{i=1}^{N_f} (\h x_\ba- y_i)^{-1}\Big]\\
&&&\times\prod_{\substack{\ba,\b b=1\\ \ba \neq \b b}}^{N_f-N_c}(x_\ba- x_{\b b})^{-1}~.
\eea
Using the factorized form of the polynomial $P(x)$, namely:
\be\label{P prod bethe roots}
P(x) = (1-q y_A^{-N_f}) \prod_{\beta=1}^{N_f}(x-\hat x_\beta)~,
\ee
where the product $\prod_\beta$ runs over all the Bethe roots, one easily derives the identities:
\bea\label{lemmas sqcd rels}
&\prod_{\beta=1}^{N_f} \h x_\beta = {y_A^{-N_f}- q \ov 1- q y_A^{-N_f}}~, \cr
&\prod_{\beta=1}^{N_f} (\h x_\beta- y_i) = (-1)^{N_f-1}  {q y_i^{N_f} y_A^{-N_f} \ov 1- q y_A^{-N_f} } \prod_{j=1}^{N_f} (1- y_i^{-1} \t y_j)~.
\eea
We also have:
\be\nn
{\prod_{a=1}^{N_c} \d_x P(\h x_a) \ov \prod_{a \neq b}^{N_c} (\h x_a- \h x_b) } = (-1)^{N_c(N_f-N_c)} (1- q y_A^{-N_f})^{2N_c-N_f}{ \prod_{\ba=1}^{N_f-N_c}\d_x P(\h x_\ba) \ov \prod_{\ba \neq \b b}^{N_f-N_c} (\h x_\ba- \h x_{\b b})}~, 
\ee
for any decomposition of the Bethe roots into dual vacua, $\{ \h x_\beta\}_{\beta=1}^{N_f} = \{ \h x\} \cup \{ \h x^D\}$. Using these relations, one can check that:
\bea
&\CH(\h x)  = \CH^D(\h x^D)  =&& (-1)^{N_f+N_c} y_A^{2N_f^2} (1- q y_A^{-N_f})^{N_c} (1- q^{-1} y_A^{-N_f})^{N_c}\\
&&& \times \prod_{i, j=1}^{N_f} (1- y_i^{-1} \t y_j)^{-1} \;  \CH^D_{\rm gauge}(\h x^D)~,
\eea
for any pair of dual Bethe vacua, in perfect agreement with the Aharony duality.

\paragraph{Flux operators across the duality.} One can similarly check that all flux operators for the $SU(N_f)\times SU(N_f) \times U(1)_A \times U(1)_T$ flavor symmetry match across the duality. As a simple example, consider the $U(1)_T$ flux operator, which in the $U(N_c)$ gauge theory simply reads:
\be
\Pi_T(x)= \prod_{a=1}^{N_c} x_a~,
\ee
while in the dual theory we have:
\be
\Pi_T^D(x)= (-q)\, \left( \prod_{\ba=1}^{N_f-N_c} x_\ba^{-1} \right) \, {1- q^{-1} y_A^{-N_f} \ov 1- q y_A^{-N_f}}~,
\ee
where the first factor corresponds to the $U(1)_T$ CS level $\Delta k_{TT}=1$, and we have used the fact that the FI terms map as $\xi_D= - \xi$ across the duality. The equality $\Pi_T(\h x)= \Pi_T(\h x^D)$ is then equivalent to the first identity in \eqref{lemmas sqcd rels}. The duality relations for the other flux operators can be checked similarly.

\subsection{Half-BPS Wilson loop algebra and dualities}
As mentioned above, one advantage of the 2D Coulomb-branch point of view is that it makes the quantum relations in the twisted chiral ring manifest; they all follow from the effective twisted superpotential. In this way, the 3D $A$-model encodes the algebra of half-BPS line operators---in particular, of half-BPS Wilson loops.

\subsubsection{The half-BPS Wilson loop quantum algebra}
If we consider the 3D theory on a circle, the twisted chiral ring operators $\CO$ in 2D uplift to the 3D half-BPS line operators $\SL$ wrapped along the $S^1$. These half-BPS operators are the lines that preserve the two supercharges $\CQ_-$ and $\t\CQ_+$  of the $A$-model. 
In a gauge theory, the correlation functions of parallel half-BPS lines on $\Sigma_g\times S^1$ can then be readily written as:
\be\label{bethe vac line exp 0}
\langle\SL_i(x) \SL_j(x)\cdots \rangle_{\Sigma_g \times S^1} =  \sum_{\h x \in \CS_{\rm BE}}  \SL_i(\h x) \SL_j(\h x)\cdots\; \CH^{g-1}(\h x)~.
\ee
This implies that they are subject to the twisted chiral ring relations:
\be\label{tcr rels gen}
\langle \SL_i(x) \SL_j(x) \cdots P(x)\rangle_{\Sigma_g \times S^1}  = 0\ ,
\ee
where $P(x)$ is any element of the ideal $\CI_\CW$ generated by the gauge flux operators, $\Pi_a(x)$, since $P(\h x)=0$ by definition of the Bethe vacua.  In particular, we can study the half-BPS Wilson lines introduced in section~\ref{subsec: 3D line obs}. Classically, the algebra of Wilson lines is generated by the symmetric polynomials in $x_a$:
\be\label{W as character}
W_{\FR}(x) = \text{Tr}_{\FR}\left( x^\rho\right)\ ,
\ee
where $\FR$ are the irreducible representations of $\GG$. Then, any Wilson line can be written as a linear combination of these operators. For example, when $\GG=U(N_c)$, the classical Wilson loop algebra  is:
\be
\mathbb{C}[x_1,\cdots, x_{N_c}, x_1^{-1},\cdots, x_{N_c}^{-1}]^{S_{N_c}}\ ,
\ee
where  $W_{\GG}=S_{N_c}$ is the Weyl group---this is the algebra of symmetric (Laurent) polynomials, or equivalently the algebra of Young tableaux.
 Quantum mechanically, we have the relations \eqref{tcr rels gen}, namely $\langle W(x) P(x)\rangle = 0$ for any Wilson line $W$, so that the quantum Wilson loop algebra takes the form:
\be
\CA = \mathbb{C}[x_1,\cdots, x_{N_c}, x_1^{-1},\cdots, x_{N_c}^{-1}]^{W_{\GG}}/\CI_\CW~,
\ee
with the relations generated by the effective twisted chiral superpotential.~\cite{Kapustin:2013hpk, Closset:2016arn}

\subsubsection{Duality relations for half-BPS Wilson loops}
An infrared duality between two theories $\CT$ and $\CT^D$ implies relations between Wilson loops expectations values in the two theories: 
\be
\langle W\rangle^{\CT} = \langle W^D \rangle^{\CT^D}~.
\ee
In general, the UV theories $\CT$ and $\CT^D$ have completely different gauge groups, and it is a non-trivial problem to identify the Wilson loop $W^D$ in $\CT^D$ dual to a given Wilson $W$ in $\CT$.
In the Bethe vacua picture, the two Wilson loops are dual if and only if:
\be
W(\h x) = W(\h x^D)\ ,
\ee
for every pairs of dual Bethe vacua \eqref{solution pair} in $\CS_{\text{BE}}$ and $\CS_{\text{BE}}^D$, respectively.

\subsubsection{Example: Aharony duality}
 Let us illustrate the above discussion with the example of 3D SQCD and its Aharony duality.~\cite{Closset:2016arn} First of all, let us define the generating function for the Wilson loops for the $U(N_c)$ gauge theory $\CT$:
\be
Q(z) = \prod_{a=1}^{N_c}(z-x_a) = \sum_{i=1}^{N_c}(-1)^i z^{N_c-i}s_i(x)\ ,
\ee
where the $s_i$'s denote for the elementary symmetric polynomials:
\be
s_i(x) = \sum_{1\leq a_1<a_2\cdots a_i\leq N_c} x_{a_1}x_{a_2}\cdots x_{a_{i}}\ ,~~ i = 0,\cdots, N_c\ ,
\ee 
which correspond to the vertical Young tableaux
\be
s_0(x)=1\ ,~~s_1(x) = {\tiny\yng(1)}\ ,~~s_2(x) = {\tiny\yng(1,1)}\ ,\cdots\ .
\ee
We also define a similar generating function for the dual $U(N_f-N_c)$ gauge theory:
\be
Q_D = \prod_{a=1}^{N_f-N_c}(z-x_{\bar a})=\sum_{i=1}^{N_f-N_c}(-1)^i z^{N_f-N_c+i}s^D_i(x)\ .
\ee 
Then, the expectation value for the characteristic polynomial \eqref{bethe aharony} satisfies the quantum relation:~\cite{Gaiotto:2013sma, Benini:2014mia}
\be\label{quantum relation}
 P(z) =  (1-q y_A^{-N_f})\, \langle Q(z) Q_D(z)\rangle~,
\ee
as follows from \eqref{P prod bethe roots}.
Here, the ``expectation value'' notation means that the Coulomb-branch variables $x_a$ and $x_\ba$ on the right-hand-side are evaluated  onto a pair of dual Bethe vacua. 
Expanding both sides of \eqref{quantum relation} in $z$, we obtain the expression $\langle s_i^{D}(x)\rangle^{\CT_D}$ explicitly in terms of a linear combination of $\langle s_i(x)\rangle^{\CT}$'s. First of all, we expand:
\be
P(x) =  (1-q y_A^{-N_f}) \sum_{i=1}^{N_f}(-1)^m z^{N_f-i} A_i(y,\t y,q)\ ,
\ee
where we define:
\be
A_i(y,\t y,q) = \frac{1}{1-q y_A^{-N_f}} \left(s_i^F(y)-q y_A^{-N_f}\t s_i^F(\t y)\right)~,
\ee
where $s_i^F(y)$ and $\t s_i^F(\t y)$ are elementary symmetric polynomials in the flavor parameters $y_i$ and $\t y_i$, respectively. The quantum relations \eqref{quantum relation} now imply:
\bea\label{wilson loop duality}
\sum_{i=1}^m s_i(x) s_{m-i}^D(x_D) = A_m(y,\t y, q)\ ,
\eea
for $m=1,\cdots, N_f$. For example, if we consider $N_c=3$ and $N_f=5$, we have $N_f = 5$ independent relations:
\bea
&{\tiny\yng(1)} +{\tiny\yng(1)}^D = A_1(y,\t y,q)\ ,\\
&{\tiny\yng(1,1)} + {\tiny\yng(1)} \times {\tiny\yng(1)}^D +{\tiny\yng(1,1)}^D = A_2(y, \t y, q)\ ,\\
&{\tiny\yng(1,1,1)} + {\tiny\yng(1,1)}\times {\tiny\yng(1)}^D + {\tiny\yng(1)}\times {\tiny\yng(1,1)}^D = A_3(y,\t y, q)\ ,\\
&{\tiny\yng(1,1,1)}\times {\tiny\yng(1)}^D + {\tiny\yng(1,1)}\times {\tiny\yng(1,1)}^D = A_4(y,\t y, q)\ ,\\
&{\tiny\yng(1,1,1)}\times {\tiny\yng(1,1)}^D = A_5(y,\t y, q)~,
\eea
where we denote the Wilson loop expectation value by its Young tableau in the obvious way.
It is straightforward to solve these equations recursively for the dual Wilson loops, which gives:
\bea\label{relation 1}
&{\tiny\yng(1)}^D = A_1 - {\tiny\yng(1)}\ ,\\
&{\tiny\yng(1,1)}^D = A_2 -  A_1\cdot {\tiny\yng(1)} + {\tiny\yng(2)}\ .
\eea
For general $N_c$ and $N_f$, we find that the dual Wilson loop  $W^D$ in the $n$-antisymmetric representation, corresponding to $n$ vertical boxes, is mapped to the direct sum of all antisymmetric Wilson loops with less than or equal to $n$ horizontal boxes, weighted with $(-1)^iA_{i}$---for instance, for $n=4$:
\be
{\tiny\yng(1,1,1,1)}^D = A_4 - A_3\cdot {\tiny\yng(1)} + A_2 \cdot{\tiny\yng(2)} - A_1\cdot {\tiny\yng(3)} + {\tiny\yng(4)}\ .
\ee
After solving the $N_f$ equations \eqref{wilson loop duality} for the $N_f-N_c$ variables $\langle s^D_i \rangle$, we are left with the relations among the original variables $\langle s_i \rangle$ only, which provides us with an explicit expression for the chiral ring relations amongst Wilson loops in $\CT$. For example, when $N_f=5, N_c = 3$, one can easily show that:
\be
{\tiny\yng(3)} = A_3 -A_2 \cdot {\tiny\yng(1)} +A_1\cdot {\tiny\yng(2)}\ ,
\ee
using the last relation in \eqref{relation 1}. For general $N_f$, $N_c$, the Wilson loops algebra is generated by the Young tableaux which fit in a box of size $N_c\times (N_f-N_c)$; there are always as many independent generators as the number of Bethe vacua. 

In the limit where all the real mass parameters of the matter multiplets are sent to infinity ($y_i,\t y_i\rightarrow 0$), we have $A_i=0\ ,\forall i>0$. Physically, this is the limit in which we integrate out all the chiral multiplets with $m_A \rightarrow \infty$ and Aharony duality reduces to level/rank duality. We indeed recover, in this limit, the known map of Wilson loops under level/rank duality, which simply transposes the Young tableaux.
Conversely, the Wilson line algebra of 3D $\CN=2$ Chern-Simons-matter theories provides a natural generalization of the Verlinde algebra in pure CS theory.

\section{Geometry-changing line operators and Seifert manifolds}\label{subsec: seifert operators}
We are now ready to discuss the supersymmetric partition function on an {\it arbitrary} Seifert manifold $\CM_3$. This section is entirely based on recent work~\cite{Closset:2017zgf, Closset:2018ghr} by Willett and the present authors. 

In the previous section, we studied the 3D $\CN=2$ theory on $\Sigma_g \times S^1$, with the A-twist along the closed Riemann surface $\Sigma_g$.  The observables of the 
 the 3D $A$-model are correlation functions on half-BPS lines, $\SL_i$, wrapped over the $S^1$. They are computed as a sum over Bethe vacua \eqref{bethe vac line exp 0}, in the infrared Coulomb branch picture, namely:
\be\label{correlation sec8}
\left\langle\SL_i\SL_j\SL_k\cdots\right\rangle_{\Sigma_g \times S^1}=  \sum_{\h x \in \CS_{\rm BE}}  \SL_i(\h x)\,  \SL_j(\h x)\,  \SL_k(\h x)\cdots\; \CH(\h x)^{g-1}~.
\ee
The lines are quasi-local operators from the 2D point of view.  On the Coulomb branch, the Wilson lines $\SL= W_\FR$ correspond to (Laurent) polynomials in $x=e^{2\pi i u}$, as in \eqref{W as character}, while the the flux operators, $\Pi_\alpha$, or the handle-gluing operator, $\CH$, are given by rational functions of $x$ (as well as of the flavor fugacities, $y_\alpha$).  The topological invariance along the $\Sigma_g$ directions implies that the correlation function \eqref{correlation sec8} does not depend on the insertion points of the lines.

\subsection{Defect line operators and Seifert geometry}
\begin{figure}[t]
\centerline{
\includegraphics[width=3cm]{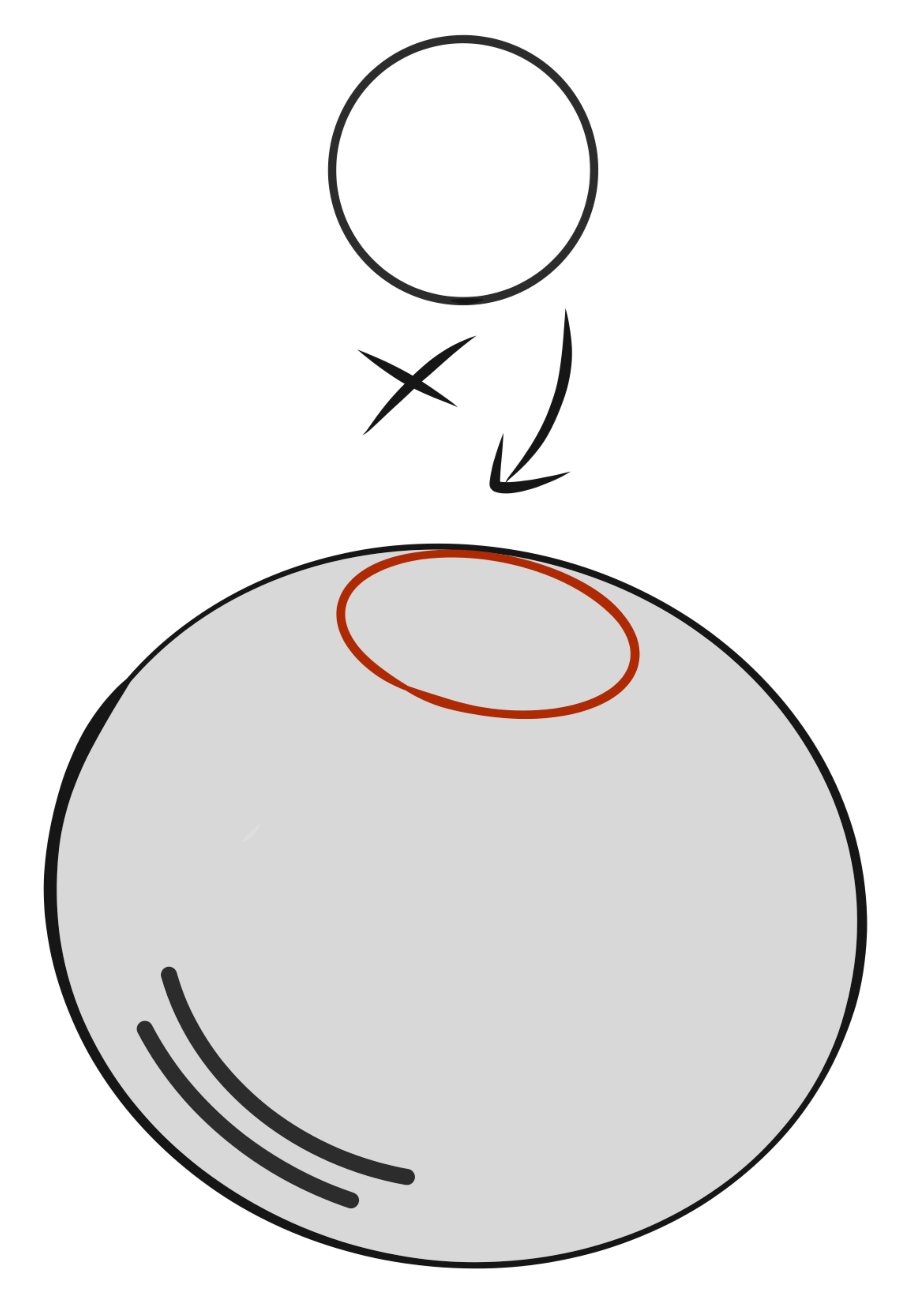}\qquad \qquad\qquad
\includegraphics[width=3cm]{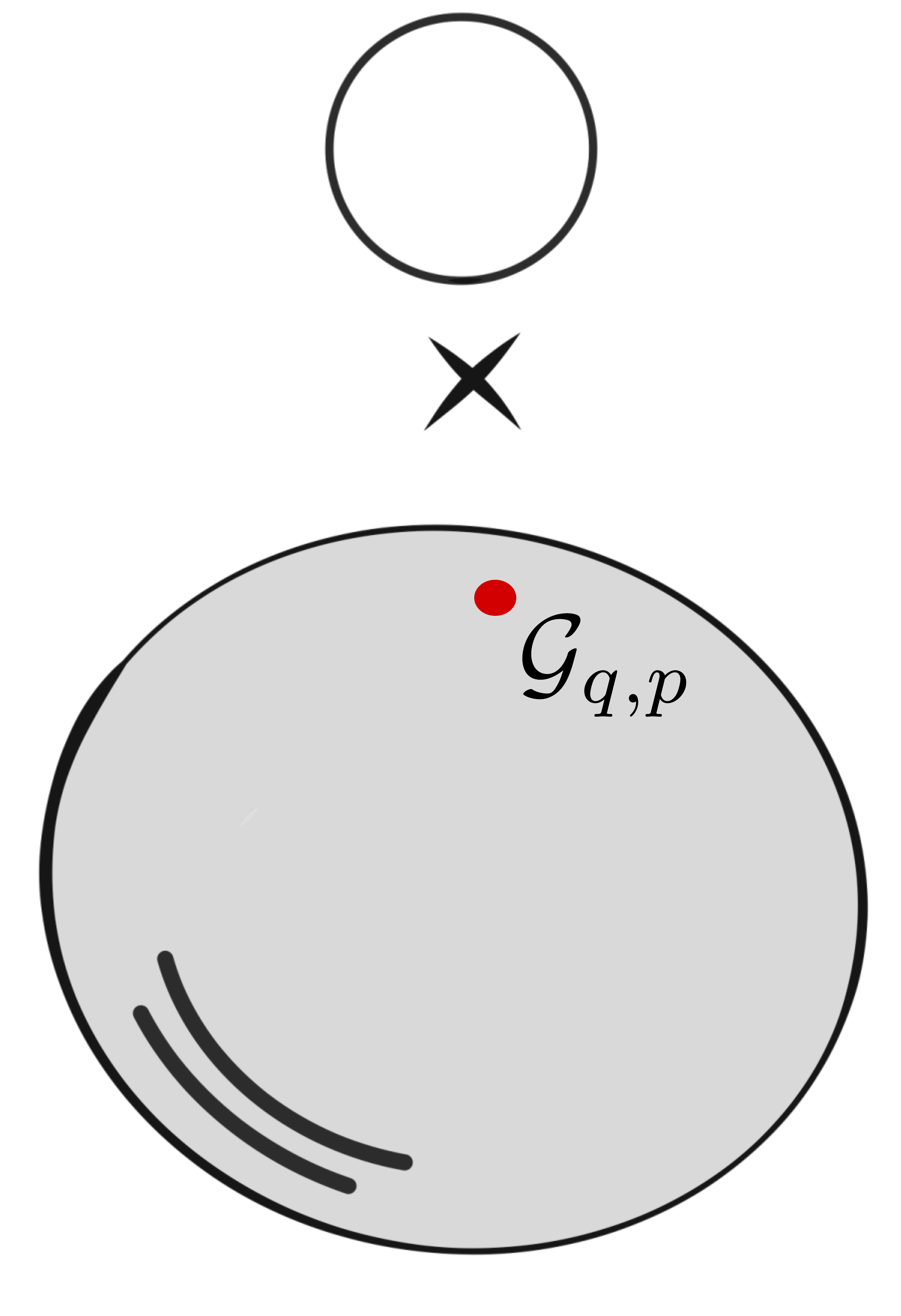}
}
\caption{{\it On the left:} Dehn surgery, cutting out a disk $D_2$ on the base, and gluing back the cap $D_2 \times S^1$ with an $SL(2, \Z)$ twist. This introduce an exceptional Seifert fiber at the center of the disk. 
{\it On the right:}  Shrinking the radius of $D_2$ to zero size, the introduction of the exceptional fiber is viewed a local defect operator, $\CG_{q,p}$, on the base (and wrapping the $S^1$). 
\label{fig:seifertGqp}}
\end{figure}

Consider now a general half-BPS Seifert geometry:
\be\label{CM3 gen 0}
\CM_3 \cong \big[\bd~; \, g~; \, (q_1, p_1)~, \cdots~, (q_n, p_n)\big]~.
\ee
As reviewed in section~\ref{subsec: 3DAtwist}, this can be obtained by simple topological surgery on the ``mother manifold" $S^2\times S^1$. First of all, as we saw above, one can increase the genus of the base using the handle-gluing operator, $\CH$. One can also perform Dehn surgery on $\Sigma_g \times S^1$ to obtain non-trivial $S^1$ fibrations. The simplest non-trivial Seifert fibration is that of a circle principal bundle of degree ${\bf d}$, which is denoted by $\CM_3 = \CM_{g, \bd}$:
\be\label{Mgd pb}
S^1 \stackrel{\bd}\longrightarrow \CM_{g, \bd} \longrightarrow \Sigma_g~.
\ee
In addition, we can add exceptional Seifert fibers of types $(q_i,p_i)$. Each $(q,p)$ fiber is obtained by Dehn surgery on a tubular neighborood of a generic $S^1$ fiber with the $SL(2, \Z)$ action:
\be
\mat{q & -t \\  p & s}~, \qquad q s + p t=1~,
\ee
as in \eqref{dehn surgery}. This introduces a $\Z_{q}$ orbifold point in the base and changes the Seifert fibration, as depicted schematically in Figure~\ref{fig:seifertGqp}.

The special case $(q,p)=(1,p)$ leaves the base invariant, corresponding to shifting the degree $\bd$ to $\bd+p$. From the point of view of the 3D $\CN=2$ theory on a Seifert manifold, the Seifert surgery operations can be   understood as the insertion of defect lines wrapped along Seifert fibers in the 3D $A$-model. From this point of view, we can compute any Seifert partition function using the general formula \eqref{correlation sec8}, once we identify the correct defect operators. For instance, we can insert all these {\it geometry-changing line defect operators} along $S^1$ on the topologically twisted $S^2$:
\be
Z_{\CM_3} = \langle {\bf 1} \rangle_{\CM_3} = \left\langle \CF^\bd \CH^g \prod_{i=1}^n \CG_{(q_i,p_i)}\right\rangle_{S_A^2\times S^1}\ .
\ee
where we have:
\begin{itemize}
\item The ``ordinary fibering operator,'' $\CF$, which shifts the degree $\bd$ by one unit.
\item The handle-gluing operator, $\CH$, which we discussed in the previous section.
\item The $(q,p)$-fibering operator, $\CG_{(q,p)}$, whose insertion at a point of $S^2$ is equivalent to adding a $(q,p)$ fiber above that point, thus introducing a $\Z_q$ orbifold point in the base.
\end{itemize}
 Since the correlation functions of line defects on $S^2\times S^1$ can be written as a sum over the Bethe vacua \eqref{bethe vacua 3D formula}, the partition function on the Seifert manifold \eqref{CM3 gen 0} takes the form:
\be\label{bethe form with G schem}
Z_{\CM_3}= \sum_{\hat u \in \CS_{\text{BE}}}\CF(\h u)^\bd \CH(\h u)^g \prod_{i=1}^N \CG_{(q_i,p_i)}(\h u)\ ,
\ee
schematically.
In the rest of this section, we make this formula completely explicit for 3D $\CN=2$ gauge theories. (For the derivation, we refer to the original reference.~\cite{Closset:2018ghr})

There are two assumptions about the 3D theory that we will make here, which are necessary for the Bethe formula \eqref{bethe form with G schem} to hold. First, we are assuming that the gauge group $\GG$ of the 3D $\CN=2$ theory is a product of {\it simply-connected} simple factors $\GG_\gamma$ and possibly of $U(N)$ factors. The generalization of these computations to completely generic gauge groups (such as, for instance, $SO(N)$ or $PSU(N)$ as opposed to $Spin(N)$ or $SU(N)$, respectively) is somewhat non-trivial, and has not yet been fully worked out in the literature. Secondly, the theory needs to be such that the Bethe vacua are discrete. This is the case for theories ``with enough flavor symmetries'' and at generic values of the flavor fugacities.  If there is a continuum of solutions to the Bethe equations (and therefore a ``quantum Coulomb branch''), these Bethe-equation methods are not directly applicable.

\subsection{The general Bethe-vacua formula for $Z_{\CM_3}$}
Consider the 3D theory  with gauge group $\GG$ and Coulomb branch parameters $u_a$, and with flavor symmetry $\GG_F$. Denote by  $\prod_\alpha U(1)_\alpha$ the maximal torus of $\GG_F$, and by $\nu_\alpha$ the  flavor parameters. The chiral multiplets are in the representation $\FR \times \FR_F$ of $\GG \times \GG_F$, with weights $(\rho, \omega)$ and $R$-charges $r_\omega$.

On the Seifert manifold $\CM_3 \cong [\bd~; \, g~; \, (q_i, p_i)\big]$, with $i=1, \cdots, n$, we can turn on background fluxes $(\m_0, \m_i)$, corresponding to ``ordinary'' and ``fractional fluxes'' on the base $\h\Sigma_{g, n}$, representing $U(1)_\alpha$ background fluxes as elements of $\Pic(\h\Sigma)$ as discussed in section~\ref{subsec:orbifold L}. We also choose a $U(1)_R$ line bundle:
\be\label{LR nuR etc}
{\bf L}_R \qquad \longleftrightarrow \qquad \nu_R\ ,~ \n_0^R\ ,~\n_i^R\ ,
\ee
 as discussed in section~\ref{subsec:LR details}. We also describe ${\bf L}_R$ in terms of the integers $2\nu_R$ and $l_i^R$ as in \eqref{param LR ii}-\eqref{sum lR zero}.

\paragraph{Seifert fibering operator.} Let us define the Seifert fibering operator:
\be\label{def GM3}
 \CG_{\CM_3}(u, \nu; \nu_R)_\m \equiv \Pi(u, \nu; \nu_R)^{\m_0} \, \CF(u, \nu; \nu_R)^{\bd}\,   \prod_{i=1}^n \CG_{(q_i, p_i)}(u, \nu; \nu_R)_{\m_i}~,
\ee
in the presence of background flux $\m$. Here, $\CG_{(q, p)}(u, \nu)_{\m}$ denotes the $(q,p)$ fibering operator which also inserts a fractional flux $\m \in \Z_q$ at the orbifold point. (The generalization to a more general flux lattice is straightforward, and will appear below.)

\paragraph{The supersymmetric Seifert partition function.} 
We then have the following explicit Bethe-vacua formula for the Seifert partition function:
\be\boxed{
Z_{\CM_3}(\nu; \nu_R)_\m =  \sum_{\h u \in \CS_{\rm BE}}   \CH(\h u, \nu; \nu_R)^{g-1}\, \CG_{\CM_3}(\h u, \nu; \nu_R)_{\m}~.}
\ee
as a function of the flavor parameters $(\nu, \m)$. We already discussed the handle-gluing operator $\CH$ and the ``ordinary'' flux operator $\Pi$ in section~\ref{subsec: 3D Amod}---namely, we have:
\bea
&\CH(u, \nu; \nu_R) = e^{2\pi i \Omega(u,\nu; \nu_R)} \det_{ab}\frac{\partial^2\CW(u, \nu; \nu_R)}{\partial u_a \partial u_b}\ .
\\
&\Pi_\alpha (u,\nu; \nu_R)=\exp\left(2\pi i \frac{\partial \CW(u, \nu; \nu_R)}{\partial \nu_\alpha}\right)~, 
\eea
in terms of the effective twisted superpotential  and of the effective dilaton of the 3D theory.
 Let us now present the formulas for the fibering operators.  In these formulas, we have $\nu_R$ in the argument to remind us that these objects depend on the choice of $U(1)_R$ line bundle. (This will be omitted in various formulas below, to avoid clutter.) The half-integer parameter $\nu_R$ that appeared in our discussion of $\CW$ and $\Omega$ in section~\ref{subsec: 3D Amod} coincides with the parameter $\nu_R$ in \eqref{LR nuR etc} describing ${\bf L}_R$ on $\CM_3$.

\subsection{The ``ordinary" fibering operator and $Z_{\CM_{g, \bd}}$}
Consider the principal circle bundle $\CM_{g, \bd}$, as in \eqref{Mgd pb}. From the 2D point of view, the introduction of the non-trivial fibration corresponds to turning on some non-zero flux:
\be
\frac{1}{2\pi} \int_{\Sigma_g} d\CC = \bd\ ,
\ee
for the 2D ``graviphoton'' $\CC_\mu$ on $\Sigma_g$. The graviphoton is the gauge field for a 2D abelian symmetry $U(1)_{\rm KK}$, which corresponds to the momentum along $S^1$, and becomes a distinguished symmetry in the low-energy 2D description.

Using the topological invariance, we can concentrate the $U(1)_{KK}$ flux at a point on the base.
The ``ordinary'' fibering operator, $\CF$, can then be understood as a 2D flux operator for $U(1)_{KK}$. Using the formula \eqref{flux op}  and the fact that the $U(1)_{KK}$ twisted mass is  $m_{\text{KK}} = \frac{1}{\beta}$, with $\beta$ the radius of $S^1$, it is easy to derive the explicit formula:~\cite{Closset:2017zgf}
\be\label{CF formula from W}
\CF(u, \nu) = \exp\left(2\pi i \left(\CW(u,\nu) - u_a \frac{\partial \CW(u,\nu)}{\partial u_a} - \nu_\alpha \frac{\partial \CW(u,\nu)}{\partial \nu_\alpha}\right)\right)\ ,
\ee
in terms of the effective twisted superpotential of the 3D theory. 
One can check that the branch cut ambiguity \eqref{ambiguity 3d} completely cancels out from this expression, so that the fibering operator is a single-valued function of the Coulomb branch variables $u_a$ and parameters $\nu_\alpha$. (It is in fact a meromorphic function, with poles coming from chiral multiplet contributions.)

\subsubsection{Large gauge transformations and flux operators} 
Under large gauge transformation of $u$ or $\nu$, the fibering operator transforms non-trivially. When shifting a single $u_a$ or $\nu_\alpha$ by  $1$, we get a power of the corresponding flux operator:
\be\label{lgt of F}
\CF(u_a+1, \nu) = \Pi_a(u,\nu)^{-1}\, \CF(u, \nu)~, \qquad 
\CF(u, \nu_\alpha +1) = \Pi_\alpha(u,\nu)^{-1}\, \CF(u, \nu)~.
\ee
This is precisely so that the $\CM_{g, \bd}$ partition is gauge invariant under large gauge transformations for flavor symmetry background vector multiplets.

This is because the $\CM_{g, \bd}$ geometry has a non-trivial second cohomology group:
\be
H^2(\CM_{g, \bd} , \Z) = \Z^{2g} \oplus \Z_{\bd}~,
\ee
and we can turn on non-trivial torsion flux, $\m \in \Z_\bd$. For a $U(1)$ symmetry with line modulus $u$, large gauge transformations of $u$ are accompanied by a shift of the torsion flux by $\bd$, according to:
\be\label{large gauge Mdg}
(\nu,\m) \rightarrow (\nu+1, \m + \bd)~.
\ee
Note that this is a special case of \eqref{lgt general shift}. The supersymmetric partition function on $\CM_{g, \bd}$ with background flavor flux reads:
\be\label{ZM3 Bethe formula}
Z_{\CM_3}(\nu; \nu_R)_\m =  \sum_{\h u \in \CS_{\rm BE}}   \CH(\h u, \nu; \nu_R)^{g-1}\, \CF(\h u, \nu; \nu_R)^\bd\, \prod_\alpha \Pi_\alpha(\h u, \nu)^{\m_\alpha}~.
\ee
Then, \eqref{lgt of F} precisely ensures gauge invariance. The handle-gluing operators and flux operator are already fully gauge invariant. For any large gauge transformation in the gauge sector, we have:
\be
\CF(\h u_a+1, \nu) =  \CF(\h u, \nu)~,
\ee
by the definition of the Bethe vacua, while for the flavor large gauge transformations the second relation in \eqref{lgt of F} ensures that \eqref{ZM3 Bethe formula} is invariant under \eqref{large gauge Mdg}.

\subsubsection{Explicit formulas for $\nu_R=0$}
When $\bd$ is even, there are two distinct spin structures we can choose on $\CM_{g, \bd}$ (for any $g$), corresponding to $\nu_R=0$ or $\nu_R= \half$ (mod $1$), and  the partition function can depend on that choice.~\cite{Toldo:2017qsh, Closset:2018ghr} The formula \eqref{CF formula from W} for the fibering operator is valid for any $\nu_R$, but we did not specify the constant term of $\CW$ (mod $1$) for $\nu_R \neq 0$, and therefore in that case we have only defined the fibering operator up to a phase. That phase can be fixed  by considering $\CF$ as a special case of the more general $(q,p)$ fibering operators that we will discuss momentarily. 

For now, however, let us consider $\nu_R=0$, for definiteness. Then, in any 3D $\CN=2$ gauge theory, the 
 CS action contributes to the fibering operator as:
\be
\CF_{\text{CS}}= \exp\Big(-\pi i \sum_{a,b}k_{ab}u_au_b - 2\pi i \sum_{a,\alpha}k_{a\alpha}u_a\nu_\alpha - \pi i \sum_{\alpha,\beta}k_{\alpha\beta}\nu_\alpha\nu_\beta + \frac{\pi i}{12}k_g\Big)\ ,
\ee
as directly follows from the definition. Note the constant-phase contribution from the gravitational CS term. 
Similarly, the fibering operator for a free chiral multiplet of charge $1$ under some $U(1)$ reads:
\be
\CF^{\Phi}(u) = \exp\left(\frac{1}{2\pi i }\text{Li}_2(e^{2\pi i u}) + u \log (1-e^{2\pi i u})\right)~,
\ee
which is the meromorphic function already defined in \eqref{CF S3 def}. Then, the full matter contribution to the fibering operator is:
\be
\CF^{\rm matter}(u, \nu) = \prod_{\rho, \omega}  \CF^\Phi( \rho(u) + \omega(\nu))~.
\ee
The vector multiplet does not contribute to $\CF$ (when $\nu_R=0$). 

\subsubsection{The round $S^3$ partition function}
A particularly important case is that of the round $S^3$ partition function. Since $\CM_{0, 1} \cong S^3$, 
we can write this partition function as the expectation value of a single fibering operator $\CF$ in the 3D $A$-model on $S^2\times S^1$:
\be\label{bethe s3}
Z_{S^3}(\nu) = \left\langle\CF\right\rangle_{S_A^2\times S^1}\\
 = \sum_{\hat u\in \CS_{\text{BE}}}\CH(\hat u,\nu)^{-1} \,  \CF(\hat u,\nu) \ .
\ee
The $R$-symmetry line bundle ${\bf L}_R$ on $S^3$ is topologically trivial. It is represented in $\Pic(S^2)$ by:
\be
L_R = L_0^{\t\n_0^R}~,\quad \qquad \t \n_0^R = \n_0^R-1~.
\ee
It is most convenient to choose $\t\n_0^R=0$, and therefore $\n_0^R=1$ and $\nu_R=1$. Consider for instance a free chiral multiplet. Its $S^3$ partition function is then written as:
\bea
&Z^\Phi_{S^3}(u) &=&\; \CH(u)^{-1} \Pi^\Phi(u+\nu_R r)^{\n_0 r} \CF^\Phi(u+ \nu_R r)\\
& &=&\;\Pi^\Phi(u+\nu_R r)^{-r+1}  \Pi^\Phi(u+\nu_R r)^{\n_0 r} \CF^\Phi(u+\nu_R r) \\
& &=&\;  \CF^\Phi(u+r-1)~,
\eea
where we used  $\nu_R=\n_0^R=1$ and $\CF(u-1)= \Pi(u) \CF(u)$ on the last line. 
The important point is that the partition function is well-defined for any $r\in \R$; this is true for the full gauge theory as well. The Bethe-formula can then be used to study SCFTs. For instance, it provides an efficient way to perform $F$-maximization.~\cite{Closset:2017zgf}

Note that, even though we are using the 3D $A$-model language, this $S^3$ partition function contains a lot of information beyond what one might have naively expected from the consideration of the A-twist. Indeed, the $S^3$ partition function is sometimes called an ``untwisted'' partition function; in some general sense, however, all 3D $\CN=2$ partition function are ``twisted.''~\cite{Closset:2014uda}

Using explicit manipulations involving the difference equations \eqref{lgt of F}, one can also derive the Bethe-equation formula \eqref{bethe s3} from the integral formula \eqref{ZS0 loc formula}.~\cite{Closset:2017zgf}

\subsection{Seifert fibering operator}
Let us now discuss the general formula for an arbitrary Seifert manifold. 
 It is convenient to rewrite the Seifert fibration \eqref{CM3 gen 0} as:
\be\label{CM3 gen 0bis}
\CM_3 \cong \big[0~; \, g~; \, (1, \bd)~, \, (q_1, p_1)~, \cdots~, (q_n, p_n)\big]~,
\ee
using unnormalized Seifert symbols, by introducing the point $x_0\in \h\Sigma$ at which the ``ordinary'' Seifert fibering operator $\CF^\bd$ sits, with:
\be
(q_0, p_0) = (1, \bd)~.
\ee
Then, we have:
\be
\CG_{(1, \bd)}(u, \nu)_{\m_0}  = \Pi(u, \nu)^{\m_0} \, \CF(u, \nu)^{\bd}
\ee
and the Seifert operator \eqref{def GM3} takes the form: 
 \be\label{def GM3 bis}
 \CG_{\CM_3}(u, \nu)_\m \equiv    \prod_{i=0}^n \CG_{(q_i, p_i)}(u, \nu)_{\m_i}~.
\ee
This is a convenient language when considering the general case. In the following, we fix $\CM_3$ and the $R$-symmetry line bundle over it. The latter is fixed in terms of $\nu_R$ and of the integers and $l_i^R$ such that:
\be\label{sum lR zero 2}
\sum_{i=0}^n l_i^R=0~,
\ee
as explained in section~\ref{subsec:LR details}. Now, we just need to discuss each $(q,p)$ fibering operator individually. We should just note that, due to subtle signs when $l_i^R \neq 0$, each $\CG_{(q,p)}$ is not completely well-defined, but,  whenever \eqref{sum lR zero 2} holds, the full Seifert fibering operator \eqref{def GM3 bis} itself is fully well-defined.~\cite{Closset:2018ghr}

\subsubsection{General properties of $\CG_{(q,p)}$}\label{subsec: prop Gqp}
The $(q,p)$ fibering operator of a 3D $\CN=2$ gauge theory gets contributions from classical CS terms, vector multiplet and chiral multiplets, as usual in such formulas. Each building block can be computed at fixed value of the fractional fluxes $\m_\alpha$ for the flavor symmetries, but also at fixed value of the fractional gauge fluxes, $\n_a$. 
The actual $(q,p)$-fibering operator entering in \eqref{def GM3 bis}  is given as {\it a sum over all the fractional gauge fluxes:}
\be\label{G as sum over frac flux}
\CG_{(q,p)}(u,\nu)_{\m} = \sum_{\n \in \Gamma_{\GG^\vee}(q)}\CG_{(q,p)}(u,\nu)_{\n,\m}\ ,
\ee
where $\CG_{(q,p)}(u, \nu)_{\n,\n_F}$ is the contribution from each fractional flux sector $(\n,\m)$. Here, the gauge flux lattice is a $\Z_q$ reduction of the GNO flux lattice:
\be\label{cochar mod q def}
\Gamma_{\GG^\vee}(q)=\big\{\n \in {\mathfrak h}\big|~ \rho(\n)\in \mathbb{Z},\; \; \forall\rho \in\Lambda_{\rm char}~;\; \n \sim \n+ q \lambda,\;\; \forall \lambda\in \Lambda_{\rm cochar}\big\}~.
\ee

\paragraph{Gauge invariance.} The fibering operator $\CG_{(q,p)}(u, \nu)_{\n,\n_F}$ must be invariant under large gauge transformations (both for gauge and flavor parameters). That is, we have:
\be\label{large gauge qp fibering}
\CG_{(q,p)}(u+1)_{\n+p} = \CG_{(q,p)}(u)_{\n}~, 
\ee
for a $U(1)$ large gauge transformation, and similarly for a more general abelian group.

\paragraph{Ordinary flux.} Consider again the $U(1)$ case. The fractional flux $\n$ mod $q$ is valued in $\Z_q$, but $q$ units of fractional fluxes are equivalent to one unit of ``ordinary'' flux, and therefore we have:
\be
\CG_{(q,p)}(u)_{\n+q} = \Pi(u) \,  \CG_{(q,p)}(u)_{\n} 
\ee
with $\Pi(u)$ the ordinary flux operator. This realizes the Picard group relations \eqref{pic orbifold}. Note that the sum over fractional fluxes \eqref{G as sum over frac flux} is well-defined on the Bethe vacua (it is independent of the way we represent $\Z_q$) precisely because the gauge flux operators are trivial on Bethe vacua,  $\Pi_a(\h u)=1$.

\paragraph{Seifert equivalence.} When working in terms of unnormalized Seifert invariants, shifting $(q, p)$ to $(q, p+q)$ is equivalent to shifting the ``degree'' $\bd$ to $\bd +1$. Therefore, we should have:
\be
\CG_{(q,p+q)}(u)_{\n}= \CF(u) \,  \CG_{(q,p)}(u)_{\n}~.
\ee

\subsubsection{The $(q,p)$-fibering operator}
Let us now give the full $(q,p)$-fibering operator as:
\be
\CG_{(q,p)}(u,\nu)_{\m} =\sum_{\n\in\Gamma_{\GG^\vee}(q)} \CG_{(q,p)}^{\text{CS}}(u,\nu)_{\n,\m}~\CG_{(q,p)}^{\text{vector}}(u)_{\n}~\CG_{(q,p)}^{\text{matter}}(u,\nu)_{\n,\m}\ .
\ee
In the following, we discuss each contribution in turn.


\paragraph{CS contributions.}
Let us first discuss the contribution from the classical CS actions. As in various examples above, we can focus on the CS terms involving the gauge group, given in terms of parameters $u$. The flavor CS terms are obtained similarly, substituting $\nu$ as appropriate. We have:
\bea
& \CG_{(q,p)}^{\text{CS}}(u)_{\n}&=&\; \left( \CG_{(q,p)}^{\rm GG}(u)_{\n}\right)^{k_{GG}}\, \left( \CG_{(q,p)}^{\rm G_1G_2}(u_I, u_J)_{\n_I, \n_I}\right)^{k_{G_IG_J}} \, \left( \CG_{(q,p)}^{\rm GR}(u)_{\n}\right)^{k_{GR}}\\
&&&\; \times \left( \CG_{(q,p)}^{\rm RR}(u)_{\n}\right)^{k_{RR}} \, \left(\CG_{(q,p)}^{\text{grav}}\right)^{k_g}~,
 \eea
schematically. For a $U(1)$ CS term, we have:
\be\label{CS GG}
\CG^{\rm GG}_{q,p}(u)_\n = (-1)^{\n (1+ t+ l^R t +2 \nu_R s)}\, \exp\left[{- {\pi i \ov q}\left(p u^2- 2 \n u + t \n^2\right)}\right]~.
\ee
Here, $\n\in \Z$ denotes the fractional flux and $l^R\in \Z$ corresponds to the parameterization \eqref{param LR ii} of the $R$-symmetry line bundle, with:
\be
\n^R=  {q-1\ov 2}+ {l^R q\ov 2}+  \nu_R p \in \Z~.
\ee
The integers  $s$ and  $t$ are such that $q s+ p t=1$, and the expression \eqref{CS GG} actually does not depend on the choice of $s, t$, for a fixed $(q,p)$. The overall sign in \eqref{CS GG} only depends on the choice of spin structure on $\CM_3$. 
The generalization is obvious. For instance, the contributions of CS term for a simply-connected simple gauge group $\GG_\gamma$ with level $k_\gamma=1$ reads:
\be
\CG_{(q,p)}^{\GG_\gamma}(u)_\n = \exp\left[-\frac{\pi i h^{ab}}{q}(pu_au_b -\n_au_b -\n_bu_a + t\n_a\n_b)\right]\ ,
\ee
with $h^{ab}$ the Killing metric.  This is independent of the spin structure. 
The mixed abelian CS term gives:
\be
\CG_{(q,p)}^{\rm G_1 G_2}(u_I,u_J)_{\n_I,\n_J} = \exp\left[-\frac{2\pi i}{q}(pu_Iu_J-\n_I u_J -\n_J u_I + t \n_I\n_J)\right]\ .
\ee
The mixed $U(1)$-$U(1)_R$ CS term can be similarly written as:
\be\label{CS GR}
\CG^{\rm GR}_{q,p}(u)_\n = \exp\left[{-{2 \pi i \ov q}\left(p u\nu_R -\n \nu_R- \n^R  u + t \n^R  \n \right)}\right]~,
\ee
while the $U(1)_R$ CS term reads:
\be\label{GCSRR}
\CG_{(q,p)}^{\text{RR}} = (-1)^{\n^R(1+t+l^R t + 2\nu_R s)}\exp\left[-\frac{\pi i}{q} (p\nu_R^2 -2\n^R\nu_R + t(\n^R)^2)\right]\ .
\ee
Finally, the gravitational CS term with level $k_g\in\mathbb{Z}$ contributes:~\footnote{Because of the square root, there is a sign ambiguity here, which drops out if we only consider $k_g$ even. Fortunately, $k_g$ is naturally even in supersymmetric theories---for instance, integrating out a chiral multiplet shifts $k_g$ by $2$.}
\be\label{grav cs fibering}
\CG_{(q,p)}^{\text{grav}} = \CG_{(q,p)}^{(0)} {\CG_{(q,p)}^{\text{RR}}}^{-\frac12}\ ,
\ee
with $\CG_{(q,p)}^{\text{RR}}$ given by \eqref{GCSRR}, and $\CG^{(0)}_{(q,p)}$ defined as:
\be
\CG^{(0)}_{(q,p)} = \exp\left[\pi i\left(\frac{p}{12q}-s(p,q)\right)\right]\ .
\ee
Here,  $s(p,q)$ is the Dedekind sum, which can be written as:
\be
s(p,q) = \frac{1}{4q}\sum_{l=1}^{q-1}\cot\left(\frac{\pi l}{q}\right)\cot\left(\frac{\pi l p}{q}\right)\ .
\ee
for $p, q$ mutually prime. Note that the $k_{RR}$ and $k_g$ CS terms contribute rather complicated pure phases, which depend on the topology of $\CM_3$ (and of ${\bf L}_R$).

\paragraph{Vector multiplet contribution.}
At fixed fractional gauge flux $\n$, the vector multiplet contributes to the $(q, p)$-fibering operator as: 
\be\label{CGqp vec}
\CG_{(q,p)}^{\text{vector}}(u)_{\n}= \left(\frac{1}{\sqrt q}\right)^{\rk}\, \left(\CG_{(q,p)}^{(0)}\right)^{\text{dim}(\GG)}\, \CG_{(q,p)}^W(u)_\n~.
\ee
Here, the last factor is a contribution from the W-bosons (in the symmetric quantization), which reads:
\be
\CG_{(q,p)}^W(u)_\n = (-1)^{2\rho_W(\n)(t+l^Rt + 2\nu_R s)} \frac{e^{-\frac{2\pi i}{q}\rho_W(u-t\n)}}{e^{-2\pi i \rho_W(u)}}\prod_{\alpha\in\Delta^+}\frac{1-e^{\frac{2\pi i}{q}\alpha(u-t\n)}}{1-e^{2\pi i \alpha(u)}}\ ,
\ee
with $\rho_W = \frac12\sum_{\alpha\in\Delta^+}\alpha$ the Weyl vector. The phase factor \eqref{CGqp vec} are understood as the contribution from the gauginos (both in the Cartan and off-diagonal) given our choice of quantization, with the non-zero UV contact terms~\eqref{kappaRR App vec}. Indeed, it is clear from \eqref{grav cs fibering} that the phase  $\CG_{(q,p)}^{(0)}$ is consistent with a contribution $\kappa_{RR}= \half$ and $\kappa_g=1$ to the effective action. Finally, a  normalization  factor  $1/\sqrt{q}$ is needed in  \eqref{CGqp vec} for each Cartan generator of the gauge group.~\cite{Closset:2018ghr}

\paragraph{Matter contribution.}
 Consider first a chiral multiplet $\Phi$ of $U(1)$ charge $Q=1$ and $R$-charge $r$. The corresponding $(q,p)$ fibering operator reads:
 \be
\CG_{(q,p)}^{\Phi} (u+\nu_R r)_{\n+\n^R r}\ , 
\ee
 in terms of the function:
 \be\label{fibering qp decomp}
\CG_{(q,p)}^\Phi(u)_\n = \Pi_{(q,p)}^\Phi(u)_\n~ \t\CG_{(q,p)}^{\Phi}(u)\ .
\ee
Here, $\t\CG_{(q,p)}^{\Phi}(u)$ is the fibering operator in the absence of the fractional flux, given by:
\be
\t\CG_{(q,p)}^{\Phi}(u)= \exp \sum_{l=0}^{q-1} \left[ \frac{p}{2\pi i}\text{Li}_2\left(e^{2\pi i \frac{u+tl}{q}}\right) + \frac{pu+l}{q} \log \left(1-e^{2\pi i\frac{u+tl}{q}}\right)\right]\ ,
\ee
with $t\in \Z$ is such that $q s+ pt=1$.  The fractional flux contribution is given by:
\be
\Pi_{(q,p)}^\Phi(u)_\n = \left(e^{2\pi i\frac{u}{q}};e^{2\pi i \frac{t}{q}}\right)_{-\n}\ ,
\ee
where $(x;q)_n$ is the $q$-Pochhammer symbol as defined in \eqref{qPochdef}.  The function \eqref{fibering qp decomp} is again meromorphic in $u$.
The full contribution from the matter fields is then simply given by:
\be
~\CG_{(q,p)}^{\text{matter}}(u,\nu)_{\n,\m}= \prod_{\rho,\omega} \CG_{(q,p)}^{\Phi} (\rho(u) + \omega(\nu)+\nu_R r_\omega)_{\rho(\n)+ \omega(\m)+\n^R r_\omega}\ .
\ee

\medskip
\noindent This completes our description of the Seifert fibering operator \eqref{def GM3 bis} for 3D $\CN=2$ gauge theories. One can easily check that all the contributions to the $(q,p)$ fibering operators, and in particular the function \eqref{fibering qp decomp}, satisfy the general properties discussed in subsection~\ref{subsec: prop Gqp}. These expression also pass numerous other consistency checks.~\cite{Closset:2018ghr}

\subsection{Examples: $S^3_b$ and $S^2_\epsilon\times S^1$ at rational values of the parameters}

Let us briefly revisit the earlier examples of partition functions on Seifert manifolds in the language of the geometry-changing operators.~\cite{Closset:2018ghr}

\paragraph{The $S^3_b$ partition function.}
As already mentioned in section \ref{subsec: 3DAtwist}, the geometry of $S^3_b$ with {\it rational} squashing parameter, $b^2\in\mathbb{Q}$, can be realized as a Seifert fibration with two orbifold points on the genus-zero base:
\be
S_b^3 \cong [0~;0~;~(q_1,p_1)\ ,(q_2,p_2)]\ ,~~~\quad q_1p_2+q_2p_1=1\ ,\quad ~~~b^2=\frac{q_1}{q_2}\ .
\ee
(We can choose both $q_1$ and $q_2$ to be positive.) This implies that the $S^3_b$ partition function with rational $b^2$ can be written as:
\be\label{bethe s3b}
Z_{S_b^3} = \sum_{\hat u\in\CS_{\text{BE}}} \CH(\hat u)^{-1} \CG_{(1,0)}(\hat u)_r~\CG_{(q_1,p_1)}(\hat u)~\CG_{(q_2,p_2)}(\hat u)\ ,
\ee
where
\be
\CG_{(q_i,p_i)} = \sum_{\n_i\in \Gamma_{\GG^\vee}(q_i)} \CG_{(q_i,p_i)}(\hat u)_{\n_i}\ ,~~~~i=1,2
\ee
are the contributions from the two exceptional fiber. The contribution from the effective R-symmetry flux $\n_0^R=1$ (so that $\t\n_R=0$) is encoded in the expression:
\be
\CG_{(1,0)}(\hat u)_r = \Pi(u)^r\ ,
\ee
for the chiral multiplets, schematically. (In the above expression, we are leaving the dependence on the flavor parameters $\nu$ implicit.) 
It is possible to show that the expression \eqref{bethe s3b} is equivalent to the contour integral expression discussed in section~\ref{sec: lens space}. Note that, using the change of variables $u = i \sqrt{q_1 q_2}\,\h\sigma$, the contour integral presentation \eqref{ZSb loc formula} can be written as:
\be
Z_{S^3_b} = \frac{(-2\pi i)^{\rk}}{|W_\GG|}\int\prod_a \frac{du_a}{2\pi i}~ e^{-2\pi i \Omega(u)}~ \t\CG_{S_b^3}(u)\ .
\ee
Here, $\t\CG_{S_b^3}(u)$ takes a form:
\be
\t\CG_{S_b^3}(u) = \CG_{(1,0)}(u)_r \t \CG_{(q_1,p_1)}(u) \t\CG_{q_2,p_2}(u)\ ,
\ee
where $\t\CG_{(q_i,p_i)}$ are the $(q_i,p_i)$ fibering operator evaluated at zero fractional flux---for instance, as in \eqref{fibering qp decomp} for the chiral multiplet. Then, using some non-trivial manipulations of the integration contour,~\cite{Closset:2018ghr} one can establish the equivalence between the expression \eqref{bethe s3b} and the $\h\sigma$-contour integral formula  \eqref{ZSb loc formula}. This also means that, when $b^2\in \Q$, the Bethe-vacua formula gives an explicit {\it evaluation formula} for the complicated squashed-$S^3$ matrix model.

\paragraph{The refined twisted index.}
Similarly, the $S^2_\epsilon\times S^1$ geometry, whose partition function computes the refined twisted index discussed in section~\ref{sec: S2 top index}, can be realised as a Seifert fibration with two orbifold points on a genus-zero base:
\be
S^2_\epsilon \times S^1 \cong [0~;~0~;~(q,p)~,(q,-p)]\ ,~~~~\quad qs +pt=1\ ,
\ee
when $\epsilon\in \mathbb{Q}$. Here the the parameter $\epsilon$ should be identified with:
\be
\epsilon = \frac{t}{q}\ ,~~~~\qquad \text{gcd}(q,t)=1\ .
\ee
This implies that the partition function \eqref{integral formula S2S1eps} can be written as:
\be\label{refined twisted index bethe}
Z_{S^2_\epsilon \times S^1} = \sum_{\hat u\in \CS_{\text{BE}}} \CH(\hat u)^{-1}~ \CG_{(1,0)}(\hat u)_{\m_0}~ \CG_{(q,p)}(\hat u)_{\m_1} ~\CG_{(q,-p)}(\hat u)_{\m_2} \ ,
\ee
where:
\bea
&\CG_{(q,p)}(\hat u)_{\m_1} = \sum_{\n_1\in\Gamma_{\GG^\vee}(q)} \CG_{(q,p)}(\hat u)_{\n_1,\m_1}\ ,\\
&\CG_{(q,-p)}(\hat u)_{\m_2} = \sum_{\n_2\in \Gamma_{\GG^\vee}(q)} \CG_{(q,-p)}(\hat u)_{\n_2,\m_2}~,
\eea
are the contribution from two exceptional fibers. The factor $\CG_{(1,0)}(\hat u)_{\n_0^F}$ encodes the contribution from the ``ordinary'' flavor symmetry flux, $\m_0$.
One can then show that the expression \eqref{refined twisted index bethe} is equivalent to the residue integral formula for the refined twisted index, after performing some non-trivial change of variables.~\cite{Closset:2018ghr}


\section{The 3D $\CN=2$ superconformal index}\label{sec: S2S1 index}
As we discussed in section~\ref{subsec: halfBPS geom}, almost all half-BPS backgrounds on closed manifolds $\CM_3$ correspond to Seifert structures on $\CM_3$, and continuous deformations thereof.  The only (and very important!) exception\footnote{Barring the existence of ``exotic'' backgrounds on $T^3$, which have not been investigated.} is the THF ``of class (v)'' on $S^2 \times S^1$, in the notation of section~\ref{subsubsec: THF class}.

In this section, we briefly discuss this $S^2\times S^1$ background and the corresponding supersymmetric partition function, to emphasize those aspects that differ from the Seifert geometries above.

\subsection{The $S^2\times S^1$ background}
Let us choose the round metric on $S^2 \times S^1$:
\be\label{S2S1 met round}
ds^2 = \beta^2 dx^2 + d\theta^2 + \sin^2 \theta d\varphi^2~,
\ee
with $x\in [0, 2 \pi)$ the $S^1$ coordinate, and $\theta \in [0, \pi]$,  $\varphi\in [0, 2\pi)$, the angular coordinates on $S^2$. We also introduced the $S^1$ radius, $\beta >0$. One can define a THF on $S^2 \times S^1$ by:
\be
\eta=\beta \cos\theta dx + \sin\theta d\theta~.
\ee
One can check that this defines a THF---in particular, we have $\eta_\mu \eta^\mu=1$. The adapted coordinates are \cite{Closset:2013vra}:
\be
\tau =\beta x - 2 \log \cos{\theta\ov 2}~, \qquad z = e^{-\beta x+ i \varphi} \sin\theta~,
\ee
on the northern patch $\theta \neq \pi$. In these coordinates, we have:
\be\label{eta S2S1}
\eta= d\tau + e^{2\tau} {\bz dz+ z d\bz \ov 4 + e^{2\tau} |z|^2}~, 
\ee
and the metric \eqref{S2S1 met round} reads:
\be
ds^2 = \eta^2+ e^1 e^{\b 1}~, \qquad e^1= {4 e^{\tau} dz\ov 4 + e^{2\tau} |z|^2}~, \qquad
e^{\b1}= {4 e^{\tau} d\b z\ov 4 + e^{2\tau} |z|^2}~,
\ee
where we introduced an adapted frame. In terms of the parameter:
\be
{\bf q}= e^{-2\pi \beta}~,
\ee
this THF on $S^2 \times S^1$ is a special case of \eqref{THF s2s1}, with ${\bf q}$ real. The other supergravity background fields read:
\be\label{sugra fields s2s1}
H=0~, \qquad V_\mu = - i \beta dx~, \qquad A_\mu^{(R)}={i\beta \ov 2} dx~,
\ee
as follows from plugging \eqref{eta S2S1} into \eqref{gen sol KSE}. Here, we chose the arbitrary functions $\kappa$ and $U_\mu$ such that the background actually preserves {\it four} supercharges---in general, it preserves only two. 
In the adapted frame, the Killing spinors are:
\be\label{ks si 1}
\zeta= \mat{0\cr 1}~, \qquad\qquad \t\zeta = \mat{\cos\theta \cr e^{i\varphi} \sin\theta}~, 
\ee
and:
\be\label{ks si 2}
\eta = \mat{1\cr 0}~, \qquad\qquad  \t\eta =\mat{-i e^{-i\varphi} \sin\theta\cr \cos\theta}~.
\ee
We also have the bilinear:
\be
\eta_\mu = {\zeta^\dagger \gamma_\mu \zeta\ov |\zeta|^2}~,\qquad \qquad K^\mu= \t\zeta \gamma^\mu \zeta= {1\ov \beta} \d_x + i \d_\varphi~.
\ee
Note that the Killing vector $K$ is intrinsically complex, in this case---it is not possible to make this $K$ real by a continuous deformation of the THF. This is in contrast with the case of the half-BPS Seifert backgrounds.

\subsection{The $S^2$ supersymmetric index}
The four curved-space supercharges, defined by \eqref{ks si 1}-\eqref{ks si 2}, generate the symmetry $SU(2|1) \times U(1)$, including the bosonic symmetry $SU(2)$ that rotates the $S^2$. One can also consider a more general background that only preserves the two Killing spinors \eqref{ks si 1}  and the $U(1) \times U(1)$ isometry generated by $K$. The resulting $S^2 \times S^1$ partition function computes a supersymmetric index:~\cite{Kim:2009wb,Imamura:2011su}
\be\label{CI 3D index}
\CI_{S^2}({\bf q}, y)= \Tr_{S^2}\Big((-1)^F {\bf q}^{{R\ov 2}+j_3} \prod_\alpha y_\alpha^{-Q_F^\alpha}\Big)~.
\ee
The (generally complex) parameters ${\bf q}$ and $y_\alpha$ are the ``geometric'' and ``flavor'' fugacties, respectively; here, $R$, $j_3$ and $Q_F^\alpha$ denote the $R$-charge, the $S^2$ angular momentum, and the $U(1)_\alpha$ flavor charge, respectively. This index is defined for any theory with a $U(1)_R$ symmetry, where the trace is over the states of the theory quantized on $S^2$. For 3D $\CN=2$ superconformal theories, the states that contribute to \eqref{CI 3D index} satisfy the half-BPS condition:
\be
\Delta = R+ j_3~.
\ee
By the state-operator correspondence, the expansion of \eqref{CI 3D index} in ${\bf q}$ counts $\CN=2$ SCFT operators weighted by their dimension $\Delta$ and by their flavor charges, starting at ${\bf q}^0=1$ for the unit operator. 
For instance, for a free chiral multiplet of unit flavor charge, at the superconformal value $r=\half$ of its $R$-charge, we have:
\be\label{CI Phi exmpl}
\CI^\Phi_{S^2}({\bf q}, y)=1+ {\bf q}^{{1\ov 4}} y^{-1} + {\bf q}^{{1\ov 2}} y^{-2} + {\bf q}^{{3\ov 4}} y^{-3} - {\bf q}^{{3\ov 4}}y + {\bf q}y^{-4} -{\bf q} + \cdots~,
\ee
for the first few terms, corresponding to the half-BPS operators:
\be
{\bf 1}~, \quad \phi~, \quad \phi^2~, \quad \phi^3~, \quad \b\psi_+~, \quad \phi^4~, \quad \phi\b\psi_+~,\quad \cdots~,
\ee
respectively.

One can further generalize the $S^2$ index by turning on non-trivial flux on $S^2$ for background gauge fields coupled to flavor symmetry currents.~\cite{Kapustin:2011jm}

\subsection{Integral formula for the index}
For UV-free gauge theories, the supersymmetric index \eqref{CI 3D index} can be computed explicitly by supersymmetric localization, as a partition function:~\cite{Kim:2009wb,Imamura:2011su,Dimofte:2011py}
\be
Z_{S^2 \times S^1}({\bf q}, y) = \CI_{S^2}({\bf q}, y)~,
\ee
on the supersymmetric background \eqref{S2S1 met round}-\eqref{sugra fields s2s1}.
 Here, we simply present the final result in a set of conventions consistent with the ones we adopted for the other 3D partition functions.
  
\subsubsection{Supersymmetric values for the gauge fields and CS contributions}
One can check that, on this $S^2 \times S^1$ background, the supersymmetric locus for a $U(1)$ vector multiplet takes the form:
\be
A_\mu = \alpha dx + {\m \ov 2}(1- \cos\theta) d\varphi~, \qquad \sigma = -{\m\ov2}~, \qquad D=0~,
\ee
  with $\alpha\in [0,1)$ and $\m\in \Z$ the flat connection on $S^1$ and the flux on $S^2$, namely:
  \be
  e^{-i \int_{S^1} A} = e^{-2\pi i \alpha} \equiv y~, \qquad\quad {1\ov 2\pi} \int_{S^2} dA= \m~,
  \ee
  where we defined the $U(1)$ flavor fugacity $y$, which corresponds to the parameter $y$ in the index.  
  For the $R$-symmetry, we have the analogous parameters:
  \be
   e^{-i \int_{S^1} A^{(R)}}= y_R= {\bf q}^{-\half}~, \qquad\quad \m_R=0~.
  \ee
  One can evaluate the supersymmetric CS terms onto these supersymmetric values. One finds:
  \be
e^{-S_{{\rm CS}, GG}} = (-y)^{k_{GG}\m}~, \qquad 
\ee
for a $U(1)$ CS term at level $k=1$.  Note the minus sign, which is determined by the fact that we are working with the periodic spin structure.~\cite{Closset:2017zgf} (One could easily consider the other spin structure as well, as explained in Ref~\protect\cite{Closset:2018ghr}, similarly to our discussion of the twisted index.)
Similarly, for a mixed $U(1)_1 \times U(1)_2$ CS term at level $k_{12}$, one finds:
\be
e^{-S_{{\rm CS}, G_1G_2}}= y_1^{k_{12} \m_2} y_2^{k_{12} \m_2}~. 
\ee
 We also have:
 \be
e^{-S_{{\rm CS}; RG}} = {\bf q}^{-{k_{RG}\m\ov 2}}~, \qquad e^{-S_{{\rm CS}; RR}}=1~, \qquad e^{-S_{{\rm CS}; {\rm grav}}}=1~.
\ee 
 Note that the $RR$ and gravitational supersymmetric CS terms contribute trivially on this background (and for this particular choice of spin structure).

 \subsubsection{One-loop determinants}
 
 \paragraph{Chiral multiplet contribution.} 
 A chiral multiplet of $R$-charge $r$ and unit $U(1)$ charge has a partition function on $S^2\times S^1$ that reads:
 \be\label{Zphi s2s1}
 Z_{S^2 \times S^1}^\Phi({\bf q}, y; r)_\m= {(y {\bf q}^{ {2-r+ \m\ov 2}}; {\bf q})_\infty\ov (y^{-1} {\bf q}^{{r+ \m \ov 2}}; {\bf q})_\infty}~,
\ee
 in terms of the ${\bf q}$-Pochhammer symbol $(x; {\bf q})_\infty= \prod_{k=0}^\infty (1- x {\bf q}^k)$. Note that, expanding in ${\bf q}$ for $r=\half$ and $\m=0$, we reproduce the index \eqref{CI Phi exmpl}.  It is easy to check the limits:
 \be
 \lim_{y \rightarrow 0}  Z_{S^2 \times S^1}^\Phi({\bf q}, y;r)_\m= 1~, \qquad 
 \lim_{y \rightarrow \infty}  Z_{S^2 \times S^1}^\Phi({\bf q}, y;r)_\m \sim (-y)^{-\m} {\bf q}^{{r-1\ov 2}\m}~.
 \ee
This confirms that \eqref{Zphi s2s1} gives $\Phi$ in the $U(1)_{-\half}$ quantization. 
For a general gauge theory with chiral multiplets sitting in a representation $\FR$, we have the matter contribution:
\be
 Z_{S^2 \times S^1}^{\rm mat}({\bf q}, x, y)_\m =\prod_{\rho, \omega\in \FR \times \FR_F} Z^\Phi_{S^2 \times S^1}({\bf q}, x^\rho y^\omega; r_\omega)_{\rho(\m)}~.
\ee
Here, $x= x_a$ denotes the gauge fugacities for $U(1)_a$ in the maximal torus of the gauge group $\GG$ (defined similarly to the flavor parameters $y_\alpha$).

 \paragraph{Vector multiplet contribution.}  The contribution from the vector multiplet can be obtained more easily by realizing that the W-bosons contribute as chiral multiplets of $R$-charge $2$, in the symmetric quantization. One finds:
\be
 Z_{S^2 \times S^1}^{\rm vec}({\bf q}, x)_\m = \prod_{\alpha \in \Delta_+} {\bf q}^{-{\alpha(\m)\ov 2}}\big(1- x^\alpha {\bf q}^{{\alpha(\m)\ov 2}}\big)\big(1- x^{-\alpha} {\bf q}^{{\alpha(\m)\ov 2}}\big)~.
\ee

 \subsubsection{The integral formula}
 For a gauge theory of the type considered so far, we have the following explicit formula for the $S^2 \times S^1$ partition function:~\cite{Kim:2009wb,Imamura:2011su, Kapustin:2011jm}
\bea\label{Zs2s1 explicit formula}
&Z_{S^2 \times S^1}({\bf q}, y) =\cr
& {1\ov |W_\GG|}\sum_{\m \in \Gamma_{\GG^\vee}}\oint \prod_{a=1}^\rk {d x_a\ov 2 \pi i x_a} \, e^{-S_{\rm CS}({\bf q}, x, y)} \,   Z_{S^2 \times S^1}^{\rm vec}({\bf q}, x)_\m  \,  Z_{S^2 \times S^1}^{\rm mat}({\bf q}, x, y)_\m ~.
\eea
 Here, $y$ denotes all the various flavor fugacities, including the FI terms. Note the sum over all GNO-quantized fluxes for $\GG$ on $S^2$. The contour of integration \eqref{Zs2s1 explicit formula} is taken to be the unit circle, $|x_a|=1$, if all the fugacities $y$ are pure phases and ${\bf q}$ is real---more generally, one can analytically continue the parameters in the integrand while deforming the contour accordingly.

  This formula gives a completely explicit form of the index, although it can be complicated to evaluate in practise, especially due to the sum over magnetic fluxes. In most applications in the literature, one expands \eqref{Zs2s1 explicit formula} at low order in ${\bf q}$ and compute the index order by order.

 \subsection{Examples: Infrared dualities}
 As for the other partition functions, we can use the $S^2$ supersymmetric index to provide non-trivial check of infrared dualities between distinct gauge theories. Let us briefly consider  some of our examples above, in this context.
 
\paragraph{$U(N)_k$ and level/rank duality.} 
 For a $U(N)_k$ supersymmetric CS theory, we have:
 \be
 Z_{S^2 \times S^1}^{{\rm vec}, U(N)}({\bf q}, x)_\m =\prod_{a>b}^N {\bf q}^{-{\m_a-\m_b\ov 2}} \left(1- x_a x_b^{-1} {\bf q}^{{\m_a-\m_b\ov2}}\right)\left(1- x_a^{-1} x_b {\bf q}^{{\m_a-\m_b\ov 2}}\right)~,
\ee
and therefore:
 \be
 Z^{U(N)_k}_{S^2 \times S^1}({\bf q}, y_T) =
 {1\ov N!}\sum_{\m\in \Z^N}\ \oint \prod_{a=1}^N \left[ {d x_a\ov 2 \pi i x_a} (-x_a)^{k\m_a}  x_a^{\n_T} y_T^{\m_a}\right]\,  Z_{S^2 \times S^1}^{{\rm vec}, U(N)}({\bf q}, x)_\m~.
 \ee
 Here, $y_T$ and $\n_T$ and the fugacity and background flux, respectively, for the topological symmetry $U(1)_T$.
 The level-rank duality relations, \eqref{levelrk susy}-\eqref{rel level susy levelrk}, imply the following identities:
 \be
  Z^{U(N)_k}_{S^2 \times S^1}({\bf q}, y_T) = (-y_T)^{-\n_T}\,  Z^{U(k-N)_{-k}}_{S^2 \times S^1}({\bf q}, y_T^{-1})~.
 \ee
  This can be checked explicitly in a number of example, though we are not aware of a general proof. For instance, in the simplest example, we have the $U(1)_1$ theory dual to a trivial theory, and indeed:
 \be\label{U11 id S2S1 index}
   Z^{U(1)_1}_{S^2 \times S^1}({\bf q}, y_T) =\sum_{\m\in \Z}\oint {dx\ov 2 \pi i x} (-1)^\m x^{\m+ n_T} y_T^\m = (-y_T)^{-\n_T}~.
 \ee

\paragraph{Other dualities.}
As another example, consider the elementary mirror symmetry \eqref{EMS duality}-\eqref{kTR kRR rel mirrsym 1}. This implies the following non-trivial  identity:
\be
\sum_{\m\in \Z} \oint  {dx\ov 2 \pi i x}  \, {(x {\bf q}^{ {2-r+ \m\ov 2}}; {\bf q})_\infty\ov (x^{-1} {\bf q}^{{r+ \m \ov 2}}; {\bf q})_\infty}\, (-x)^\m y_T^\m x^{\n_T}  = {\bf q}^{{r\ov 2}\n_T}   {(y_T {\bf q}^{ {r+1+ \n_T\ov 2}}; {\bf q})_\infty\ov (y_T^{-1} {\bf q}^{{1-r+ \n_T \ov 2}}; {\bf q})_\infty}~.
\ee
This formally reduces to \eqref{U11 id S2S1 index} in the limit $y_T \rightarrow \infty$ (and $x \rightarrow 0$).

One can similarly write down the complicated-looking identities implied by a general Aharony duality. In the special case of $U(1)$ SQED dual to the XYZ model, one known proof by direct evalution of the integral~\cite{Kapustin:2011jm} involves the use of highly non-trivial Ramanujan identities. In the general case, a proof of the relevant integral identities was provided in Ref.~\citen{Hwang:2017kmk}, building on earlier work.~\cite{Hwang:2012jh, Hwang:2015wna}, by studying the effective vortex quantum mechanics in each flux sector.

\section*{Acknowledgements} We are especially grateful to Brian Willett for past and current collaboration on the recent work on Seifert manifolds reviewed here, as well as for numerous insightful discussions. C.C. is also indebted to Francesco Benini, Stefano Cremonesi, Thomas Dumitrescu, Guido Festuccia, Zohar Komargodski, Daniel S. Park and Nathan Seiberg for past collaborations and discussions on some of the material presented in these notes.
C.C. is a Royal Society University Research Fellow and a Research Fellow at St John's College, Oxford. 
The work of H.K. is supported by ERC Consolidator Grant 682608 Higgs bundles: Supersymmetric Gauge Theories and Geometry (HIGGSBNDL).

\bibliographystyle{ws-ijmpa}
\bibliography{bib_3D_SUSY_loc}

\end{document}